\documentclass[a4paper,12pt]{article}
\pdfoutput=1 % if your are submitting a pdflatex (i.e. if you have
\usepackage{jheppub_X} % for details on the use of the package, please
\usepackage[T1]{fontenc} % if needed
\usepackage{amsmath}
\usepackage{feynmp-auto}
\usepackage{slashed}
\usepackage{bm}
\usepackage{enumerate}

\usepackage{longtable}
\usepackage{multirow}
\renewcommand{\arraystretch}{1.25}
\usepackage{hhline}
\usepackage{rotating}
\usepackage{array}
\usepackage{float}
\usepackage{afterpage}
\usepackage{anyfontsize}
\usepackage{enumitem}

\usepackage[dvipsnames,table]{xcolor}

\usepackage{arydshln}
\usepackage{blkarray}
\usepackage{physics}
\usepackage{cleveref}
\usepackage{mathtools}
\usepackage{xspace}
\usepackage{setspace}

\usepackage[most]{tcolorbox}
\usepackage{bbold}

\newcommand\Tstrut{\rule{0pt}{3.5ex}}         % = `top' strut
\newcommand\Bstrut{\rule[-2ex]{0pt}{0pt}}   % = `bottom' strut

\def\hs{\rule{1pt}{0pt}} % horizontal space

\def\G{F}

\crefname{table}{Table}{Tables}
\crefname{equation}{Eq.}{Eqs.}
\crefname{appendix}{App.}{Apps.}
\crefname{section}{Sec.}{Secs.}
\crefname{figure}{Fig.}{Figs.}

  {}% \end{boldenv}

\allowdisplaybreaks

\def\beq{\begin{equation}}
\def\eeq#1{\label{#1}\end{equation}}
\def\eeqn{\end{equation}}

\def\beqa{\begin{eqnarray}}
\def\eeqa#1{\label{#1}\end{eqnarray}}
\def\eeqan{\end{eqnarray}}

\def\bseq{\begin{subequations}}
\def\eseq#1{\label{#1}\end{subequations}}
\def\eseqn{\end{subequations}}

\def\li#1{{\color{OliveGreen}{#1}}} % Lorentz index
\def\ci#1{{\color{BrickRed}{#1}}} % color index
\def\wi#1{{\color{Cerulean}{#1}}} % SU(2) weak index
\def\wid{\wi{\mathbb{1}}}
\def\weps{\wi{\bm{\epsilon}}}
\def\epsmat{{\bm{\epsilon}}}
\def\cid{\ci{\mathbb{1}}}

\def\L{{\cal L}}
\def\O{{\cal O}}
\def\R{\bm{R}}
\def\S{{\cal S}}

\def\c{\text{c}}

\def\tr{\mathrm{tr}}
\def\Psl{\slashed{P}}

\def\ie{\emph{i.e.}}
\def\eg{\emph{e.g.}}

\def\mat#1{\boldsymbol{#1}}

\providecommand{\MSbar}{\ensuremath{\overline{\text{MS}}}\xspace}
\def\MS{M}
\def\mH{m}
\def\Sdet{\text{Sdet}}

\newcommand{\s}{\hspace{0.8pt}}

\preprint{CALT-TH-2020-047
}

\title{\Large Functional Prescription for EFT Matching}

\author[1]{Timothy~Cohen,}
\author[1]{Xiaochuan~Lu,}
\author[2]{and Zhengkang~Zhang\hspace{1pt}}

\affiliation[1]{\fontsize{9}{9}\selectfont \,Institute for Fundamental Science, University of Oregon, Eugene, OR 97403}
\affiliation[2]{\fontsize{9}{9}\selectfont \,Walter Burke Institute for Theoretical Physics, California Institute of Technology, Pasadena, CA 91125}

\emailAdd{tcohen@uoregon.edu}
\emailAdd{xlu@uoregon.edu}
\emailAdd{zkzhang@caltech.edu}

\abstract{
We simplify the one-loop functional matching formalism to develop a streamlined prescription.
The functional approach is conceptually appealing: all calculations are performed within the UV theory at the matching scale, and no prior determination of an Effective Field Theory (EFT) operator basis is required.
Our prescription accommodates any relativistic UV theory that contains generic interactions (including derivative couplings) among scalar, fermion, and vector fields.
As an example application, we match the singlet scalar extended Standard Model (SM) onto SMEFT.
}

\begin{document}
\maketitle
\flushbottom
\setcounter{page}{2}
%\newpage

\begin{spacing}{1.1}
\parskip=0ex
%%%%%%%%%%%%%%%%%%%%%%%%%%%%%%%%%%%%%%%%%%%%%%%%%%%%%%%%%%%%%%%%%%%%%%%%%%%%%%%%
\section{Introduction}
\label{sec:intro}
%%%%%%%%%%%%%%%%%%%%%%%%%%%%%%%%%%%%%%%%%%%%%%%%%%%%%%%%%%%%%%%%%%%%%%%%%%%%%%%%

Effective Field Theory (EFT) approaches have wide-ranging applications across many areas of physics, and are especially useful when one encounters a system that has a large hierarchy of dimensionful scales, see \eg\ Refs.~\cite{Rothstein:2003mp,Skiba:2010xn,Petrov:2016azi,deFlorian:2016spz,Brivio:2017vri,Manohar:2018aog,Neubert:2019mrz,Cohen:2019wxr,Penco:2020kvy} for reviews.
An EFT provides a more transparent expression of a theory's IR dynamics with the added benefit that one can systematically sum IR logarithms using renormalization group techniques.
Such frameworks can be useful purely from the bottom up: one specifies the dynamical degrees of freedom along with their transformations under a set of symmetries, and identifies a small power counting parameter to organize the operator expansion.
In this sense EFTs are ``model independent,'' and as such they provide a compelling approach for classifying observables to facilitate comparisons against data.
On the other hand, the EFT paradigm is also useful when applied from the top down.
In scenarios where the (more) fundamental UV description of the system is calculable, one can ``match'' it onto an EFT by ``integrating out'' the heavy states.
This relates the Wilson coefficients in the EFT to the microscopic parameters of the UV theory, and enables the interpretation of experimental measurements and/or constraints on the Wilson coefficients in the context of specific UV models.

Our focus here is on the methodology for matching a UV theory onto an EFT in this top-down approach.
Concretely, we consider a UV theory $\L_\text{UV}[\varphi]$ with a mass hierarchy among the fields $\varphi$:
\vspace{-2pt}
\begin{equation}
\varphi = \left(\Phi, \phi\right) \,, \qquad\text{with}\qquad m_\Phi\gg m_\phi \,,
\vspace{-2pt}
\end{equation}
where we are denoting the heavy (light) fields with $\Phi$ ($\phi$).
We would like to integrate out the heavy fields $\Phi$ to obtain $\L_\text{EFT}[\phi]$.
In this case, the EFT power counting is simply set by the mass ratio $m_\phi/m_\Phi$.
More generally, the discussion that follows may be extended to other cases where the power counting parameter is set by a kinematic restriction, provided there is a clear separation between ``hard'' and ``soft'' modes.

\begin{figure}[t]
\centering
\includegraphics[width=0.95\linewidth]{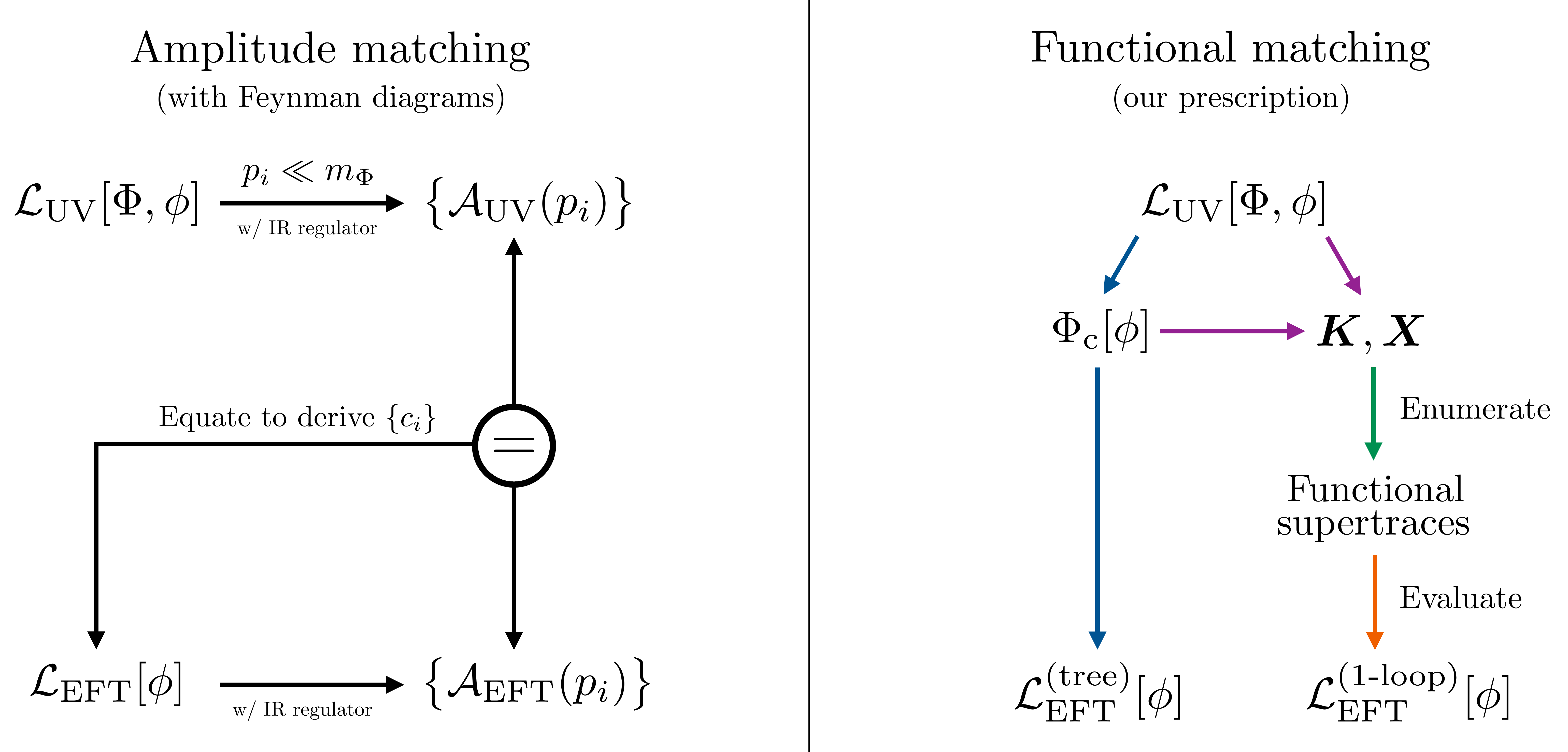}\\[10pt]
\caption{This figure contrasts the procedure one follows for two different approaches to EFT matching.
Amplitude matching [left] requires first working out a basis of EFT operators, and then determining their coefficients $\{c_i\}$ by equating a carefully curated set of low-energy amplitudes that must be computed twice, first using the UV theory and then again using the EFT.
Functional matching [right] provides a more direct route from $\L_\text{UV}$ to $\L_\text{EFT}$, which requires neither constructing an EFT operator basis in advance nor computing low-energy amplitudes.
This paper establishes a concise, readily accessible, four-step prescription (represented by the four colors) for functional matching up to one loop order, as summarized in~\cref{sec:Recipe}.
\label{fig:feynman_vs_functional}}
\end{figure}

A familiar strategy to derive $\L_\text{EFT}[\phi]$ is to match low-energy amplitudes between the UV theory and the EFT, as illustrated in the left panel of Fig.~\ref{fig:feynman_vs_functional}.
In this approach, one must first work out all the EFT operators, leaving only their coefficients $\{c_i\}$ to be determined, and then identify a set of amplitudes to compute (typically via Feynman diagrams) that can be used to solve for all these coefficients.
This procedure is computationally expensive, and typically requires significant human intervention.
Furthermore, it critically relies on performing amplitude calculations, which is conceptually a separate task and requires keeping track of IR details.

%%%
%%%

In this work, we use functional methods to tackle the problem of EFT matching.
Instead of matching individual amplitudes, the idea is to equate their generating functionals, the one-(light-)particle-irreducible (1(L)PI) effective actions:
\begin{equation}
\Gamma_\text{EFT}[\phi] = \Gamma_\text{L,\,UV}[\phi] \,.
\label{eqn:MatchingCondition}
\end{equation}
At tree-level, this yields the familiar result:
\begin{equation}
\left.
\begin{aligned}
\Gamma_\text{L,\,UV}^\text{(tree)}[\phi] &= \S_\text{UV}[\Phi,\phi] \bigr|_{\Phi=\Phi_\text{c}[\phi]} \\[6pt]
\Gamma_\text{EFT}^\text{(tree)}[\phi] &= \S_\text{EFT}^\text{(tree)}[\phi]
\end{aligned}
\;\right\} \quad\Longrightarrow \quad
\L_\text{EFT}^\text{(tree)}[\phi] = \L_\text{UV} [\Phi,\phi] \big|_{\Phi=\Phi_\text{c}[\phi]} \,,
\end{equation}
where $\S \equiv\int \dd^d x\, \L$ denotes the action, and $\Phi_\text{c}[\phi]$ solves the classical equations of motion (EOMs) for the heavy fields:
\begin{equation}
\frac{\delta \S_\text{UV}[\varphi]}{\delta\Phi} \bigg|_{\Phi=\Phi_c[\phi]} = 0 \,.
\vspace{-4pt}
\end{equation}
Obviously, solving the EOMs provides a more direct route to obtain $\L_\text{EFT}^\text{(tree)}[\phi]$ than computing amplitudes.

The efficacy of functional matching extends beyond tree level.
Critically, at one loop, Eq.~\eqref{eqn:MatchingCondition} allows us to systematically solve the matching condition once and for all, and to derive an expression for $\L_\text{EFT}^\text{(1-loop)}$ directly in terms of $\L_\text{UV}$.
This is achieved by using the method of regions~\cite{Beneke:1997zp, Smirnov:2002pj} to split the UV 1LPI effective action into hard and soft region contributions:
\begin{equation}
\Gamma_\text{L,UV}^\text{(1-loop)}[\phi] = \Gamma_\text{L,\,UV}^\text{(1-loop)}[\phi]\Bigr|_\text{hard} + \Gamma_\text{L,\,UV}^\text{(1-loop)}[\phi]\Bigr|_\text{soft} \,,
\label{eqn:matchGamma}
\end{equation}
obtained by expanding all loop integrands assuming the loop momentum $q\sim m_\Phi\gg m_\phi$ and $q\sim m_\phi\ll m_\Phi$, respectively, before performing the integration using dimensional regularization.
On the other hand, the EFT 1PI effective action receives contributions from both operators with one-loop-generated matching coefficients used at tree (classical) level and one-loop amplitudes computed with the tree-level EFT operators:
\begin{equation}
\Gamma_\text{EFT}^\text{(1-loop)}[\phi] = \S_\text{EFT}^\text{(1-loop)}[\phi] + \left( \text{1-loop contributions from } \L_\text{EFT}^\text{(tree)}[\phi] \,\right) \,.
\label{eqn:GammaEFT}
\end{equation}
One can show that the second term in \cref{eqn:matchGamma} is identical to the second term in \cref{eqn:GammaEFT}~\cite{Henning:2016lyp,Fuentes-Martin:2016uol,Zhang:2016pja}.
The matching condition therefore becomes
\begin{equation}
\int \dd^d x\,\L_\text{EFT}^\text{(1-loop)}[\phi] = \Gamma_\text{L,\,UV}^\text{(1-loop)}[\phi]\Bigr|_\text{hard} \,.
\label{eq:S_EFT_loop}
\end{equation}
The intuition here is that a highly virtual loop whose momentum is outside the EFT regime ($q\sim m_\Phi\gg m_\phi$) should be encoded by local operators within the EFT.
Importantly, when using Eq.~\eqref{eq:S_EFT_loop}, one does not have to guess what effective operators will be generated by integrating out the heavy states, and can fully disentangle the tasks of ``matching'' from the IR aspects of amplitude calculations.

Despite these advantages, some technical aspects of one-loop functional matching have only been firmly established recently, as demonstrated in the contexts of the Standard Model EFT (SMEFT) \cite{Henning:2014gca, Henning:2014wua, Drozd:2015rsp, Ellis:2016enq, Wells:2017vla, Ellis:2017jns, Summ:2018oko, Kramer:2019fwz, Ellis:2020ivx, Angelescu:2020yzf}, Higgs EFT (HEFT)~\cite{Cohen:2020xca}, as well as non-relativistic EFTs such as the Heavy Quark Effective Theory (HQET)~\cite{Cohen:2019btp}.
The goal of this paper is to make these technical advances more easily accessible by devising a streamlined prescription.
Our new formulation, summarized in the right panel of~\cref{fig:feynman_vs_functional}, highlights the simplicity and efficiency at the core of the functional approach.
It can be applied to integrating out \emph{any} perturbative UV states in a relativistic theory, regardless of the interaction structure.

The rest of this paper is organized as follows.
We begin in \cref{sec:Supertraces} by setting up the framework for functional matching calculations.
As we will show, the key objects are a set of {\it functional supertraces} that take very specific forms.
One-loop matching is essentially reduced to ($i$) \emph{enumerating} the relevant supertraces, and ($ii$) \emph{evaluating} them to simultaneously obtain the effective operators and their coefficients.
In \cref{sec:Enumerating}, we develop step ($i$) and show how the infinite series of functional supertraces can be organized graphically in the spirit of Ref.~\cite{Zhang:2016pja} (though the graphs presented here are technically different).
In \cref{sec:Evaluating}, we explain step ($ii$).
This is usually the most tedious step, but given its algorithmic nature, we have developed a \texttt{Mathematica} package that automates the process using the covariant derivative expansion (CDE) technique~\cite{Gaillard:1985uh, Chan:1986jq, Cheyette:1987qz}; see our forthcoming paper~\cite{TraceCode} for details.
All these steps can be summarized into a simple practical prescription, which we present in \cref{sec:Recipe} as the central result of this work.
To demonstrate the prescription in detail, we reproduce the results for an example of phenomenological interest in \cref{sec:Singlet}: matching the singlet scalar extend SM onto SMEFT up to dimension six, which was first studied comprehensively in Ref.~\cite{Jiang:2018pbd} and later cross checked by Ref.~\cite{Haisch:2020ahr}.
Finally, we conclude in \cref{sec:Outlook} and discuss some future directions.

The streamlined approach presented here improves upon previous works on one-loop functional matching.
The interested reader can find a brief comparison to the recent literature in \cref{appsec:Relation}.
Technical details for the matching example in \cref{sec:Singlet} are provided in~Apps.~\ref{appsec:Xmatrix}, \ref{appsec:TraceList}, and \ref{sec:SingletSummary}.

%%%%%%%%%%%%%%%%%%%%%%%%%%%%%%%%%%%%%%%%%%%%%%%%%%%%%%%%%%%%%%%%%%%%%%%%%%%%%%%%
\section{One-loop Matching From Functional Supertraces}
\label{sec:Supertraces}
%%%%%%%%%%%%%%%%%%%%%%%%%%%%%%%%%%%%%%%%%%%%%%%%%%%%%%%%%%%%%%%%%%%%%%%%%%%%%%%%

Similar to the classic Coleman-Weinberg potential calculation, the one-loop 1LPI effective action, $\Gamma_\text{L,\,UV}^\text{(1-loop)}$ in \cref{eq:S_EFT_loop} yields the logarithm of a functional superdeterminant:\footnote{A superdeterminant ``Sdet'' is a generalization of the regular determinant by stipulating an inverse power for the eigenvalues in fermionic blocks of the matrix. Similarly, a supertrace ``STr'' generalizes the regular trace by assigning a minus sign for fermionic blocks of a matrix.}
\begin{equation}
\int \dd^d x \,\L_\text{EFT}^\text{(1-loop)}[\phi]
\;=\; \frac{i}{2}\, \log\text{Sdet} \Biggl(-\frac{\delta^2 \S_\text{UV}}{\delta\varphi^2}\biggr|_{\Phi=\Phi_\text{c}[\phi]}\Biggr)\Biggr|_\text{hard} \,,
\label{eqn:LEFTStr}
\end{equation}
where $\Phi_\c[\phi]$ is the solution to the heavy fields' EOMs.
The functional derivative here generally consists of an inverse propagator part and an interaction part, such that
\begin{equation}
\int \dd^d x \,\L_\text{EFT}^\text{(1-loop)}[\phi]
= \frac{i}{2}\,\log\text{Sdet} \bigl(\bm{K} - \bm{X}\bigr)\Big|_\text{hard} = \frac{i}{2}\,\text{STr} \log \bigl( \bm{K} - \bm{X}\bigr)\Big|_\text{hard} \,.
\label{eq:K_minus_X}
\end{equation}
We will explain the derivation of this equation shortly, and discuss the forms of the ``inverse propagator matrix'' $\bm{K}$ and ``interaction matrix'' $\bm{X}$ (see Eqs.~\eqref{eq:Kphi} and \eqref{eq:Xmat_expand} below).
Writing $\bm{K}-\bm{X} = \bm{K}\, (\bm{1}-\bm{K}^{-1}\bm{X})$ and Taylor expanding $\log(\bm{1}-\bm{K}^{-1}\bm{X})$, we obtain our central formula for one-loop matching:
\vspace{5pt}
\begin{center}
	\begin{minipage}{0.98\textwidth}
		\begin{tcolorbox}[colback=white]
			\vspace{-12pt}
			\begin{equation}
			\hspace{-8pt}\int \dd^d x \,\L_\text{EFT}^\text{(1-loop)}[\phi] = \frac{i}{2}\,\text{STr}\log \bm{K} \Bigr|_\text{hard}
			- \frac{i}{2}\sum_{n=1}^\infty \frac{1}{n} \,\text{STr} \Bigl[\bigl(\bm{K}^{-1}\bm{X}\bigr)^n\Bigr]\Bigr|_\text{hard} \,.
			\label{eq:separate_STr}
			\end{equation}
		\end{tcolorbox}
	\end{minipage}
\end{center}
\vspace{5pt}
This expresses the EFT Lagrangian as a sum over two different types of supertraces, which we shall call ``log-type'' and ``power-type,'' respectively.
As we will see, the $\bm{X}$ matrix is derived from taking the second derivative of three- and higher-point interactions in the UV theory, so it contains terms with at least one power of $\phi$ and has a (canonical) operator dimension $\ge 1$; meanwhile, each $\bm{K}^{-1}$ contributes an operator dimension $\ge 0$.
This means that only a finite number of terms in the infinite series of power-type supertraces (the sum over $n$ in~\cref{eq:separate_STr}) contribute to EFT operators up to a certain dimension.
Hence, we can truncate the series according to the desired order in the EFT Lagrangian, \eg\ up to dimension six for the SMEFT application in \cref{sec:Singlet}.

In the rest of this section, we fill in the steps from Eq.~\eqref{eqn:LEFTStr} to Eq.~\eqref{eq:K_minus_X}, and discuss the general forms of the $\bm{K}$ and $\bm{X}$ matrices.

\subsection*{Field multiplet}

The functional derivative in Eq.~\eqref{eqn:LEFTStr} should be taken with respect to all the independent fields that appear in the path integral measure.
For example, a complex scalar $s$ and a Dirac fermion $f$ are each represented by a pair of fields with conjugate quantum numbers:
\begin{equation}
\varphi_s = \mqty(s \\ s^*) \,,\qquad
\varphi_f = \mqty(f \\ f^\c) \,.
\label{eqn:RealComponents}
\end{equation}
Here $f^\c\equiv -i\gamma^2 f^*$ is the charge conjugated fermion; note that both $f$ and $f^\c$ are (4-component) Dirac spinors.
Meanwhile, it is convenient to define a set of conjugate fields $\bar\varphi$. For example, for a complex scalar and a Dirac fermion, we define
\begin{equation}
\bar\varphi_s \equiv
\begin{pmatrix}
\,s^\dagger & & s^T\,
\end{pmatrix}
= \varphi_s^T
\begin{pmatrix}
\;0\; & \;1\;\; \\
\;1\; & \;0\;\;
\end{pmatrix}
\,,\qquad
\bar\varphi_f \equiv
\begin{pmatrix}
\,\bar f & & \bar f^\c\,
\end{pmatrix}
= \varphi_f^T
\begin{pmatrix}
0 & i\gamma^0\gamma^2 \\
i\gamma^0\gamma^2 & 0
\end{pmatrix}\,.
\end{equation}
On the other hand, if $s$ were a real scalar (or vector) and $f$ were a Majorana fermion, we would have $\varphi_s=s$, $\bar\varphi_s = s^T$ and $\varphi_f =f$, $\bar\varphi_f=\bar f$.
Generally, $\bar\varphi$ contains the same independent fields as $\varphi$, but with different ordering, and we can write
\begin{equation}
\bar\varphi = \varphi^T \bm{R}\,, \qquad\text{with}\qquad\;\; \bigl|\,\Sdet\R\,\bigr| = 1\,.
\end{equation}

\subsection*{Inverse propagator matrix $\bm{K}$}

When written in terms of $\bar\varphi$ and $\varphi$, the kinetic and mass terms in the (relativistic) UV Lagrangian take the familiar block-diagonal form:
\begin{equation}
\L_\text{UV}\supset %\L_{\bm{K}} =
\frac12\, \bar\varphi\, \bm{K}\, \varphi =  \frac12\,\sum_i \bar\varphi_i\, K_i\, \varphi_i \,,
\label{eq:L_K}
\end{equation}
with\footnote{If there is kinetic or mass mixing between the fields in the UV theory, we first rotate it away. Also, for a non-renormalizable UV theory, which is an EFT itself, there could be terms like $\bar\varphi_i\, D^k \varphi_i$ with $k>2$. However, they can be traded for terms with fewer powers of covariant derivatives via a basis change using the EOMs, so that $K_i$ can still be written in  the form of Eq.~\eqref{eq:Kphi}.}
\begin{equation}
K_i =
\begin{cases}
P^2 - m_i^2 & \big(\text{spin-}0\big) \\[3pt]
\Psl - m_i & \big(\text{spin-}\frac{1}{2}\big) \\[3pt]
-\eta^{\mu\nu} (P^2 - m_i^2) + \bigl(1-\frac{1}{\xi}\bigr) P^\mu P^\nu & \big(\text{spin-}1\big)
\end{cases}\quad.
\label{eq:Kphi}
\end{equation}
Here, we have introduced the notation
\begin{align}
P_\mu\equiv iD_\mu \,.
\end{align}
This is the Hermitian version of a covariant derivative, as can be seen from $(A P_\mu B )^\dagger = (A\, iD_\mu B )^\dagger = \left(-iD_\mu B^\dagger\right) A^\dagger \overset{\text{IBP}}{=} B^\dagger iD_\mu A^\dagger = B^\dagger P_\mu A^\dagger$, where $A$ and $B$ are arbitrary operators, and we have used integration by parts (IBP).
When $\varphi_i$ represents a pair of conjugate fields, as in the case of a complex scalar $s$ or a Dirac fermion $f$ in Eq.~\eqref{eqn:RealComponents}, a $2\times 2$ identity matrix in this field space is implicitly understood in Eq.~\eqref{eq:Kphi}; the kinetic and mass terms are written in a symmetric way between the two fields:
\begin{subequations}
\begin{eqnarray}
|D_\mu s|^2 -m_s^2 |s|^2
&=&
\frac{1}{2}\hs s^\dagger (P^2-m_s^2) \hs s  +\frac{1}{2}\hs s^T (P^2-m_s^2) \hs s^*
= \frac{1}{2}\hs \bar\varphi_s (P^2-m_s^2) \hs\mathbb{1}\hs \varphi_s\, ,\qquad\quad\\[5pt]
\bar f\hs (i\slashed{D} -m_f)\, f
&=& \frac{1}{2}\, \bar f\hs (\Psl -m_f)\, f +\frac{1}{2}\, \bar f^\c\hs (\Psl -m_f)\, f^\c = \frac{1}{2} \,\bar\varphi_f\hs (\Psl -m_f)\,\mathbb{1}\,\varphi_f \,,
\end{eqnarray}
\end{subequations}
where IBP has been used to make each $P_\mu$ act to the right.

Note that we work with four-component spinors for the spin-$\frac{1}{2}$ case, hence the appearance of $\Psl\equiv \gamma^\mu P_\mu$. This obviously applies to Majorana fermions and Weyl fermions that form Dirac pairs. The case of chiral fermions can also be accommodated by introducing auxiliary fields as their Dirac partners, which we discuss in detail in \cref{sec:Singlet}. In the spin-1 case, $\xi$ is the gauge fixing parameter. In practice, it is convenient to choose $\xi=1$, so that $K_i$ for a spin-1 field takes the same form as for a spin-0 field. We will adopt this gauge throughout this paper.

\subsection*{Interaction matrix $\bm{X}$}

In order to define the interaction matrix $\bm{X}$, let us go back to Eq.~\eqref{eqn:LEFTStr} and compute the second variation:
\begin{equation}
\delta^2 \L_\text{UV} = 2\,\L_\text{UV}[\varphi+\delta\varphi] \Big|_{\O(\delta\varphi^2)} \equiv \delta\bar\varphi\, (\bm{K} -\bm{X}_\text{UV})\, \delta\varphi
= \delta\varphi^T \R\,(\bm{K}-\bm{X}_\text{UV}) \,\delta\varphi\,.
\label{eqn:LXvariation}
\end{equation}
Here $\varphi$ is the classical background field and $\delta\varphi$ captures its quantum fluctuations.
Note that for a gauge field, we gauge fix $\delta\varphi$, while maintaining the gauge invariance for $\varphi$, as is standard when using the background field method.
From Eq.~\eqref{eq:L_K}, we anticipate the appearance of $\bm{K}$ in Eq.~\eqref{eqn:LXvariation}; the rest is then collected into the UV interaction matrix $\bm{X}_\text{UV}$.
Since the functional superdeterminant in Eq.~\eqref{eqn:LEFTStr} is evaluated with $\Phi=\Phi_\c[\phi]$, we define
\begin{equation}
\bm{X} \equiv \bm{X}_\text{UV} \bigr|_{\Phi=\Phi_\c[\phi]} \,,
\label{eq:X_mat_def}
\end{equation}
which then only depends on the light background fields $\phi$.
Note that there is no distinction between quantities before and after setting $\Phi=\Phi_\c[\phi]$ for the inverse propagator part, since $\bm{K}$ does not depend on the heavy background fields.
At this point, we can substitute Eqs.~\eqref{eqn:LXvariation} and \eqref{eq:X_mat_def} into \cref{eqn:LEFTStr} and obtain
\begin{align}
\int \dd^d x \,\L_\text{EFT}^\text{(1-loop)}[\phi]
&= \frac{i}{2}\,\log\text{Sdet} \Bigl[-\R\,\bigl(\bm{K} - \bm{X}\bigr)\Bigr]\Big|_\text{hard} \,,
\label{eqn:STrlogQ}
\end{align}
which yields \cref{eq:K_minus_X} up to an irrelevant constant.

Next, we discuss the structure of the $\bm{X}$ matrix.
Without loss of generality, it can be cast in the form:
\beq
\bm{X}(\phi, P_\mu) = \bm{U}[\phi] + \bigl(P_\mu \bm{Z}^\mu[\phi] +\bm{\bar{Z}}^\mu[\phi] P_\mu\bigr) + \cdots \,.
\eeq{eq:Xmat_expand}
As operators, the explicit factors of $P_\mu$ that multiply $\bm{Z}^\mu$ and $\bm{\bar{Z}}^\mu$ should be understood as ``open'' covariant derivatives that act on everything to their right (same for the $P_\mu$'s in \cref{eq:Kphi} above).
These are in contrast with ``closed'' covariant derivatives that can be written as commutators: $[P_\mu,\phi]=i(D_\mu\phi)$ with $D_\mu$ acting on $\phi$ alone.
The $\bm{U}$, $\bm{Z}^\mu$ and $\bm{\bar{Z}}^\mu$ matrices, written here as functionals of $\phi$, can contain closed covariant derivatives:
\begin{align}
\bm{U}[\phi] &= \bm{U}\big( \,\phi\,,\; [P_\mu,\phi]\,,\; [P_\mu, [P_\nu,\phi]]\,,\;\dots \,\big) \notag\\[3pt]
&= \bm{U}\big( \,\phi\,,\; i\,(D_\mu\phi)\,,\; i^2\,(D_\mu D_\nu\phi)\,,\; \dots \,\big) \,.
\end{align}

We emphasize that \cref{eq:Xmat_expand} is an expansion in the number of open covariant derivatives, and we have only written out the first two orders explicitly. Additional functionals of $\phi$ appear in the higher order terms represented by ``$\dots$''.
Technically, this expansion is not unique, since a closed covariant derivative can always be rewritten in terms of open covariant derivatives: $[P_\mu,\phi]=P_\mu\phi -\phi P_\mu$.
Typically, the calculation is more involved when more open covariant derivatives appear, so it is desirable to write the $\bm{X}$ matrix in a form that has the fewest possible open covariant derivatives.

In many practical matching calculations, the UV theory does not contain any derivative interactions, and we simply have $\bm{X}(\phi, P_\mu) = \bm{U}[\phi]$.
More generally, derivative interactions often involve a relatively small subset of fields in the UV theory, so we still have $X_{ij}=U_{ij}[\phi]$ for many blocks of the $\bm{X}$ matrix.
However, we reiterate that the utility of functional methods (in particular the prescription presented in this work) does not rely on the series in Eq.~\eqref{eq:Xmat_expand} truncating after the first or second order; derivative interactions with any number of open covariant derivatives are all accommodated.

%%%%%%%%%%%%%%%%%%%%%%%%%%%%%%%%%%%%%%%%%%%%%%%%%%%%%%%%%%%%%%%%%%%%%%%%%%%%%%%%
\section{Enumerating Supertraces}
\label{sec:Enumerating}
%%%%%%%%%%%%%%%%%%%%%%%%%%%%%%%%%%%%%%%%%%%%%%%%%%%%%%%%%%%%%%%%%%%%%%%%%%%%%%%%

Starting from the central formula \cref{eq:separate_STr}, the remaining tasks are clear. We need to {\it enumerate} the functional supertraces that contribute to the specific matching calculation of interest and then {\it evaluate} them, the subjects of this and the next section, respectively.

\subsection*{Log-type supertraces}
\label{sec:EnumLog}

Since the $\bm{K}$ matrix is block-diagonal, the first term in \cref{eq:separate_STr} becomes a simple sum over the $K_i$ blocks given in \cref{eq:Kphi}, each corresponding to a field in the UV theory.
For the light fields $\phi$, isolating the hard region contribution yields scaleless loop integrals, which vanish in dimensional regularization.
For the heavy fields $\Phi$, on the other hand, it is the soft region contribution that yields vanishing scaleless integrals.
Thus,
\begin{equation}
\frac{i}{2}\,\text{STr}\log \bm{K} \Bigr|_\text{hard} = \frac{i}{2}\sum_{i\,\in\,\{\Phi\}} \text{STr}\log K_i\,\Bigr|_\text{hard}  = \frac{i}{2}\sum_{i\,\in\,\{\Phi\}} \text{STr}\log K_i \,.
\label{eqn:Type1detail}
\end{equation}
Moreover, only the heavy fields that are charged under the EFT gauge group need to be included; otherwise, if $P_\mu=i\partial_\mu$, the supertrace would evaluate to a constant.
Therefore, enumerating the log-type supertraces amounts to identifying the heavy fields $\Phi$ in the UV theory that are charged under the EFT gauge group.

%%%%%%%%%%%%%%%%%%%%%%%%%%%%%%%%%%%%%%%%%%%%%%%%%%%%%%%%%%%%%%%%%%%%%%%%%%%%%%%%
\subsection*{Power-type supertraces}
\label{sec:EnumPower}
%%%%%%%%%%%%%%%%%%%%%%%%%%%%%%%%%%%%%%%%%%%%%%%%%%%%%%%%%%%%%%%%%%%%%%%%%%%%%%%%

The second term in~\cref{eq:separate_STr} can be written in terms of the blocks of the $\bm{K}$ and $\bm{X}$ matrices. Taking into account that $\bm{K}$ is block-diagonal, we have
\begin{align}
-\frac{i}{2}\frac{1}{n}\,\text{STr} \Bigl[\bigl(\bm{K}^{-1}\bm{X}\bigr)^n\Bigr] = -\frac{i}{2}\frac{1}{n} \sum_{i_1, \cdots, i_n}\text{STr} \bigg[ \frac{1}{K_{i_1}} X_{i_1 i_2}\, \frac{1}{K_{i_2}} X_{i_2 i_3} \cdots \frac{1}{K_{i_n}} X_{i_n i_1} \bigg] \,.
\label{eqn:Type2detail}
\end{align}
The structure on the RHS admits an intuitive graphical representation.
We draw lines for ``propagators'' $\frac{1}{K_i}$ (we remind the reader that these are functional operators, not the momentum space Feynman propagators), and nodes for interactions $X_{ij}$.
Concretely, we define
\beq
\parbox[c][60pt][c]{60pt}{\centering
	\begin{fmffile}{general}
		\begin{fmfgraph*}(40,40)
			\fmfsurround{x12,p1,xn1,pn,pd,x3d,p3,x23,p2}
			\fmf{plain,left=0.22}{pn,xn1,p1,x12,p2,x23,p3}
			\fmf{dots,left=0.22}{p3,x3d,pd,pn}
			\fmfv{decor.shape=circle,decor.filled=empty,decor.size=2.5thick}{x12,x23,xn1}
			\fmfv{label=$\varphi_{i_1}$,l.d=2pt}{p1}
			\fmfv{label=$\varphi_{i_2}$,l.d=2pt}{p2}
			\fmfv{label=$\varphi_{i_3}$,l.d=2pt}{p3}
			\fmfv{label=$\varphi_{i_n}$,l.d=2pt}{pn}
		\end{fmfgraph*}
	\end{fmffile}
} \,\, \equiv\, -\frac{i}{2}\,\frac{1}{r}\, \text{STr} \bigg[
\frac{1}{K_{i_1}} X_{i_1 i_2}\, \frac{1}{K_{i_2}} X_{i_2 i_3} \cdots \frac{1}{K_{i_n}} X_{i_n i_1}
\bigg]  \biggr|_\text{hard} \,,
\label{eqn:CovariantGraph}
\eeqn
where the indices $i_1, \cdots, i_n$ are {\it not} summed over.
Here $\frac{1}{r}$ is a symmetry factor accounting for a possible $\mathbb{Z}_{r}$ symmetry of the graph under rotation.
For a generic set $\{i_1,\cdots,i_n\}$, the $n$ cyclic permutations are distinct, and the terms that they represent in the sum in \cref{eqn:Type2detail} all yield identical results upon evaluating the supertrace, so the $\frac{1}{n}$ prefactor is fully canceled, and $r=1$.
On the other hand, if the graph has a non-trivial $\mathbb{Z}_{r}$ symmetry under rotation, there would only be $\frac{n}{r}$ distinct cyclic permutations, leaving a prefactor $\frac{1}{r}$ in \cref{eqn:CovariantGraph}.

Enumerating the power-type supertraces therefore amounts to enumerating distinct graphs of the form in \cref{eqn:CovariantGraph}. Note that there is only one graph topology, so this enumeration is quite simple.
We just need to keep track of the minimum operator dimension contained in each $X_{ij}$, and draw graphs where the sum of these numbers does not exceed the desired maximum EFT operator dimension (\eg\ six).
The minimum operator dimension of $X_{ij}$ is determined by $U_{ij}[\phi]$, $Z_{ij}^\mu[\phi]$, $\bar{Z}_{ij}^\mu[\phi]$, etc.\ following the expansion in Eq.~\eqref{eq:Xmat_expand}, without counting open covariant derivatives $P_\mu$ (which, just like $K_i^{-1}$, will yield a series of terms starting at dimension zero upon evaluation).
Also, each graph must contain at least one heavy propagator, since a loop involving only light fields yields scaleless integrals upon isolating the hard region.

%%%%%%%%%%%%%%%%%%%%%%%%%%%%%%%%%%%%%%%%%%%%%%%%%%%%%%%%%%%%%%%%%%%%%%%%%%%%%%%%
\section{Evaluating Supertraces}
\label{sec:Evaluating}
%%%%%%%%%%%%%%%%%%%%%%%%%%%%%%%%%%%%%%%%%%%%%%%%%%%%%%%%%%%%%%%%%%%%%%%%%%%%%%%%

We now move on to the next step, evaluating functional supertraces.
This is an isolated problem that can be solved in a variety of ways.
For example, one can appeal to traditional momentum-space Feynman diagrams, see \emph{e.g.}\ Secs.~9.5, 11.4 and 16.6 of Ref.~\cite{Peskin:1995ev}.
On the other hand, the covariant derivative expansion (CDE) provides a more efficient approach, and we will use it here, mostly following Ref.~\cite{Henning:2014wua} and App.~B of Ref.~\cite{Cohen:2019btp}.
We aim to provide a high-level summary in this section, and refer the reader to these references for technical details.
In particular, we focus on using simple examples to illustrate what kind of results to expect from a CDE evaluation.
More involved supertraces are evaluated in the same manner.
The tedium of supertrace evaluation grows rapidly as the calculation extends to higher operator dimensions, and/or supertraces with more complicated structures.
To facilitate this process, and to make functional matching fully accessible to matching practitioners who are not necessarily familiar with the technical details of CDE, we have authored a \texttt{Mathematica} package that automates the CDE evaluation of all functional supertraces relevant for one-loop matching between relativistic theories, to be presented in a forthcoming paper~\cite{TraceCode}.

\subsection*{Log-type supertraces}
\label{sec:EvalLog}

The evaluation of log-type supertraces in \cref{eqn:Type1detail} is universal across all UV theories.
From \cref{eq:Kphi}, we see that there are essentially only two scenarios (taking $\xi=1$ for the spin-1 case): $\text{STr}\log\left(P^2-m^2\right)$ and $\text{STr}\log\left(\slashed P-m\right)$. These can be directly evaluated with CDE techniques, and will yield an infinite series of effective operators.
Since the $K_i$'s only depend on covariant derivatives $P_\mu=iD_\mu=i\partial_\mu+g_a G_\mu^a T^a$ (with $a$ summed over the gauge group generators), the resulting EFT operators can only involve the gauge field strength of light vectors,
\begin{equation}
\G_{\mu\nu} \equiv -i \comm{P_\mu}{P_\nu} = g_a G_{\mu\nu}^a T^a \,,
\label{eqn:Gdef}
\end{equation}
and their covariant derivatives.
Each $P_\mu=iD_\mu$ has dimension one and each $\G_{\mu\nu}$ has dimension two, so the operator dimension truncation is straightforward.
Here we show the results up to dimension six, while noting that the same CDE procedure can be applied to derive operators at dimension eight and higher (represented by ``$\dots$''):
\begin{subequations}\label{eqn:LogTypes}
\begin{align}
\frac{i}{2}\, \text{STr} \log \left( P^2 - m^2 \right) &= \int \dd^d x\, \frac{1}{16\s\pi^2}\,\tr \Bigg\{ \left( \log\frac{m^2}{\mu^2} \right) \frac{1}{24} \G_{\mu\nu} \G^{\mu\nu} \notag\\[3pt]
&\hspace{-40pt} + \frac{1}{m^2} \left[ -\frac{1}{120} \left(D^\mu F_{\mu\nu}\right)\left(D_\rho F^{\rho\nu}\right) -\frac{1}{180}\,i\, F_\mu{}^\nu F_\nu{}^\rho F_\rho{}^\mu \right] + \dots
\Bigg\} \,, \label{eqn:STrK1} \\[20pt]
\frac{i}{2}\, \text{STr} \log \left( \slashed{P} - m \right) &= \int \dd^d x\, \frac{1}{16\s\pi^2}\,\tr \Bigg\{ \left( \log\frac{m^2}{\mu^2} \right) \frac{1}{24} \G_{\mu\nu} \G^{\mu\nu} \notag\\[3pt]
&\hspace{-40pt} + \frac{1}{m^2} \left[ -\frac{1}{60} \left(D^\mu F_{\mu\nu}\right)\left(D_\rho F^{\rho\nu}\right)  + \frac{1}{360}\,i\, F_\mu{}^\nu F_\nu{}^\rho F_\rho{}^\mu \right] + \dots
\Bigg\} \,. \label{eqn:STrK2}
\end{align}
\end{subequations}
Here and throughout this paper, we use dimensional regularization with the $\MSbar$ scheme; $\mu$ is the renormalization scale, and we have dropped the $\frac{1}{\epsilon}$ poles and associated finite terms accompanying the logarithms that will be cancelled by the $\MSbar$ counterterms.
We have assumed physical spin-statistics relations when addressing the ``super'' aspect of the traces, \ie, $\left(P^2-m^2\right)$ comes from commuting fields and $\left(\slashed P - m\right)$ comes from anticommuting fields.
An exception is the Faddeev-Popov ghosts, which are anticommuting Lorentz scalars whose inverse propagator is $(P^2-m^2)$; in this case, one should multiply the RHS of \cref{eqn:STrK1} by an extra minus sign.

Note that the operators $(P^2-m^2)$ and $(\Psl-m)$ (as well as their logarithms) are acting on the field $\Phi(x)$.
Therefore, as matrices they are acting on the giant vector space labeled by both the components of $\Phi$ and the spacetime coordinate $x$, namely their direct product space.
In \cref{eqn:LogTypes}, we have carried out the (more complicated) trace over the infinite-dimensional subspace labeled by $x$; this is the ``functional part'' of the supertrace.
The remaining trace ``$\tr$'' in the results is taken over the finite-dimensional space formed by all the components of $\Phi$, including its components in the field multiplet $\varphi$, spin indices, gauge indices, etc.
Concretely, this remaining trace is over three sets of $\Phi$ indices and can be schematically written as
\begin{equation}
\tr = \tr_\varphi \times \tr_\text{Lorentz} \times \tr_G \,.
\label{eq:tr_def}
\end{equation}
The first two traces are over the components in the field multiplet $\varphi$ and the Lorentz representation space, respectively.
These are trivial in the present case: the operators on the RHS of \cref{eqn:LogTypes} are proportional to the identity element in each of these spaces, so these traces simply count the number of independent fields in the path integral measure $n_\varphi$ and the number of Lorentz components $n_\text{Lorentz}$.
For example, $n_\varphi = 1$ for a real scalar or a Majorana fermion, and $n_\varphi = 2$ for a complex scalar or a Dirac fermion; $n_\text{Lorentz}$ equals 1, 4,  and  $d=4-\epsilon$ for scalars, fermions and vectors, respectively.
The third trace, $\tr_G$, is over internal gauge indices, and is evaluated with the covariant derivatives and field strengths inheriting their representations from $\Phi$.
For the common case of a simple Lie group with associated gauge coupling $g$, we have $\G_{\mu\nu}=g\,G_{\mu\nu}^a T_\Phi^a$, and therefore
\begin{subequations}\label{eqn:trGs}
	\begin{align}
	\tr_G\bigl(F_{\mu\nu}F^{\mu\nu}\bigr) &= C_\Phi \,g^2 \,G_{\mu\nu}^a G^{a\mu\nu} \,,\\[8pt]
	\tr_G\bigl[ \left(D^\mu F_{\mu\nu}\right) \left(D_\rho F^{\rho\nu}\right) \bigr] &= C_\Phi \,g^2\bigl(D^\mu G_{\mu\nu}^a\bigr)^2\,,\\[5pt]
	\tr_G\bigl( iF_\mu{}^\nu F_\nu{}^\rho F_\rho{}^\mu \bigr) &= -C_\Phi \,\frac12\, g^3 f^{abc}{G^a_\mu}^\nu {G^b_\nu}^\rho {G^c_\rho}^\mu \,,
	\end{align}
\end{subequations}
where $C_\Phi$ is the group invariant defined by $\tr_G(T_\Phi^a T_\Phi^b)=C_\Phi\,\delta^{ab}$.

\begin{table}[t!]
	\begin{small}
	\renewcommand{\arraystretch}{1.5}
	\setlength{\arrayrulewidth}{.3mm}
	\setlength{\tabcolsep}{0.8 em}
	\begin{center}
		\begin{tabular}{c|ccc}
			\hline
			\rowcolor[HTML]{EFEFEF}
			& \multicolumn{3}{c}{{\bf Operator coefficients $\times\; 16\s\pi^2$}} \\
			\rowcolor[HTML]{EFEFEF}
			\multirow{-2}{*}{\parbox{90pt}{\rule{0pt}{3ex} \centering{\bf Integrate out\\[2pt] a heavy ...}}}
			& $\tr_G\bigl(F_{\mu\nu}F^{\mu\nu}\bigr)$ & $\tr_G\bigl[ \left(D^\mu F_{\mu\nu}\right)\left(D_\rho F^{\rho\nu}\right)\bigr]$ & $\tr_G\bigl( iF_\mu{}^\nu F_\nu{}^\rho F_\rho{}^\mu \bigr)$ \Bstrut\\		
			\hline
			real scalar    & $\frac{1}{24}\,\log\frac{m^2}{\mu^2}$ & $-\frac{1}{120}\,\frac{1}{m^2}$ & $-\frac{1}{180}\,\frac{1}{m^2}$ \Tstrut\Bstrut\\
			complex scalar & $\frac{1}{12}\,\log\frac{m^2}{\mu^2}$ & $-\frac{1}{60}\,\frac{1}{m^2}$  & $-\frac{1}{90}\,\frac{1}{m^2}$  \Tstrut\Bstrut\\
			Majorana fermion   & $\frac{1}{6}\,\log\frac{m^2}{\mu^2}$  & $-\frac{1}{15}\,\frac{1}{m^2}$  & $\frac{1}{90}\,\frac{1}{M^2}$   \Tstrut\Bstrut\\
			Dirac fermion  & $\frac{1}{3}\,\log\frac{m^2}{\mu^2}$  & $-\frac{2}{15}\,\frac{1}{m^2}$  & $\frac{1}{45}\,\frac{1}{m^2}$   \Tstrut\Bstrut\\
			real vector         & $\frac16\left(\log\frac{m^2}{\mu^2}+\frac12\right)$ & $-\frac{1}{30}\,\frac{1}{m^2}$ & $-\frac{1}{45}\,\frac{1}{M^2}$ \Tstrut\Bstrut\\
			ghost          & $-\frac{1}{12}\,\log\frac{m^2}{\mu^2}$& $\frac{1}{60}\,\frac{1}{m^2}$ & $\frac{1}{90}\,\frac{1}{m^2}$   \Tstrut\Bstrut\\
			\hline
		\end{tabular}
	\end{center}
	\end{small}
\caption{Universal results for log-type supertraces up to dimension six.}
\label{tab:LogTypeEval}\vspace{20pt}
\end{table}

We summarize the coefficients of these EFT operators (up to a common factor of $\frac{1}{16\s\pi^2}$) that result from evaluating log-type supertraces for the various types of fields in \cref{tab:LogTypeEval}, as a convenient reference.
As a technical note, in the real vector case, when multiplying the coefficient of the $\G_{\mu\nu}\G^{\mu\nu}$ operator inside the curly brackets in Eq.~\eqref{eqn:STrK1}, $\frac{1}{24}\log\frac{m^2}{\mu^2}$, by $n_\text{Lorentz}=d=4-\epsilon$, we obtain a finite piece that results from $\epsilon$ multiplying the $\frac{1}{\epsilon}$ pole.
The latter has not been written out explicitly in \cref{eqn:STrK1}, but can be easily restored using the canonical substitution %$\log\frac{m^2}{\mu^2} \to -\frac{2}{\epsilon}+ \log\frac{m^2}{\mu^2}$.
\begin{equation}
-\log\frac{m^2}{\mu^2} \to \frac{2}{\epsilon}- \log\frac{m^2}{\mu^2}\,.
\end{equation}

\newpage
\subsection*{Power-type supertraces}
\label{sec:PowerTypeEval}

The power-type supertraces, \cref{eqn:CovariantGraph}, involve the interaction matrix $\bm{X}$, whose detailed expression is derived from the UV theory.
With the expansion in \cref{eq:Xmat_expand}, a power-type supertrace becomes a sum of terms that are ready to be evaluated using CDE.
To illustrate this, let us consider a simple example with two spin-0 propagators, $\varphi_i$ with a heavy mass $M$ and $\varphi_j$ with zero mass:
\begin{equation}
K_i = P^2-M^2 \,,\qquad\text{and}\qquad\;\;
K_j = P^2 \,,
\end{equation}
with the following interaction structure:
\begin{equation}
X_{ij} = U_{ij} + \bar{Z}^\mu_{ij} P_\mu \equiv U_1 + \bar{Z}^\mu P_\mu  \,,\quad\text{and}\quad\;\;
X_{ji} = U_{ji} + P_\mu Z^\mu_{ji} \equiv U_2 + P_\mu Z^\mu \,.
\end{equation}
In this case, we obtain a sum of four supertraces:%\\[-5pt]
\begin{fmffile}{SSexample}
\begin{flalign}
&\parbox[c][60pt][c]{60pt}{\centering
	\begin{fmfgraph*}(40,40)
	\fmfsurround{u2,u1}
	\fmf{plain,left=1,label=$\varphi_i$,l.d=4pt}{u2,u1}
	\fmf{plain,left=1,label=$\varphi_j$,l.d=4pt}{u1,u2}
	\fmfv{decor.shape=circle,decor.filled=empty,decor.size=2.5thick}{u1,u2}
	\end{fmfgraph*}
}= -\frac{i}{2}\, \text{STr} \biggl( \frac{1}{K_i} X_{ij}\, \frac{1}{K_j} X_{ji} \biggr)\biggr|_\text{hard} & \label{eq:Str_example} \\[15pt]
&=-\frac{i}{2}\, \text{STr} \biggl( \frac{1}{P^2-M^2} U_1\, \frac{1}{P^2} U_2 \biggr)\biggr|_\text{hard} -\frac{i}{2}\, \text{STr} \biggl( \frac{1}{P^2-M^2} U_1\, \frac{1}{P^2} P_\mu Z^\mu \biggr)\biggr|_\text{hard} &\notag\\[15pt]
&\hspace{12pt} -\frac{i}{2}\, \text{STr} \biggl( \frac{1}{P^2-M^2} \bar{Z}^\mu P_\mu\, \frac{1}{P^2} U_2 \biggr)\biggr|_\text{hard} -\frac{i}{2}\, \text{STr} \biggl( \frac{1}{P^2-M^2} \bar{Z}^\mu P_\mu\, \frac{1}{P^2} P_\nu Z^\nu \biggr)\biggr|_\text{hard} \,. & \notag
\end{flalign}
\end{fmffile}%\noindent
%
%\\[4pt]
Each of these supertraces can be directly evaluated using CDE without further specifying the quantities $U_1$, $U_2$, $\bar{Z}^\mu$, $Z^\mu$.
As in~\cref{eqn:LogTypes} above, we obtain a series of EFT operators from each supertrace with successively higher powers of covariant derivatives:%\\[-5pt]
%
%\begin{small}
\begin{subequations}\label{eqn:PowerTypes}
\begin{flalign}
-\frac{i}{2}\, &\text{STr} \Bigl(\frac{1}{P^2-M^2} \, U_1\, \frac{1}{P^2}\, U_2 \Bigr) \Bigr|_\text{hard} &\notag\\[10pt]
&\hspace{-10pt} = \int\dd^dx\,\frac{1}{16\s\pi^2}\, \tr \Bigg\{ \frac12 \left(1-\log\frac{M^2}{\mu^2}\right) U_1 U_2 + \frac{1}{4M^2} \left(D^\mu U_1\right) \left(D_\mu U_2\right) + \dots \Bigg\} \,,&
\\[20pt]
-\frac{i}{2}\, &\text{STr} \Bigl(\frac{1}{P^2-M^2} \, U_1\, \frac{1}{P^2}\, P_\mu Z^\mu \Bigr) \Bigr|_\text{hard} &\notag\\[8pt]
&\hspace{-10pt} = \int\dd^dx\,\frac{1}{16\s\pi^2}\, \tr \Bigg\{ \frac14 \left(\frac12-\log\frac{M^2}{\mu^2}\right) i \,U_1 \left( D_\mu Z^\mu \right) + \dots \Bigg\} \,,&\\[30pt]
-\frac{i}{2}\, &\text{STr} \Bigl(\frac{1}{P^2-M^2}\, \bar{Z}^\mu P_\mu\, \frac{1}{P^2}\, U_2 \Bigr) \Bigr|_\text{hard} &\notag\\[8pt]
&\hspace{-10pt} = \int\dd^dx\,\frac{1}{16\s\pi^2}\, \tr \Bigg\{ -\frac14 \left(\frac12-\log\frac{M^2}{\mu^2}\right) i\left(D_\mu \bar{Z}^\mu\right) U_2 + \dots \Bigg\} \,,&\\[30pt]
-\frac{i}{2}\, &\text{STr} \Bigl(\frac{1}{P^2-M^2} \,\bar{Z}^\mu P_\mu\, \frac{1}{P^2}\, P_\nu Z^\nu \Bigr) \Bigr|_\text{hard} &\notag\\[8pt]
&\hspace{-10pt} = \int\dd^dx\,\frac{1}{16\s\pi^2}\, \tr \Bigg\{ \frac18 M^2\left(\frac32-\log\frac{M^2}{\mu^2}\right) \bar{Z}^\mu Z_\mu &\notag\\[5pt]
&\hspace{24pt} - \frac18 \,i\,\bar{Z}^\mu \G_{\mu\nu} Z^\nu - \frac{1}{24} \left(\frac56 - \log\frac{M^2}{\mu^2} \right) \left( D^\mu \bar{Z}^\nu \right) \left( D_\mu Z_\nu \right) &\notag\\[5pt]
&\hspace{24pt} + \frac{1}{12} \biggl(\frac13 - \log\frac{M^2}{\mu^2} \biggr) \Bigl[ \bigl( D_\mu \bar{Z}^\mu \bigr) \bigl( D_\nu Z^\nu \bigr) + \bigl( D_\nu \bar{Z}^\mu \bigr) \bigl( D_\mu Z^\nu \bigr) \Bigr] + \dots \Bigg\} \,.&
\end{flalign}
\end{subequations}
%\end{small}\noindent
%
%\\[-5pt]
On the RHS of these equations, we have shown terms with up to two covariant derivatives; higher derivative terms in ``$\dots$'', corresponding to higher dimensional EFT operators, can be similarly derived.
The operator dimension is bounded from below by the minimum operator dimensions carried by $U_1$, $U_2$, $Z_\mu$ and $\bar Z_\mu$; this explains why, when enumerating power-type supertraces, we count the dimensions of $U$, $Z$, $\bar Z$ but not open covariant derivatives $P_\mu$ or propagators $K_i^{-1}$.
Note that the CDE algorithm puts all covariant derivatives into commutators~\cite{TraceCode,Henning:2014wua}, so the results involve gauge field strengths $\G_{\mu\nu}\equiv-i\,[P_\mu, P_\nu]$ and closed covariant derivatives like $\left(D_\mu U_{1}\right) \equiv -i\,\comm{P_\mu}{U_{1}}$.

The procedure above carries over to all other power-type supertraces.
Generally, we can apply CDE to evaluate any supertrace over a product of covariant propagators $\frac{1}{K_i}$ (which can have any spin and can be either heavy or light), open covariant derivatives $P_\mu$, and generic functionals of the light fields $U_{ij}[\phi]$, $Z^\mu_{ij}[\phi]$, $\bar{Z}^\mu_{ij}[\phi]$, etc.
The result will be a series of operators similar to \cref{eqn:PowerTypes}.
Then the remaining straightforward tasks are to substitute in explicit expressions for $U_{ij}[\phi]$, $Z^\mu_{ij}[\phi]$, $\bar{Z}^\mu_{ij}[\phi]$, etc.\ derived from the specific UV theory, and to carry out the remaining trace ``$\tr$'' as defined in \cref{eq:tr_def}.
In this way, we arrive at the operators in the one-loop EFT Lagrangian together with their coefficients.

\newpage
%%%%%%%%%%%%%%%%%%%%%%%%%%%%%%%%%%%%%%%%%%%%%%%%%%%%%%%%%%%%%%%%%%%%%%%%%%%%%%%%
\section{Summary: Prescription for Functional Matching}
\label{sec:Recipe}
%%%%%%%%%%%%%%%%%%%%%%%%%%%%%%%%%%%%%%%%%%%%%%%%%%%%%%%%%%%%%%%%%%%%%%%%%%%%%%%%

We can summarize the procedure discussed in the previous sections into the following practical prescription for functional matching up to one-loop order:
%%%
\begin{enumerate}
  \item {\bf Derive heavy EOM(s) and $\bm{\L_\text{EFT}^\text{(tree)}}$.}
  Starting with the UV Lagrangian $\L_\text{UV}[\Phi,\phi]$, derive the EOMs for the heavy fields $\Phi$ that one wishes to integrate out.
  Solve these EOMs and substitute the solution $\Phi_\c[\phi]$ (expanded in inverse powers of the heavy masses) into $\L_\text{UV}$, to obtain the tree-level EFT for the light fields: $\L_\text{EFT}^\text{(tree)}[\phi] = \L_\text{UV} [\Phi,\phi]\bigr|_{\Phi=\Phi_\c[\phi]}$.
  \item {\bf Derive $\bm{K}$ and $\bm{X}$ matrices.}
  Write the UV field multiplet $\varphi=\{\Phi,\phi\}$ in terms of the independent fields in the path integral measure, as in \eg\ Eq.~\eqref{eqn:RealComponents}.
  Take the second variation of the UV action with respect to $\varphi$ to extract the inverse propagator matrix $\bm{K}$ and interaction matrix $\bm{X}$ (where $\Phi$ is set to $\Phi_\c[\phi]$), as explained in \cref{eq:L_K,eq:Kphi,eqn:LXvariation,eq:X_mat_def}.
  These enter the two types of functional supertraces (log-type and power-type) in \cref{eq:separate_STr}, from which the one-loop EFT Lagrangian $\L_\text{EFT}^\text{(1-loop)}[\phi]$ will be derived.
  \item {\bf Enumerate supertraces.}
  For log-type supertraces, simply enumerate the heavy fields charged under the EFT gauge group.
  For power-type supertraces, identify the minimum operator dimension of each block $X_{ij}$ of the $\bm{X}$ matrix (excluding open covariant derivatives in the counting), and
  enumerate distinct graphs of the form in \cref{eqn:CovariantGraph} with at least one heavy propagator, where the sum of the miminum operator dimensions of the $X_{ij}$ nodes does not exceed the desired operator dimension truncation of the EFT Lagrangian.
  \item {\bf Evaluate supertraces to obtain $\bm{\L_\text{EFT}^\text{(1-loop)}}$.}
  Apply CDE to evaluate the supertraces, \eg\ as implemented in our package~\cite{TraceCode}.
  For log-type supertraces, the results are universal; see \cref{eqn:LogTypes} and Table~\ref{tab:LogTypeEval}.
  For power-type supertraces, first work in terms of generic $\bm{U}[\phi]$, $\bm{Z}_\mu[\phi]$, $\bm{\bar Z}_\mu[\phi]$, etc.\ up to the desired EFT operator dimension as in \cref{eq:Str_example,eqn:PowerTypes}, and then substitute in the concrete expressions derived in Step~2 for the specific UV theory under consideration to carry out the remaining trace defined in Eq.~\eqref{eq:tr_def}.
  Add up the results from evaluating all supertraces enumerated in Step~3 to obtain $\L_\text{EFT}^\text{(1-loop)}[\phi]$.
\end{enumerate}
%%%
These four steps are illustrated by arrows with different colors in the right panel of Fig.~\ref{fig:feynman_vs_functional}.
Following this prescription, one can derive the EFT Lagrangian up to one-loop order directly from any perturbative UV theory (renormalizable or not).
In the next section, we provide a detailed pedagogical example to demonstrate the prescription at work, and explain some of the more technical aspects and subtleties that one encounters when matching functionally.

%%%%%%%%%%%%%%%%%%%%%%%%%%%%%%%%%%%%%%%%%%%%%%%%%%%%%%%%%%%%%%%%%%%%%%%%%%%%%%%%
\section{Example: Singlet Scalar Extended Standard Model}
\label{sec:Singlet}
%%%%%%%%%%%%%%%%%%%%%%%%%%%%%%%%%%%%%%%%%%%%%%%%%%%%%%%%%%%%%%%%%%%%%%%%%%%%%%%%

Let us consider a UV theory where the SM is extended by a heavy singlet scalar $S$. Including all renormalizable interactions between $S$ and the SM fields, we have
\beq
\L_\text{UV} = \L_\text{SM} + \frac{1}{2}(\partial_\mu S)^2 -\frac{1}{2} \MS^2 S^2 -A |H|^2S -\frac{1}{2}\kappa |H|^2S^2 -\frac{1}{3!} \mu_S^{} S^3 -\frac{1}{4!} \lambda_S S^4 \,,
\eeq{eq:LUV_singlet}
where $H$ is the SM Higgs doublet, and our conventions for the SM Lagrangian are as follows:
\begin{align}
\L_\text{SM} =\;& |D_\mu H|^2 +\sum_{f=q,u,d,l,e} \bar f\, i\slashed{D} f -\frac{1}{4} G^{\ci{A}}_{\mu\nu} G^{\ci{A}\mu\nu} -\frac{1}{4} W^{\wi{I}}_{\mu\nu} W^{\wi{I}\mu\nu} -\frac{1}{4} B_{\mu\nu} B^{\mu\nu}   \notag\\[5pt]
&  -\mH^2|H|^2 -\frac{1}{2}\lambda_H |H|^4 -\Bigl(
\bar q \,\mat{y}_u u\, \widetilde H
+\bar q \,\mat{y}_d\, d \,H
+\bar l\,\mat{y}_e\, e \,H+\text{h.c.} \Bigr) \,,\label{eqn:LagSM}
\end{align}
where $\widetilde H\equiv \epsmat H^*$ with $\epsmat = i\sigma^2=\begin{psmallmatrix}
0 & 1 \\[2pt] -1 & \;0\;
\end{psmallmatrix}$\,, and the Yukawa couplings $\mat{y}_u$, $\mat{y}_d$, $\mat{y}_e$ are $3\times3$ matrices in generation space.
We will often write $\ci{SU(3)_C}$ and $\wi{SU(2)_L}$ indices in color for clarity.

We will match this theory onto SMEFT up to dimension six.
This example has been adopted as a benchmark for one-loop SMEFT matching calculations in the recent literature: Ref.~\cite{Ellis:2017jns} used functional methods (with a slightly different formulation than the present work, see \cref{appsec:Relation}) to obtain the scalar sector contribution, Ref.~\cite{Jiang:2018pbd} computed additional EFT operators using Feynman diagrams, while Ref.~\cite{Haisch:2020ahr} presented the full matching calculation using Feynman diagrams.
While repeating the complete calculation using our approach, we will encounter many interesting aspects of one-loop functional matching, such as mixed heavy-light loops, mixed bosonic-fermionic loops, and derivative interactions.

\subsection*{Step 1: Derive heavy EOM and $\bm{\L_\text{EFT}^\text{(tree)}}$}

The EOM for the heavy field $S$ is
\begin{equation}
\frac{\delta \S_\text{UV}}{\delta S} = -A|H|^2 + \left( P^2-\MS^2 -\kappa|H|^2\right) S -\frac{1}{2}\mu_S^{} S^2 -\frac{1}{3!}\lambda_S S^3 = 0\,.
\end{equation}
To solve this equation order by order, we write the solution $S_\c$ as
\begin{equation}
S_\c = S_\c^{(2)} + S_\c^{(4)} + S_\c^{(6)} + \dots \,,
\label{eq:Sc}
\end{equation}
where $S_\c^{(n)}$ contains operators with mass dimension $n$ multiplied by prefactors that scale as $\MS^{1-n}$. Collecting terms in the EOM with operator dimensions 2, 4 and 6, we obtain
\begin{subequations}
\begin{align}
0 &= -A|H|^2 - \MS^2 S_\c^{(2)} \,,\\[10pt]
0 &= \left( P^2 -\kappa|H|^2\right) S_\c^{(2)} -\MS^2 S_\c^{(4)} -\tfrac{1}{2}\mu_S^{} \Big[S_\c^{(2)}\Big]^2 \,,\\[10pt]
0 &= \left( P^2 -\kappa|H|^2\right) S_\c^{(4)} -\MS^2 S_\c^{(6)} -\mu_S^{} S_\c^{(2)} S_\c^{(4)} -\tfrac{1}{3!}\lambda_S \Big[S_\c^{(2)}\Big]^3 \,.
\end{align}
\end{subequations}
Therefore,
\begin{subequations}
\begin{align}
S_\c^{(2)} &= -\tfrac{A}{\MS^2} |H|^2 \,, \\[8pt]
S_\c^{(4)} &= \tfrac{A}{\MS^4} \left[ \left(\partial^2 |H|^2\right) + \left(\kappa - \tfrac{\mu_S^{}A}{2\MS^2}\right) |H|^4 \right] \,, \\[8pt]
S_\c^{(6)} &= -\tfrac{A}{\MS^6} \bigg\{ \left(\kappa-\tfrac{\mu_S^{} A}{\MS^2}\right) |H|^2 \left(\partial^2 |H|^2\right) +\left[\left(\kappa-\tfrac{\mu_S^{} A}{\MS^2}\right)\left(\kappa-\tfrac{\mu_S^{} A}{2\MS^2}\right) -\tfrac{\lambda_S A^2}{6\MS^2} \right] |H|^6 \notag\\[3pt]
&\hspace{60pt} + \partial^2 \left[ \left(\partial^2 |H|^2\right) + \left(\kappa-\tfrac{\mu_S^{} A}{2\MS^2}\right) |H|^4 \right] \bigg\} \,.
\end{align}
\end{subequations}
Note that the term $\partial^2[\dots ]$ in $S_\c^{(6)}$ is a total derivative with operator dimension six, so it cannot contribute to any EFT operators up to dimension six.

The tree-level EFT Lagrangian is obtained by substituting $S_\c$ into $\L_\text{UV}$. Up to dimension six, we have
\beq
\L_\text{EFT}^\text{(tree)} = \L_\text{SM} +\tfrac{A^2}{2\MS^2} |H|^4 -\tfrac{A^2}{2\MS^4} |H|^2 \bigl(\partial^2 |H|^2\bigr) -\tfrac{A^2}{2\MS^4} \bigl(\kappa -\tfrac{\mu_S^{} A}{3\MS^2}\bigr) |H|^6\,.
\label{eq:L_EFT_tree_singlet}
\eeqn
~\\[-20pt]

\subsection*{Step 2: Derive $\bm{K}$ and $\bm{X}$ matrices}

To take functional variations of the UV action, we need to write the field multiplet $\varphi$ in terms of the independent fields in the path integral measure. In the present case, we have
\begin{subequations}
\label{eq:singlet_field_multiplet}
\begin{align}
\varphi_i \;&\in\;
\left\{
\varphi_S \,,\quad
\varphi_H \,,\quad
\varphi_q \,,\quad
\varphi_u \,,\quad
\varphi_d \,,\quad
\varphi_l \,,\quad
\varphi_e \,,\quad
\varphi_G%^{\li{\nu}\ci{B}}
\,,\quad
\varphi_W%^{\li{\nu}\wi{J}}
\,,\quad
\varphi_B%^{\li{\nu}}
\right\} \,, \\[10pt]
\bar\varphi_i \;&\in\;
\left\{
\bar\varphi_S \,,\quad
\bar\varphi_H \,,\quad
\bar\varphi_q \,,\quad
\bar\varphi_u \,,\quad
\bar\varphi_d \,,\quad
\bar\varphi_l \,,\quad
\bar\varphi_e \,,\quad
\bar\varphi_G%^{\li{\mu}\ci{A}}
\,,\quad
\bar\varphi_W%^{\li{\mu}\wi{I}}
\,,\quad
\bar\varphi_B%^{\li{\mu}}
\right\} \,,
\end{align}
\end{subequations}
where
\begin{subequations}
\begin{alignat}{4}
\varphi_S &= S \,,&\qquad
\varphi_H &=
\begin{pmatrix}
H \\ H^*
\end{pmatrix} \,,&\qquad
\varphi_f &=
\begin{pmatrix}
f \\ f^\c
\end{pmatrix} \,,&\qquad
\varphi_V &= V \,, \\[8pt]
\bar\varphi_S &= S \,,&\qquad
\bar\varphi_H &=
\begin{pmatrix}
H^\dagger & & H^T
\end{pmatrix} \,,&\qquad
\bar\varphi_f &=
\begin{pmatrix}
\bar f & & \bar f^\c
\end{pmatrix} \,,&\qquad
\bar\varphi_V &= V\,,
\end{alignat}
\end{subequations}
with $f = q, u, d, l, e$, and $V = G, W, B$.
We have omitted the ghosts fields that accompany the SM gauge fields, as their only interactions are with the gauge field fluctuations, which do not contribute to one-loop matching onto operators involving physical fields.

To treat the SM chiral fermions in a simple way, we introduce a set of auxiliary chiral fermions (denoted with prime) as their Dirac partners, so that $f$ and $f^\c$ are Dirac fermion fields, with the following Weyl components:\footnote{Here we use the same symbol for a Dirac field and its Weyl components. This will not be confusing in what follows as we will not need to write out the Weyl components explicitly in our calculation, and $f = q, u, d, l, e$ always refer to Dirac fields when they appear.}
\begin{subequations}
\label{eq:f_Weyl_components}
\begin{align}
&
q = \begin{pmatrix}
q_a \\ q^{\prime\dagger\dot a}
\end{pmatrix} \,,\qquad
q^\c = \begin{pmatrix}
q^\prime_a \\ q^{\dagger \dot a}
\end{pmatrix} \,,\\[5pt]
&
u = \begin{pmatrix}
u_a^\prime \\ u^{\dagger\dot a}
\end{pmatrix} \,,\qquad
u^\c = \begin{pmatrix}
u_a \\ u^{\prime\dagger \dot a}
\end{pmatrix}\,,
\end{align}
\end{subequations}
and similarly for $d$, $l$ and $e$.
Note that the positions of unprimed physical fields and primed auxiliary fields are swapped for the left-handed vs.\ right-handed SM fermions ($q$, $l$ vs.\ $u$, $d$, $e$).
With these auxiliary Weyl components introduced, projection operators $\frac{1}{2}(1\pm\gamma^5)$ need to be properly inserted in the Yukawa interactions in \cref{eqn:LagSM} to isolate the physical-chirality fermions.

With the field multiplet in \cref{eq:singlet_field_multiplet}, the inverse propagator matrix $\bm{K}$ takes the standard block-diagonal form with entries given by \cref{eq:Kphi}.
The interaction matrix $\bm{X}$ follows from varying the UV Lagrangian as in \cref{eqn:LXvariation}, and setting $S=S_\c$ given by Eq.~\eqref{eq:Sc}.
We provide a few examples of this calculation, and collect the explicit expressions for the $\bm{X}$ matrix entries in \cref{appsec:Xmatrix}, in the interest of providing a useful reference for future SMEFT matching calculations, since the majority of $\bm{X}$ matrix entries are derived from the SM Lagrangian.

It is worth noting that, for most of the $\bm{X}$ matrix blocks, the series in \cref{eq:Xmat_expand} truncates after the first order, $X_{ij}(\phi, P_\mu)=U_{ij}[\phi]$.
The only exceptions are blocks between the SM Higgs $H$ and the electroweak gauge bosons $W$, $B$, where open covariant derivatives appear and the series truncates after the next order; this serves as a concrete example of functional matching involving derivative interactions.

\subsection*{Step 3: Enumerate supertraces}

There are no log-type supertraces, since the only heavy field $S$ is a gauge singlet. To enumerate the power-type supertraces, we first list the minimum operator dimensions of the non-vanishing $\bm{X}$ matrix blocks as follows:
\beq
\text{dim} (\bm{X}) =
\begin{blockarray}{*{11}{c}}
	& S & H & q & u & d & l & e & G & W & B \\
	\begin{block}{c(*{10}{c})}
		S & 2 & 1 & & & & & & & & \\
		H &1 & 2 & \frac{3}{2} & \frac{3}{2} & \frac{3}{2} & \frac{3}{2} & \frac{3}{2} & & 1 & 1 \\
		q & & \frac{3}{2} & & 1 & 1 & & & \frac{3}{2} & \frac{3}{2}  & \frac{3}{2}  \\
		u & & \frac{3}{2} & 1 &  &  & & & \frac{3}{2} &   & \frac{3}{2}  \\
		d & & \frac{3}{2} & 1 &  &  & & & \frac{3}{2} &   & \frac{3}{2}  \\
		l & & \frac{3}{2} &  &  &  & & 1 &  & \frac{3}{2}  & \frac{3}{2}  \\
		e & & \frac{3}{2} &  &  &  & 1 & & &   & \frac{3}{2}  \\
		G & &  & \frac{3}{2} & \frac{3}{2} & \frac{3}{2} & & & 2 & & \\
		W & & 1  & \frac{3}{2} &  &  & \frac{3}{2} & & & 2 & 2 \\
		B & & 1  & \frac{3}{2} & \frac{3}{2} & \frac{3}{2} & \frac{3}{2} & \frac{3}{2} & & 2 & 2 \\
	\end{block}
\end{blockarray}\;\;.
\label{eq:dim_X}
\eeqn

The next step is to enumerate the graphs of the form shown in \cref{eqn:CovariantGraph} that contribute up to operator dimension six.
We will follow the usual convention, using dashed lines for scalars, solid lines for fermions, and wavy lines for vectors.
We double the dashed line for the heavy scalar $S$ to distinguish it from the light scalar $H$.
To make the operator dimension counting transparent, we will label the nodes in the graphs with their minimum operator dimensions --- the sum of these numbers in each graph should be $\le 6$.
Correspondingly, we will label $X_{ij}$, $U_{ij}$, etc.\ with superscripts to indicate their minimum operator dimensions; for example, $U_{SH}^{[1]}$ indicates that $U_{SH}$ starts with operator dimension one.
We will represent an $X_{ij}$ node as $U_{ij}$ when the two are equal; otherwise, we will first express the graph in terms of $X_{ij}$, and then expand the supertrace according to \cref{eq:Xmat_expand}.

Having set up the notation, we now systematically enumerate graphs with increasing numbers of propagators.
Since a graph must have at least one heavy propagator, we begin with an $S$ propagator in each case, and then complete the loop in all possible ways according to the nonvanishing blocks of the $\bm{X}$ matrix, as shown in Eq.~\eqref{eq:dim_X}.

\paragraph{1-propagator graph.}
There is only one graph with a single $S$ propagator:
\begin{equation}
\parbox[c][60pt][c]{60pt}{\centering
	\begin{fmffile}{S}
	\begin{fmfgraph*}(40,40)
	\fmfsurround{o3,o2,o1,u1}
	\fmf{dbl_dashes,left=0.414}{u1,o1,o2,o3,u1}
	\fmfv{decor.shape=circle,decor.filled=empty,decor.size=2.5thick,label={\scriptsize 2},l.d=6pt}{u1}
	\end{fmfgraph*}
	\end{fmffile}
}
\;\;= -\tfrac{i}{2}\,\text{STr} \Bigl[\tfrac{1}{P^2-\MS^2}\, U_{SS}^{[2]}\Bigr]\Bigr|_\text{hard}
\,.\label{eq:Tr_S}
\end{equation}

\paragraph{2-propagator graphs.}
The second propagator can be either $S$ or $H$, so we have
\begin{eqnarray}
\parbox[c][60pt][c]{60pt}{\centering
	\begin{fmffile}{SS}
	\begin{fmfgraph*}(40,40)
	\fmfsurround{u2,u1}
	\fmf{dbl_dashes,left=1}{u1,u2,u1}
	\fmfv{decor.shape=circle,decor.filled=empty,decor.size=2.5thick,label={\scriptsize 2},l.d=6pt}{u1,u2}
	\end{fmfgraph*}
	\end{fmffile}
}
\;\;&=& -\tfrac{i}{2}\,\tfrac{1}{2}\,\text{STr} \Bigl[\bigl(\tfrac{1}{P^2-\MS^2}\, U_{SS}^{[2]}\bigr)^2\Bigr]\Bigr|_\text{hard}
\,,\label{eq:Tr_SS}\\[6pt]
\parbox[c][60pt][c]{60pt}{\centering
	\begin{fmffile}{SH}
	\begin{fmfgraph*}(40,40)
	\fmfsurround{u2,u1}
	\fmf{dbl_dashes,left=1}{u1,u2}
	\fmf{dashes,left=1}{u2,u1}
	\fmfv{decor.shape=circle,decor.filled=empty,decor.size=2.5thick,label={\scriptsize 1},l.d=6pt}{u1,u2}
	\end{fmfgraph*}
	\end{fmffile}
}
\;\;&=& -\tfrac{i}{2}\,\text{STr} \Bigl[\tfrac{1}{P^2-\MS^2}\, U_{S H}^{[1]} \,\tfrac{1}{P^2-\mH^2} \,U_{HS}^{[1]}\Bigr] \Bigr|_\text{hard}
\,.
\label{eq:Tr_SH}
\end{eqnarray}
Note the symmetry factor $\frac{1}{2}$ in Eq.~\eqref{eq:Tr_SS}, due to the graph's $\mathbb{Z}_2$ symmetry under rotation.

\paragraph{3-propagator graphs.}
With 3 propagator, we can draw an $SSS$ loop, an $SSH$ loop, and an $SHH$ loop:
\begin{eqnarray}
\parbox[c][60pt][c]{60pt}{\centering
	\begin{fmffile}{SSS}
	\begin{fmfgraph*}(40,40)
	\fmfsurround{u3,u2,u1}
	\fmf{dbl_dashes,left=0.577}{u1,u2,u3,u1}
	\fmfv{decor.shape=circle,decor.filled=empty,decor.size=2.5thick,label={\scriptsize 2},l.d=6pt}{u1,u2,u3}
	\end{fmfgraph*}
	\end{fmffile}
}
\;\;&=& -\tfrac{i}{2}\,\tfrac{1}{3}\,\text{STr} \Bigl[\bigl(\tfrac{1}{P^2-\MS^2}\, U_{SS}^{[2]}\bigr)^3\Bigr]\Bigr|_\text{hard}
\,,\label{eq:Tr_SSS}\\[6pt]
\parbox[c][60pt][c]{60pt}{\centering
	\begin{fmffile}{SSH}
	\begin{fmfgraph*}(40,40)
	\fmfsurround{u3,u2,u1,o1}
	\fmf{dbl_dashes,left=0.414}{u1,u2,u3}
	\fmf{dashes,left=0.414}{u3,o1,u1}
	\fmfv{decor.shape=circle,decor.filled=empty,decor.size=2.5thick,label={\scriptsize 1},l.d=6pt}{u1,u3}
	\fmfv{decor.shape=circle,decor.filled=empty,decor.size=2.5thick,label={\scriptsize 2},l.d=6pt}{u2}
	\end{fmfgraph*}
	\end{fmffile}
}
\;\;&=& -\tfrac{i}{2}\,\text{STr} \Bigl[\tfrac{1}{P^2-\MS^2}\, U_{SS}^{[2]} \,\tfrac{1}{P^2-\MS^2}\, U_{S H}^{[1]} \,\tfrac{1}{P^2-\mH^2} \,U_{HS}^{[1]}\Bigr]\Bigr|_\text{hard}
\,,\label{eq:Tr_SSH}\\[6pt]
\parbox[c][60pt][c]{60pt}{\centering
	\begin{fmffile}{SHH}
	\begin{fmfgraph*}(40,40)
	\fmfsurround{u3,o1,u2,u1}
	\fmf{dbl_dashes,left=0.414}{u2,o1,u3}
	\fmf{dashes,left=0.414}{u3,u1,u2}
	\fmfv{decor.shape=circle,decor.filled=empty,decor.size=2.5thick,label={\scriptsize 1},l.d=6pt}{u2,u3}
	\fmfv{decor.shape=circle,decor.filled=empty,decor.size=2.5thick,label={\scriptsize 2},l.d=6pt}{u1}
	\end{fmfgraph*}
	\end{fmffile}
}
\;\;&=& -\tfrac{i}{2}\,\text{STr} \Bigl[\tfrac{1}{P^2-\MS^2} \,U_{S H}^{[1]}\,\tfrac{1}{P^2-\mH^2}\, U_{HH}^{[2]} \,\tfrac{1}{P^2-\mH^2}\, U_{HS}^{[1]} \Bigr]\Bigr|_\text{hard}
\,.\label{eq:Tr_SHH}
\end{eqnarray}
Again, note the symmetry factor $\frac{1}{3}$ in Eq.~\eqref{eq:Tr_SSS}.

\paragraph{4-propagator graphs.}
We can draw a loop with four $S$ propagators, but it has a minimum operator dimension of 8 and will not contribute to the EFT Lagrangian up to dimension six.
So we need at least one non-$S$ propagator.
First, we restrict ourselves to the scalar sector and use just $S$ and $H$.
The possibilities are an $SSSH$ loop, an $SSHH$ loop, an $SHSH$ loop, and an $SHHH$ loop (among which only the $SHSH$ loop has a nontrivial symmetry factor of $\frac{1}{2}$):
\begin{eqnarray}
\parbox[c][60pt][c]{60pt}{\centering
	\begin{fmffile}{SSSH}
	\begin{fmfgraph*}(40,40)
	\fmfsurround{u4,u3,u2,u1,o2,o1}
	\fmf{dbl_dashes,left=0.268}{u1,u2,u3,u4}
	\fmf{dashes,left=0.268}{u4,o1,o2,u1}
	\fmfv{decor.shape=circle,decor.filled=empty,decor.size=2.5thick,label={\scriptsize 1},l.d=6pt}{u1,u4}
	\fmfv{decor.shape=circle,decor.filled=empty,decor.size=2.5thick,label={\scriptsize 2},l.d=6pt}{u2,u3}
	\end{fmfgraph*}
	\end{fmffile}
}
\;\;&=& -\tfrac{i}{2}\,\text{STr} \Bigl[\bigl(\tfrac{1}{P^2-\MS^2}\, U_{SS}^{[2]}\bigr)^2 \,\tfrac{1}{P^2-\MS^2}\, U_{S H}^{[1]} \,\tfrac{1}{P^2-\mH^2} \,U_{HS}^{[1]}\Bigr]\Bigr|_\text{hard}
\,,\label{eq:Tr_SSSH}\\[6pt]
\parbox[c][60pt][c]{60pt}{\centering
	\begin{fmffile}{SSHH}
	\begin{fmfgraph*}(40,40)
	\fmfsurround{u4,u3,u2,u1}
	\fmf{dbl_dashes,left=0.414}{u2,u3,u4}
	\fmf{dashes,left=0.414}{u4,u1,u2}
	\fmfv{decor.shape=circle,decor.filled=empty,decor.size=2.5thick,label={\scriptsize 1},l.d=6pt}{u2,u4}
	\fmfv{decor.shape=circle,decor.filled=empty,decor.size=2.5thick,label={\scriptsize 2},l.d=6pt}{u1,u3}
	\end{fmfgraph*}
	\end{fmffile}
}
\;\;&=& -\tfrac{i}{2}\,\text{STr} \Bigl[\tfrac{1}{P^2-\MS^2}\, U_{SS}^{[2]}\,\tfrac{1}{P^2-\MS^2} \,U_{S H}^{[1]}\tfrac{1}{P^2-\mH^2}\, U_{HH}^{[2]} \,\tfrac{1}{P^2-\mH^2}\, U_{HS}^{[1]} \Bigr]\Bigr|_\text{hard}
\,,\qquad\label{eq:Tr_SSHH}\\[6pt]
\parbox[c][60pt][c]{60pt}{\centering
	\begin{fmffile}{SHSH}
	\begin{fmfgraph*}(40,40)
	\fmfsurround{o4,u4,o3,u3,o2,u2,o1,u1}
	\fmf{dbl_dashes,left=0.2}{u1,o1,u2}
	\fmf{dashes,left=0.2}{u2,o2,u3}
	\fmf{dbl_dashes,left=0.2}{u3,o3,u4}
	\fmf{dashes,left=0.2}{u4,o4,u1}
	\fmfv{decor.shape=circle,decor.filled=empty,decor.size=2.5thick,label={\scriptsize 1},l.d=6pt}{u1,u2,u3,u4}
	\end{fmfgraph*}
	\end{fmffile}
}
\;\;&=& -\tfrac{i}{2} \tfrac{1}{2}\,\text{STr} \Bigl[\bigl(\tfrac{1}{P^2-\MS^2}\, U_{S H}^{[1]} \,\tfrac{1}{P^2-\mH^2} \,U_{HS}^{[1]}\bigr)^2\Bigr] \Bigr|_\text{hard}
\,,\label{eq:Tr_SHSH}\\[6pt]
\parbox[c][60pt][c]{60pt}{\centering
	\begin{fmffile}{SHHH}
	\begin{fmfgraph*}(40,40)
	\fmfsurround{u4,o2,o1,u3,u2,u1}
	\fmf{dashes,left=0.268}{u4,u1,u2,u3}
	\fmf{dbl_dashes,left=0.268}{u3,o1,o2,u4}
	\fmfv{decor.shape=circle,decor.filled=empty,decor.size=2.5thick,label={\scriptsize 1},l.d=6pt}{u3,u4}
	\fmfv{decor.shape=circle,decor.filled=empty,decor.size=2.5thick,label={\scriptsize 2},l.d=6pt}{u1,u2}
	\end{fmfgraph*}
	\end{fmffile}
}
\;\;&=& -\tfrac{i}{2}\,\text{STr} \Bigl[ \tfrac{1}{P^2-\MS^2}\, U_{S H}^{[1]} \,\bigl(\tfrac{1}{P^2-\mH^2} \, U_{HH}^{[2]}\bigr)^2 \tfrac{1}{P^2-\mH^2} \,U_{HS}^{[1]}\Bigr]\Bigr|_\text{hard}
\,.\label{eq:Tr_SHHH}
\end{eqnarray}

With four propagators, fields other than $S$ and $H$ can also enter the loop.
After attaching two $H$ propagators to both ends of an $S$ propagator, we can complete the loop with a SM fermion $f=q,u,d,l,e$ or electroweak vector $V=W, B$ as the fourth propagator:
\begin{eqnarray}
\parbox[c][60pt][c]{60pt}{\centering
	\begin{fmffile}{SHfH}
	\begin{fmfgraph*}(40,40)
	\fmfsurround{u4,o2,o1,u3,u2,u1}
	\fmf{dbl_dashes,left=0.268}{u3,o1,o2,u4}
	\fmf{dashes,left=0.268}{u4,u1}
	\fmf{plain,left=0.268}{u1,u2}
	\fmf{dashes,left=0.268}{u2,u3}
	\fmfv{decor.shape=circle,decor.filled=empty,decor.size=2.5thick,label={\scriptsize 1},l.d=6pt}{u3,u4}
	\fmfv{decor.shape=circle,decor.filled=empty,decor.size=2.5thick,label={\scriptsize $\frac{3}{2}$},l.d=6pt}{u1,u2}
	\end{fmfgraph*}
	\end{fmffile}
}
\;\;&=& -\tfrac{i}{2}\,\text{STr} \Bigl[ \tfrac{1}{P^2-\MS^2}\, U_{S H}^{[1]} \,\tfrac{1}{P^2-\mH^2} \, U_{Hf}^{[3/2]}\,\tfrac{1}{\Psl}\,U_{fH}^{[3/2]} \tfrac{1}{P^2-\mH^2} \,U_{HS}^{[1]}\Bigr]\Bigr|_\text{hard}
\,,\qquad \label{eq:Tr_SHfH}\\[15pt]
\parbox[c][60pt][c]{60pt}{\centering
	\begin{fmffile}{SHVH}
	\begin{fmfgraph*}(40,40)
	\fmfsurround{u4,o2,o1,u3,u2,u1}
	\fmf{dbl_dashes,left=0.268}{u3,o1,o2,u4}
	\fmf{dashes,left=0.268}{u4,u1}
	\fmf{wiggly,left=0.268}{u1,u2}
	\fmf{dashes,left=0.268}{u2,u3}
	\fmfv{decor.shape=circle,decor.filled=empty,decor.size=2.5thick,label={\scriptsize 1},l.d=6pt}{u1,u2,u3,u4}
	\end{fmfgraph*}
	\end{fmffile}
}
\;\;&=& -\tfrac{i}{2}\,\text{STr} \Bigl[ \tfrac{1}{P^2-\MS^2}\, U_{S H}^{[1]} \,\tfrac{1}{P^2-\mH^2} \, X_{HV}^{\li{\nu}[1]}\,\tfrac{-\eta_{\li{\nu\mu}}}{P^2}\,X_{VH}^{\li{\mu}[1]} \tfrac{1}{P^2-\mH^2} \,U_{HS}^{[1]}\Bigr]\Bigr|_\text{hard} \notag\\[10pt]
&&\hspace{-85pt} = -\tfrac{i}{2}\,\text{STr} \Bigl[ \tfrac{1}{P^2-\MS^2}\, U_{S H}^{[1]} \,\tfrac{1}{P^2-\mH^2} \, U_{HV}^{\li{\nu}[2]}\,\tfrac{-\eta_{\li{\nu\mu}}}{P^2}\,U_{VH}^{\li{\mu}[2]} \tfrac{1}{P^2-\mH^2} \,U_{HS}^{[1]}\Bigr]\Bigr|_\text{hard} \label{eq:Tr_SHVH-1} \\[10pt]
&&\hspace{-75pt} -\tfrac{i}{2}\,\text{STr} \Bigl[ \tfrac{1}{P^2-\MS^2}\, U_{S H}^{[1]} \,\tfrac{1}{P^2-\mH^2} \, P_\rho Z_{HV}^{\rho\li{\nu}\,[1]}\,\tfrac{-\eta_{\li{\nu\mu}}}{P^2}\,U_{VH}^{\li{\mu}[2]} \tfrac{1}{P^2-\mH^2} \,U_{HS}^{[1]}\nonumber\\
&&\hspace{-30pt} + \tfrac{1}{P^2-\MS^2}\, U_{S H}^{[1]} \,\tfrac{1}{P^2-\mH^2} \, U_{HV}^{\li{\nu}[2]}\,\tfrac{-\eta_{\li{\nu\mu}}}{P^2}\,\bar Z_{VH}^{\rho\li{\mu}\,[1]} P_\rho \tfrac{1}{P^2-\mH^2} \,U_{HS}^{[1]}\Bigr]\Bigr|_\text{hard} \label{eq:Tr_SHVH-2}\\[10pt]
&&\hspace{-75pt} -\tfrac{i}{2}\,\text{STr} \Bigl[ \tfrac{1}{P^2-\MS^2}\, U_{S H}^{[1]} \,\tfrac{1}{P^2-\mH^2} \, P_\rho Z_{HV}^{\rho\li{\nu}\,[1]}\,\tfrac{-\eta_{\li{\nu\mu}}}{P^2}\,\bar Z_{VH}^{\tau\li{\mu}\,[1]} P_\tau \tfrac{1}{P^2-\mH^2} \,U_{HS}^{[1]}\Bigr]\Bigr|_\text{hard}
\,,\label{eq:Tr_SHVH-3}
\end{eqnarray}
where we have written out the \li{Lorentz} indices carried by $V$.
Note that the two terms in Eq.~\eqref{eq:Tr_SHVH-2} are Hermitian conjugates of each other, so we only need to compute one of them explicitly.

\paragraph{5-propagator graphs.}
With five propagators, many possibilities are eliminated by the requirement that the sum of the nodes' minimum operator dimensions should be $\le 6$.
Again starting within the scalar sector, we find only two possibilities:
\begin{eqnarray}
\parbox[c][60pt][c]{60pt}{\centering
	\begin{fmffile}{SSHSH}
	\begin{fmfgraph*}(40,40)
	\fmfsurround{o4,u5,u4,u3,o2,u2,o1,u1}
	\fmf{dbl_dashes,left=0.2}{u3,u4,u5}
	\fmf{dashes,left=0.2}{u5,o4,u1}
	\fmf{dbl_dashes,left=0.2}{u1,o1,u2}
	\fmf{dashes,left=0.2}{u2,o2,u3}
	\fmfv{decor.shape=circle,decor.filled=empty,decor.size=2.5thick,label={\scriptsize 1},l.d=6pt}{u1,u2,u3,u5}
	\fmfv{decor.shape=circle,decor.filled=empty,decor.size=2.5thick,label={\scriptsize 2},l.d=6pt}{u4}
	\end{fmfgraph*}
	\end{fmffile}
}
\;\;&=& -\tfrac{i}{2}\,\text{STr} \Bigl[\tfrac{1}{P^2-\MS^2}\, U_{SS}^{[2]} \,\bigl(\tfrac{1}{P^2-\MS^2}\, U_{S H}^{[1]} \,\tfrac{1}{P^2-\mH^2} \,U_{HS}^{[1]}\bigr)^2\Bigr]\Bigr|_\text{hard} \,,
\label{eq:Tr_SSHSH}\\[6pt]
\parbox[c][60pt][c]{60pt}{\centering
	\begin{fmffile}{SHSHH}
	\begin{fmfgraph*}(40,40)
	\fmfsurround{o3,u5,o2,u4,u3,u2,o1,u1}
	\fmf{dbl_dashes,left=0.2}{u4,o2,u5}
	\fmf{dashes,left=0.2}{u5,o3,u1}
	\fmf{dbl_dashes,left=0.2}{u1,o1,u2}
	\fmf{dashes,left=0.2}{u2,u3,u4}
	\fmfv{decor.shape=circle,decor.filled=empty,decor.size=2.5thick,label={\scriptsize 1},l.d=6pt}{u1,u2,u4,u5}
	\fmfv{decor.shape=circle,decor.filled=empty,decor.size=2.5thick,label={\scriptsize 2},l.d=6pt}{u3}
	\end{fmfgraph*}
	\end{fmffile}
}
\;\;&=& -\tfrac{i}{2}\,\text{STr} \Bigl[\tfrac{1}{P^2-\mH^2}\, U_{HH}^{[2]} \,\bigl(\tfrac{1}{P^2-\mH^2}\, U_{HS}^{[1]} \,\tfrac{1}{P^2-\MS^2} \,U_{S H}^{[1]}\bigr)^2\Bigr]\Bigr|_\text{hard} \,.
\label{eq:Tr_SHSHH}
\end{eqnarray}
Including fermions, we find one additional graph:
\beq
\parbox[c][60pt][c]{60pt}{\centering
	\begin{fmffile}{SHffH}
		\begin{fmfgraph*}(40,40)
			\fmfsurround{u5,o3,o2,o1,u4,u3,u2,u1}
			\fmf{dbl_dashes,left=0.2}{u4,o1,o2,o3,u5}
			\fmf{dashes,left=0.2}{u5,u1}
			\fmf{plain,left=0.2}{u1,u2,u3}
			\fmf{dashes,left=0.2}{u3,u4}
			\fmfv{decor.shape=circle,decor.filled=empty,decor.size=2.5thick,label={\scriptsize 1},l.d=6pt}{u2,u4,u5}
			\fmfv{decor.shape=circle,decor.filled=empty,decor.size=2.5thick,label={\scriptsize $\frac{3}{2}$},l.d=6pt}{u1,u3}
		\end{fmfgraph*}
	\end{fmffile}
}
\;\;= -\tfrac{i}{2}\,\text{STr} \Bigl[ \tfrac{1}{P^2-\MS^2}\, U_{S H}^{[1]} \,\tfrac{1}{P^2-\mH^2} \, U_{Hf_1}^{[3/2]}\,\tfrac{1}{\Psl}\,U_{f_1f_2}^{[1]}\,\tfrac{1}{\Psl}\,U_{f_2H}^{[3/2]} \tfrac{1}{P^2-\mH^2} \,U_{HS}^{[1]}\Bigr]\Bigr|_\text{hard}
\,,\label{eq:Tr_SHffH}
\eeqn
where $f_1, f_2$ are summed over $q,u,d,l,e$.

One may also draw graphs with vector propagators, such as $SSHVH$, $SHVHH$, $SHfVH$ with $V=W,B$.
However, to keep the total operator dimension $\le 6$, we must take the dimension-one $Z$ ($\bar Z$) part of $X_{HV}$ ($X_{VH}$), not the dimension-two $U$ part.
Furthermore, in each case, the first order term in the CDE, which involves just $U$, $Z$, $\bar Z$ but not covariant derivatives, already saturates the six operator dimensions, so the result must contain the matrix product $U_{SH} Z^{\rho\li{\nu}}_{HV}$ or $\bar Z^{\rho\li{\mu}}_{VH} U_{HS}$.
One can easily confirm that $U_{SH} Z^{\rho\li{\nu}}_{HV}=\bar Z^{\rho\li{\mu}}_{VH} U_{HS}=0$ from Eqs.~\eqref{eq:U_SH}, \eqref{eq:Z_HW} and \eqref{eq:Z_HB}.
Therefore, all the additional graphs with vector propagators vanish at the dimension-six level.

\paragraph{6-propagator graphs.}
With six propagators, which come with six nodes, we have no choice but to select only from the ``1'' entries in Eq.~\eqref{eq:dim_X}; in this way we saturate the six operator dimensions.
Starting from an $S$ propagator, there is no way to get to the ``1''s in the fermion-fermion blocks.
Meanwhile, the $HV$ and $VH$ blocks are excluded because they would result in $U_{SH} Z^{\rho\li{\nu}}_{HV} $ or $\bar Z^{\rho\li{\mu}}_{VH} U_{HS}$, both of which vanish as discussed above.
We are thus left with only one possibility:
\beq
\parbox[c][60pt][c]{60pt}{\centering
	\begin{fmffile}{SHSHSH}
		\begin{fmfgraph*}(40,40)
			\fmfsurround{u6,u5,u4,u3,u2,u1}
			\fmfset{dash_len}{2mm} % default is 3mm, apparently the change is discrete and 2mm and 2.9mm are the same.
			\fmf{dbl_dashes,left=0.268,l.d=4pt}{u4,u5}
			\fmf{dashes,left=0.268,l.d=2pt}{u5,u6}
			\fmf{dbl_dashes,left=0.268}{u6,u1}
			\fmf{dashes,left=0.268}{u1,u2}
			\fmf{dbl_dashes,left=0.268}{u2,u3}
			\fmf{dashes,left=0.268}{u3,u4}
			\fmfv{decor.shape=circle,decor.filled=empty,decor.size=2.5thick,label={\scriptsize 1},l.d=6pt}{u1,u2,u3,u4,u5,u6}
		\end{fmfgraph*}
	\end{fmffile}
}
\;\;= -\tfrac{i}{2} \tfrac{1}{3}\,\text{STr} \Bigl[\bigl(\tfrac{1}{P^2-\MS^2}\, U_{S H}^{[1]} \,\tfrac{1}{P^2-\mH^2} \,U_{HS}^{[1]}\bigr)^3\Bigr] \Bigr|_\text{hard}\,.
\eeq{eq:Tr_SHSHSH}
Again note the symmetry factor $\frac{1}{3}$.

Beyond six propagators, any graph one can draw contributes to EFT operators with dimension $>6$.
So we have completed the enumeration of supertraces that appear for matching the singlet scalar extended SM onto SMEFT up to dimension six.
We have obtained 18 supertraces, shown in Eqs.~\eqref{eq:Tr_S}-\eqref{eq:Tr_SHSHSH}.
We are now ready to move on to the last step, evaluating these 18 supertraces.

\subsection*{Step 4:  Evaluate supertraces to obtain $\bm{\L_\text{EFT}^\text{(1-loop)}}$}

As explained in Sec.~\ref{sec:Evaluating}, we follow a two-step procedure to convert the 18 power-type supertraces into effective operators in the EFT Lagrangian. First, we apply CDE to Eqs.~\eqref{eq:Tr_S}-\eqref{eq:Tr_SHSHSH} assuming generic $U$, $Z$, and $\bar Z$. These results are summarized in Eqs.~\eqref{eq:mt_1}-\eqref{eq:mt_6} in \cref{appsec:TraceList}. We then substitute in the concrete expressions of $U$, $Z$, and $\bar Z$ derived in Step 2 and collected in \cref{appsec:Xmatrix}, and carry out the remaining trace ``$\tr$'' defined in Eq.~\eqref{eq:tr_def}.

For completeness, we present the various contributions to the one-loop EFT Lagrangian in a format that makes it transparent which equation in \cref{appsec:TraceList} is used to evaluate which supertrace.
This should allow the interested reader to fill in the intermediate steps by carefully working out the matrix algebra.
Note that we use ``$\Rightarrow$'' (rather than ``$=$'') to mean that the LHS of each equation is equal to the spacetime integral $\int \dd^dx$ of the expression on the RHS:
\begin{flalign}
\eqref{eq:Tr_S} &\overset{\eqref{eq:mt_1}}{\Longrightarrow}
\tfrac{1}{16\s\pi^2} \tfrac{1}{2} \Bigl( 1-\log\tfrac{\MS^2}{\mu^2}\Bigr) \bigg\{
(\kappa \MS^2 -\mu_S^{} A) |H|^2
+ \Bigl[ \tfrac{\lambda_S A^2}{2\MS^2} +\tfrac{\mu_S^{} A}{\MS^2} \Bigl(\kappa -\tfrac{\mu_S^{} A}{2\MS^2}\Bigr) \Bigr] |H|^4
&\notag\\[5pt]
&\hspace{72pt}
-\tfrac{1}{\MS^2}\Bigl[
\tfrac{\lambda_S A^2}{\MS^2} \Bigl(\kappa-\tfrac{2\mu_S^{} A}{3\MS^2}\Bigr) +\tfrac{\mu_S^{} A}{\MS^2} \Bigl(\kappa-\tfrac{\mu_S^{} A}{\MS^2} \Bigr) \Bigl(\kappa-\tfrac{\mu_S^{} A}{2\MS^2} \Bigr)
\Bigr] |H|^6 
&\notag\\[5pt]
&\hspace{72pt}
-\tfrac{1}{\MS^2}\Bigl[
\tfrac{\lambda_S A^2}{\MS^2} +\tfrac{\mu_S^{} A}{\MS^2} \Bigl( \kappa -\tfrac{\mu_S^{} A}{\MS^2}\Bigr)
\Bigr] |H|^2 \bigl(\partial^2 |H|^2\bigr)
\bigg\}\,,&\label{eq:Tr_S_evaluate}
\end{flalign}\vspace{-30pt}

\begin{flalign}
\eqref{eq:Tr_SS} &\overset{\eqref{eq:mt_2-1}}{\Longrightarrow}
\tfrac{1}{16\s\pi^2} \tfrac{1}{4} \Bigl(\kappa-\tfrac{\mu_S^{} A}{\MS^2} \Bigr) \bigg\{
\Bigl(\kappa-\tfrac{\mu_S^{} A}{\MS^2} \Bigr) \Bigl( -\log\tfrac{\MS^2}{\mu^2}\Bigr) |H|^4
&\notag\\[5pt]
&\hspace{72pt}
+\tfrac{1}{\MS^2}\Bigl[ \tfrac{\lambda_S A^2}{\MS^2} +\tfrac{2\mu_S^{} A}{\MS^2}\Bigl(\kappa-\tfrac{\mu_S^{} A}{2\MS^2} \Bigr)\Bigr] \Bigl( -\log\tfrac{\MS^2}{\mu^2}\Bigr) |H|^6 
&\notag\\[5pt]
&\hspace{72pt}
-\tfrac{1}{\MS^2}\Bigl[ \tfrac{1}{6}\kappa -\tfrac{2\mu_S^{} A}{\MS^2} \Bigl( \tfrac{1}{12}-\log\tfrac{\MS^2}{\mu^2}\Bigr)\Bigr] |H|^2 \bigl(\partial^2|H|^2\bigr)
\bigg\}\,,&
\end{flalign}\vspace{-30pt}

\begin{flalign}
\eqref{eq:Tr_SH} &\overset{\eqref{eq:mt_2-2}}{\Longrightarrow}
\tfrac{1}{16\s\pi^2} \bigg\{ \Bigl(1-\log\tfrac{\MS^2}{\mu^2}\Bigr) \Bigl[
\Bigl(1+\tfrac{\mH^2}{\MS^2} +\tfrac{\mH^4}{\MS^4}\Bigr) A^2 |H|^2
&\notag\\
&\hspace{135pt}
-\Bigl(1+\tfrac{\mH^2}{\MS^2} \Bigr) \tfrac{2\kappa A^2}{\MS^2} |H|^4
+ \tfrac{\kappa A^2}{\MS^4} \Bigl( 3\kappa-\tfrac{\mu_S^{} A}{\MS^2}\Bigr) |H|^6
\Bigr]
&\notag\\[5pt]
&\hspace{72pt}
+\tfrac{A^2}{\MS^2} \Bigl[\tfrac{1}{2} +\tfrac{\mH^2}{\MS^2}\Bigl(\tfrac{5}{2}-\log\tfrac{\MS^2}{\mu^2}\Bigr)\Bigr] |D_\mu H|^2
-\tfrac{\kappa A^2}{\MS^4} |H|^2 |D_\mu H|^2
&\notag\\[5pt]
&\hspace{72pt}
+\tfrac{2\kappa A^2}{\MS^4}\Bigl(\tfrac{5}{4}-\log\tfrac{\MS^2}{\mu^2}\Bigr) |H|^2 \bigl(\partial^2 |H|^2\bigr)
+\tfrac{A^2}{6\MS^4} \bigl|D^2 H\bigr|^2
&\notag\\[5pt]
&\hspace{72pt}
-\tfrac{A^2}{12\MS^4} \Bigl[\, ig_2(D^\mu H)^\dagger \sigma^{\wi{I}} (D^\nu H) \,W^{\wi{I}}_{\mu\nu}
+ig_1(D^\mu H)^\dagger (D^\nu H) \,B_{\mu\nu}\Bigr]
&\notag\\[5pt]
&\hspace{72pt}
-\tfrac{A^2}{12\MS^4} \Bigl(\tfrac{7}{3}-\log\tfrac{\MS^2}{\mu^2}\Bigr)
\Bigl[ \,ig_2 \bigl(H^\dagger \sigma^{\wi{I}} \overleftrightarrow{D}^{\hspace{-1pt}\mu} H\bigr) \bigl(D^\nu W^{\wi{I}}_{\mu\nu} \bigr)
&\notag\\
&\hspace{200pt}
+ig_1\bigl(H^\dagger \overleftrightarrow{D}^{\hspace{-1pt}\mu} H\bigr) \bigl(\partial^\nu B_{\mu\nu} \bigr)
\Bigr]
&\notag\\[5pt]
&\hspace{72pt}
+\tfrac{A^2}{16\MS^4} \Bigl[ \,g_2^2 \bigl(H^\dagger \sigma^\wi{I}\sigma^\wi{J} H\bigr) \,W^{\wi{I}}_{\mu\nu} W^{\wi{J}\mu\nu}
+g_1^2\, |H|^2 B_{\mu\nu} B^{\mu\nu}
&\notag\\
&\hspace{200pt}
+2 g_1 g_2 \bigl(H^\dagger \sigma^{\wi{I}} H\bigr) W^{\wi{I}}_{\mu\nu} B^{\mu\nu}
\Bigr]
\bigg\}\,,&\label{eq:Tr_SH_evaluate}
\end{flalign}\vspace{-30pt}

\begin{flalign}
\eqref{eq:Tr_SSS} &\overset{\eqref{eq:mt_3-1}}{\Longrightarrow}
\tfrac{1}{16\s\pi^2} \tfrac{1}{\MS^2} \bigg\{ -\tfrac{1}{12} \Bigl( \kappa -\tfrac{\mu_S^{} A}{\MS^2} \Bigr)^3\, |H|^6 \bigg\} \,,&
\end{flalign}\vspace{-30pt}

\begin{flalign}
\eqref{eq:Tr_SSH} &\overset{\eqref{eq:mt_3-2}}{\Longrightarrow}
\tfrac{1}{16\s\pi^2}\tfrac{A^2}{\MS^2} \bigg\{ -\Bigl(\kappa-\tfrac{\mu_S^{} A}{\MS^2}\Bigr) \Bigl[1+\tfrac{\mH^2}{\MS^2}\Bigl(2-\log\tfrac{\MS^2}{\mu^2}\Bigr)\Bigr] |H|^4
&\notag\\[5pt]
&\hspace{72pt}
+ \tfrac{2}{\MS^2}\Bigl[\Bigl(\kappa-\tfrac{\mu_S^{} A}{2\MS^2}\Bigr)^2 -\tfrac{\lambda_S A^2}{4\MS^2} \Bigr] |H|^6 
+\tfrac{1}{3\MS^2} \Bigl(\kappa-\tfrac{4\mu_S^{} A}{\MS^2}\Bigr) |H|^2 \bigl(\partial^2 |H|^2\bigr) 
&\notag\\[5pt]
&\hspace{72pt}
-\tfrac{1}{2\MS^2}\Bigl(\kappa-\tfrac{\mu_S^{} A}{\MS^2}\Bigr) |H|^2 |D_\mu H|^2 \bigg\} \,,&
\end{flalign}\vspace{-30pt}

\begin{flalign}
\eqref{eq:Tr_SHH} &\overset{\eqref{eq:mt_3-3}}{\Longrightarrow}
\tfrac{1}{16\s\pi^2}\tfrac{A^2}{\MS^2} \bigg\{ \Bigl(1+\tfrac{2\mH^2}{\MS^2}\Bigr)
\Bigl( 3\lambda_H-\tfrac{A^2}{\MS^2}\Bigr)
\Bigl(1-\log\tfrac{\MS^2}{\mu^2}\Bigr) |H|^4 
&\notag\\[5pt]
&\hspace{72pt}
- \tfrac{1}{\MS^2}\Bigl[ 6\kappa\lambda_H -\tfrac{A^2}{2\MS^2} \Bigl(7\kappa-\tfrac{\mu_S^{} A}{\MS^2}\Bigr) \Bigr]
\Bigl(1-\log\tfrac{\MS^2}{\mu^2}\Bigr) |H|^6 
&\notag\\[5pt]
&\hspace{72pt}
-\tfrac{1}{\MS^2} \Bigl[
\tfrac{1}{2}\lambda_H
-\tfrac{A^2}{\MS^2} \Bigl( \tfrac{3}{2}-\log\tfrac{\MS^2}{\mu^2}\Bigr)
\Bigr] |H|^2 \bigl(\partial^2 |H|^2\bigr) 
&\notag\\[5pt]
&\hspace{72pt}
+ \tfrac{\lambda_H}{2\MS^2}\Bigl( \tfrac{3}{2}-\log\tfrac{\MS^2}{\mu^2}\Bigr) \bigl(\partial_\mu |H|^2\bigr)^2
-\tfrac{\lambda_H}{\MS^2}\, |H|^2\bigl[H^\dagger(D^2 H) + (D^2 H)^\dagger H\bigr] 
&\notag\\[5pt]
&\hspace{72pt}
+\tfrac{1}{\MS^2} \Bigl(\lambda_H -\tfrac{A^2}{\MS^2}\Bigr)
\Bigl(\tfrac{5}{2}-\log\tfrac{\MS^2}{\mu^2}\Bigr) |H|^2 |D_\mu H|^2
\bigg\} \,,&
\end{flalign}\vspace{-30pt}

\begin{flalign}
\eqref{eq:Tr_SSSH} &\overset{\eqref{eq:mt_4-1}}{\Longrightarrow} \tfrac{1}{16\s\pi^2} \tfrac{A^2}{\MS^4} \bigg\{ \tfrac{1}{2} \Bigl( \kappa -\tfrac{\mu_S^{} A}{\MS^2} \Bigr)^2 \,|H|^6 \bigg\} \,,&
\end{flalign}\vspace{-30pt}

\begin{flalign}
\eqref{eq:Tr_SSHH} &\overset{\eqref{eq:mt_4-2}}{\Longrightarrow}
\tfrac{1}{16\s\pi^2}\tfrac{A^2}{\MS^4} \bigg\{ -\Bigl( \kappa - \tfrac{\mu_S^{} A}{\MS^2} \Bigr) \Bigl( 3\lambda_H -\tfrac{A^2}{\MS^2} \Bigr) \Bigl(2-\log\tfrac{\MS^2}{\mu^2}\Bigr) |H|^6 \bigg\} \,,&
\end{flalign}\vspace{-30pt}

\begin{flalign}
\eqref{eq:Tr_SHSH} &\overset{\eqref{eq:mt_4-3}}{\Longrightarrow}
\tfrac{1}{16\s\pi^2} \tfrac{A^4}{\MS^4} \bigg\{
-\Bigl[\Bigl(2 -\log \tfrac{\MS^2}{\mu^2}\Bigr) +\tfrac{4\mH^2}{\MS^2}\Bigl(\tfrac{3}{2} -\log \tfrac{\MS^2}{\mu^2}\Bigr)\Bigr] |H|^4
&\notag\\[5pt]
&\hspace{72pt}
+\tfrac{4\kappa}{\MS^2} \Bigl(2 -\log \tfrac{\MS^2}{\mu^2}\Bigr) |H|^6
+\tfrac{1}{2\MS^2} \bigl(\partial_\mu |H|^2\bigr)^2
&\notag\\[5pt]
&\hspace{72pt}
+\tfrac{1}{\MS^2}\Bigl(\tfrac{8}{3} -\log \tfrac{\MS^2}{\mu^2}\Bigr) |H|^2 |D_\mu H|^2
&\notag\\[5pt]
&\hspace{72pt}
+\tfrac{1}{2\MS^2}\Bigl(\tfrac{14}{3} -\log \tfrac{\MS^2}{\mu^2}\Bigr)|H|^2\bigl[H^\dagger(D^2 H) + (D^2 H)^\dagger H\bigr]
\bigg\}
\,,&
\end{flalign}\vspace{-30pt}

\begin{flalign}
\eqref{eq:Tr_SHHH} &\overset{\eqref{eq:mt_4-4}}{\Longrightarrow}
\tfrac{1}{16\s\pi^2} \tfrac{A^2}{\MS^4} \bigg\{ \Bigl( 3\lambda_H -\tfrac{A^2}{\MS^2} \Bigr)^2 \Bigl(1-\log\tfrac{\MS^2}{\mu^2}\Bigr) |H|^6 \bigg\} \,,&
\end{flalign}\vspace{-30pt}

\begin{flalign}
\eqref{eq:Tr_SHfH} &\overset{\eqref{eq:mt_4-5}}{\Longrightarrow}
\tfrac{1}{16\s\pi^2}\tfrac{A^2}{\MS^4} \bigg\{
\tfrac{1}{4}\Bigl(\tfrac{5}{2}-\log\tfrac{\MS^2}{\mu^2}\Bigr) \Bigl[
H^\dagger_{\wi{\alpha}}\,\bigl( \bar q_{\wi{\beta}}^{}\, \mat{y}_d^{} \mat{y}_d^\dagger\gamma^\mu q_\wi{\alpha}^{}
+\bar l_{\wi{\beta}}^{}\, \mat{y}_e^{} \mat{y}_e^\dagger\gamma^\mu l_\wi{\alpha}^{}\bigr) (iD_\mu H)_{\wi{\beta}}^{}
&\notag\\[5pt]
&\hspace{100pt}
+(iD_\mu H)^\dagger_{\wi{\alpha}}\,\bigl( \bar q_{\wi{\beta}}^{}\, \mat{y}_d^{} \mat{y}_d^\dagger\gamma^\mu q_\wi{\alpha}^{}
+\bar l_{\wi{\beta}}^{}\, \mat{y}_e^{} \mat{y}_e^\dagger\gamma^\mu l_\wi{\alpha}^{}\bigr) H_{\wi{\beta}}^{}
&\notag\\[5pt]
&\hspace{100pt}
-2\, \widetilde{H}^\dagger (iD_\mu H)
\bigl(\bar u \,\mat{y}_u^\dagger \mat{y}_d^{} \gamma^\mu d \bigr)
-2\, H^\dagger (iD_\mu \widetilde{H})
\bigl(\bar d \,\mat{y}_d^\dagger \mat{y}_u^{} \gamma^\mu u \bigr)
&\notag\\[5pt]
&\hspace{100pt}
+\widetilde{H}^\dagger_{\wi{\alpha}} \bigl( \bar q_{\wi{\beta}}^{}\, \mat{y}_u^{} \mat{y}_u^\dagger\gamma^\mu q_\wi{\alpha}^{} \bigr) (iD_\mu \widetilde{H}_{\wi{\beta}}^{})
+(iD_\mu \widetilde{H})^\dagger_{\wi{\alpha}} \bigl( \bar q_{\wi{\beta}}^{}\, \mat{y}_u^{} \mat{y}_u^\dagger\gamma^\mu q_\wi{\alpha}^{} \bigr)  \widetilde{H}_{\wi{\beta}}^{}
&\notag\\[5pt]
&\hspace{100pt}
+\bigl(H^\dagger \,i\overleftrightarrow{D}_{\hspace{-2pt}\mu} H\bigr)
\bigl( \bar u \,\mat{y}_u^\dagger \mat{y}_u^{} \gamma^\mu u
-\bar d \,\mat{y}_d^\dagger \mat{y}_d^{} \gamma^\mu d
-\bar e \,\mat{y}_e^\dagger \mat{y}_e^{} \gamma^\mu e \bigr) 
\Bigr] 
&\notag\\[5pt]
&\hspace{72pt}
+\tfrac{1}{4}\Bigl(\tfrac{1}{2}-\log\tfrac{\MS^2}{\mu^2}\Bigr)\Bigl[
\widetilde{H}^\dagger_{\wi{\alpha}} \bigl( \bar q_{\wi{\beta}}\, \mat{y}_u^{} \mat{y}_u^\dagger \,i \overleftrightarrow{\slashed{D}} q_{\wi{\alpha}}\bigr)\widetilde{H}_{\wi{\beta}}
&\notag\\[5pt]
&\hspace{100pt}
+|H|^2 \bigl(
\bar u\, \mat{y}_u^\dagger \mat{y}_u^{} \,i \overleftrightarrow{\slashed{D}} u
+\bar d\, \mat{y}_d^\dagger \mat{y}_d^{} \,i \overleftrightarrow{\slashed{D}} d
+\bar e\, \mat{y}_e^\dagger \mat{y}_e^{} \,i \overleftrightarrow{\slashed{D}} e
\bigr)
&\notag\\[5pt]
&\hspace{100pt}
+H^\dagger_{\wi{\alpha}} \bigl( \bar q_{\wi{\beta}}\, \mat{y}_d^{} \mat{y}_d^\dagger\, i \overleftrightarrow{\slashed{D}}  q_\wi{\alpha}
+\bar l_{\wi{\beta}}\, \mat{y}_e^{} \mat{y}_e^\dagger\, i \overleftrightarrow{\slashed{D}}  l_\wi{\alpha}\bigr) H_\wi{\beta}
\Bigr] \bigg\}
\,,&\label{eq:Tr_SHfH_evaluate}
\end{flalign}\vspace{-30pt}

\begin{flalign}
\eqref{eq:Tr_SHVH-1} &\overset{\eqref{eq:mt_4-6}}{\Longrightarrow}
\tfrac{1}{16\s\pi^2}\tfrac{A^2}{\MS^4} \bigg\{ \tfrac{1}{8} \Bigl(1-\log\tfrac{\MS^2}{\mu^2}\Bigr)
\Bigl[ g_2^2 \bigl(H^\dagger \sigma^\wi{I} \overleftrightarrow{D}_{\hspace{-2pt}\mu} H\bigr)^2
+g_1^2 \bigl(H^\dagger \overleftrightarrow{D}_{\hspace{-2pt}\mu} H\bigr)^2
\Bigr] \bigg\} \,,&
\end{flalign}\vspace{-30pt}

\begin{flalign}
\eqref{eq:Tr_SHVH-2} &\overset{\eqref{eq:mt_4-7}}{\Longrightarrow}
\tfrac{1}{16\s\pi^2}\tfrac{A^2}{\MS^4} \bigg\{ \tfrac{1}{8} \Bigl(\tfrac{1}{2}-\log\tfrac{\MS^2}{\mu^2}\Bigr)
\Bigl[ g_2^2 \bigl(H^\dagger \sigma^\wi{I} \overleftrightarrow{D}_{\hspace{-2pt}\mu} H\bigr)^2
+g_1^2 \bigl(H^\dagger \overleftrightarrow{D}_{\hspace{-2pt}\mu} H\bigr)^2
\Bigr] \bigg\} \,,&
\end{flalign}\vspace{-30pt}

\begin{flalign}
\eqref{eq:Tr_SHVH-3} &\overset{\eqref{eq:mt_4-8}}{\Longrightarrow}
\tfrac{1}{16\s\pi^2} \tfrac{A^2}{\MS^4} \biggl\{\tfrac{1}{8}\, \Bigl( 1-\log\tfrac{\MS^2}{\mu^2}\Bigr)
\Bigl[
g_2^2 \bigl(H^\dagger \sigma^\wi{I} \overleftrightarrow{D}_{\hspace{-2pt}\mu} H\bigr)^2
+g_1^2 \bigl(H^\dagger \overleftrightarrow{D}_{\hspace{-2pt}\mu} H\bigr)^2
\Bigr] \biggr\} \,,&
\end{flalign}\vspace{-30pt}

\begin{flalign}
\eqref{eq:Tr_SSHSH} &\overset{\eqref{eq:mt_5-1}}{\Longrightarrow}
\tfrac{1}{16\s\pi^2} \tfrac{A^4}{\MS^6} \bigg\{ 2\, \Bigl(\kappa-\tfrac{\mu_S^{} A}{\MS^2}\Bigr) \Bigl( \tfrac{5}{2} -\log\tfrac{\MS^2}{\mu^2}\Bigr)\, |H|^6 \bigg\} \,,&
\end{flalign}\vspace{-30pt}

\begin{flalign}
\eqref{eq:Tr_SHSHH} &\overset{\eqref{eq:mt_5-2}}{\Longrightarrow}
\tfrac{1}{16\s\pi^2} \tfrac{A^4}{\MS^6} \bigg\{ -4\, \Bigl(3\lambda_H -\tfrac{A^2}{\MS^2}\Bigr)\, |H|^6 \bigg\} \,,&
\end{flalign}\vspace{-30pt}

\begin{flalign}
\eqref{eq:Tr_SHffH} &\overset{\eqref{eq:mt_5-3}}{\Longrightarrow}
\tfrac{1}{16\s\pi^2}\tfrac{A^2}{\MS^4} \bigg\{ \tfrac{1}{4} (4-\epsilon) \Bigl(\tfrac{3}{2}-\log\tfrac{\MS^2}{\mu^2}\Bigr) \Big[
|H|^2 \bigl(\bar q\, \mat{y}_u^{} \mat{y}_u^\dagger \mat{y}_u^{} u \,\widetilde{H} + \widetilde{H}^\dagger \,\bar u\, \mat{y}_u^\dagger \mat{y}_u^{} \mat{y}_u^\dagger\, q \bigr)
&\notag\\[3pt]
&\hspace{145pt}
+|H|^2 \bigl(\bar q \,\mat{y}_d^{} \mat{y}_d^\dagger \mat{y}_d^{} d \,H + H^\dagger\, \bar d\, \mat{y}_d^\dagger \mat{y}_d^{} \mat{y}_d^\dagger \,q \bigr)
&\notag\\[3pt]
&\hspace{145pt}
+|H|^2 \bigl(\bar l \,\mat{y}_e^{} \mat{y}_e^\dagger \mat{y}_e^{} e \,H + H^\dagger\, \bar e\, \mat{y}_e^\dagger \mat{y}_e^{} \mat{y}_e^\dagger \,l \bigr) \Big] \bigg\}
\,,&\label{eq:Tr_SHffH_evaluate}
\end{flalign}\vspace{-30pt}

\begin{flalign}
\eqref{eq:Tr_SHSHSH} &\overset{\eqref{eq:mt_6}}{\Longrightarrow} \tfrac{1}{16\s\pi^2} \tfrac{A^6}{\MS^8} \bigg\{ 4\, \Bigl(\tfrac{11}{6} -\log \tfrac{\MS^2}{\mu^2}\Bigr) |H|^6 \bigg\} \,.&\label{eq:Tr_SHSHSH_evaluate}
\end{flalign}

In these equations, we have left most of the internal indices implicit when their contraction is obvious; in Eq.~\eqref{eq:Tr_SHfH_evaluate}, however, we have written out the $\wi{SU(2)_L}$ fundamental indices $\wi{\alpha}$, $\wi{\beta}$ explicitly in several terms for clarity.
Note that when the evaluation yields operators that involve fermions, bilinears of both the original fields $q,u,d,l,e$ and the charge conjugated fields $q^\c,u^\c,d^\c,l^\c,e^\c$ appear.
We have rewritten the fermion bilinears involving charge conjugated fields in terms of the original fields via $\bar f_1^\c\, \Gamma f_2^\c = \pm\,\bar f_2\,\Gamma f_1$, where the $+$ ($-$) sign applies when $\Gamma$ is a product of an even (odd) number of $\gamma$ matrices.
Also, we have dropped the chiral projection operators $\frac{1\pm\gamma^5}{2}$ when writing the final results, with the understanding that all auxiliary fields (\eg\ the primed fields in Eq.~\eqref{eq:f_Weyl_components}) are set to zero.
Finally, we have denoted the $SU(2)_L$ and $U(1)_Y$ gauge couplings by $g_2$ and $g_1$, respectively, and adopted the standard notation $\overleftrightarrow{D}_{\hspace{-2pt}\mu}$ when writing some of the operators, {\it e.g.}\ $H^\dagger \overleftrightarrow{D}_{\hspace{-2pt}\mu} H \equiv  H^\dagger (D_\mu H) -(D_\mu H)^\dagger H$.

This completes the application of the prescription detailed in \cref{sec:Recipe} for matching the singlet scalar extended SM onto SMEFT up to one loop and dimension six.
In~\cref{sec:SingletSummary}, we further rewrite the results of this calculation in a way that makes them more amenable to comparisons with the literature; Tables~\ref{tab:renormalizable}-\ref{tab:fermions} in that appendix provide an organized summary of the results.

%%%%%%%%%%%%%%%%%%%%%%%%%%%%%%%%%%%%%%%%%%%%%%%%%%%%%%%%%%%%%%%%%%%%%%%%%%%%%%%%
\section{Outlook}
\label{sec:Outlook}
%%%%%%%%%%%%%%%%%%%%%%%%%%%%%%%%%%%%%%%%%%%%%%%%%%%%%%%%%%%%%%%%%%%%%%%%%%%%%%%%

We have presented a concise prescription for systematically matching a UV theory onto an EFT up to one-loop order.
Our prescription is based on functional methods augmented by covariant derivative expansion (CDE) techniques.
The functional approach has the conceptual benefit that all aspects of the calculation are performed within the UV theory at the matching scale, which avoids the need to carefully keep track of many IR details that cannot contribute to the EFT Wilson coefficients.
By streamlining the formalism, we were able to reframe functional matching calculations as a four-step procedure, as summarized in Sec.~\ref{sec:Recipe} and illustrated in the right panel of Fig.~\ref{fig:feynman_vs_functional}.
At the core of our approach are the simple graphical enumeration and CDE evaluation of a set of functional supertraces.
The evaluation step can be treated in isolation; in a forthcoming paper~\cite{TraceCode}, we will provide a \texttt{Mathematica} package that automates the evaluation of any supertrace that can appear when integrating out heavy particles in relativistic theories.
Our point of view is that the calculation of one-loop matching coefficients for this general class of theories (including the very important application of SMEFT) is now completely straightforward and accessible.

Many interesting directions for future investigations remain.
The most obvious is to simply apply this formalism to other beyond the SM examples.
Since the output of this calculation is typically in a non-standard operator basis, there is also an opportunity to integrate this technology with automated approaches to changing basis such as Ref.~\cite{Gripaios:2018zrz}.
We have not yet explored the supertrace building block structures for EFTs arising from other low energy limits, such as a non-relativistic example like HQET.
Finally, it would be exciting to extend our prescription beyond one-loop fixed order~\cite{Jackiw:1974cv}, \eg\ to incorporate renormalization group improvements~\cite{Manohar:2020nzp}.

%%%%%%%%%%%%%%%%%%%%%%%%%%%%%%%%%%%%%%%%%%%%%%%%%%%%%%%%%%%%%%%%%%%%%%%%%%%%%%%%
\acknowledgments
%%%%%%%%%%%%%%%%%%%%%%%%%%%%%%%%%%%%%%%%%%%%%%%%%%%%%%%%%%%%%%%%%%%%%%%%%%%%%%%%
%
We are very grateful to Dave Sutherland for detailed comments on the draft.
X.L.\ is grateful to Brian Henning and Hitoshi Murayama for fruitful discussions on CDE and functional matching.
Z.Z.\ has likewise benefited from collaboration with Sebastian Ellis, J\'er\'emie Quevillon, Pham Ngoc Hoa Vuong and Tevong You on several past works on functional matching.
T.C.\ and X.L.\ are supported by the U.S. Department of Energy, under grant number DE-SC0011640.
The work of Z.Z.\ was supported in part by  the U.S.\ Department of Energy, Office of Science, Office of High Energy Physics, under Award Number DE-AC02-05CH11231.
%\paragraph{Note added.}

%%%%%%%%%%%%%%%%%%%%%%%%%%%%%%%%%%%%%%%%%%%%%%%%%%%%%%%%%%%%%%%%%%%%%%%%%%%%%%%%
\appendix
\section*{Appendices}
\addcontentsline{toc}{section}{\protect\numberline{}Appendices}%
\renewcommand*{\thesubsection}{\Alph{subsection}}
\numberwithin{equation}{subsection}
%%%%%%%%%%%%%%%%%%%%%%%%%%%%%%%%%%%%%%%%%%%%%%%%%%%%%%%%%%%%%%%%%%%%%%%%%%%%%%%%

%%%%%%%%%%%%%%%%%%%%%%%%%%%%%%%%%%%%%%%%%%%%%%%%%%%%%%%%%%%%%%%%%%%%%%%%%%%%%%%%
\subsection{Comparison With Previous Approaches}
\label{appsec:Relation}
%%%%%%%%%%%%%%%%%%%%%%%%%%%%%%%%%%%%%%%%%%%%%%%%%%%%%%%%%%%%%%%%%%%%%%%%%%%%%%%%

In the main text of this paper, our goal was to introduce our prescription in the most straightforward way to emphasize its simplicity and accessibility.
We have therefore avoided technical comparisons to the literature, and in particular how our prescription relates to previous approaches. This appendix aims to provide such a discussion.

The use of functional methods and CDE for one-loop matching calculations dates back to the 1980s~\cite{Gaillard:1985uh,Chan:1986jq,Cheyette:1987qz}.
More recently, interest in calculating precision electroweak and Higgs observables in SMEFT has led to a revival of these methods~\cite{Henning:2014gca,Henning:2014wua}.
In particular, following the CDE approach of Gaillard~\cite{Gaillard:1985uh} and Cheyette~\cite{Cheyette:1987qz}, Ref.~\cite{Henning:2014wua} presented universal results of integrating out heavy particles with degenerate masses.
This idea of universality in one-loop matching calculations was further emphasized in Ref.~\cite{Drozd:2015rsp}, which extended the results of Ref.~\cite{Henning:2014wua} to the nondegenerate case, and initiated the Universal One-Loop Effective Action (UOLEA) program.
It was soon realized, however, that the calculations in Refs.~\cite{Henning:2014wua,Drozd:2015rsp} must be further extended to take into account contributions from mixed heavy-light loops~\cite{delAguila:2016zcb,Boggia:2016asg}.
This was initially done in Refs.~\cite{Henning:2016lyp,Ellis:2016enq}, and was later on simplified in Refs.~\cite{Fuentes-Martin:2016uol,Zhang:2016pja}.
In particular, Ref.~\cite{Zhang:2016pja} developed a diagrammatic framework, dubbed ``covariant diagrams,'' to facilitate the CDE calculation for generic interaction structures among spin-0, spin-$\frac{1}{2}$ and spin-1 fields in the UV theory.
At this point, it is fair to claim that one-loop functional matching was a fully solved problem.
Afterward, efforts to explicitly work out additional terms in the relativistic UOLEA continued~\cite{Ellis:2017jns,Kramer:2019fwz,Ellis:2020ivx,Angelescu:2020yzf}, while functional matching also found applications in other contexts, including HEFT~\cite{Cohen:2020xca} and HQET~\cite{Cohen:2019btp}.

In the following paragraphs, we discuss how our new prescription relates to some of the key results in the literature summarized above.

\subsubsection*{Relation to Ref.~\cite{Henning:2016lyp}}

Our one-loop matching formula \cref{eq:S_EFT_loop}, and subsequently \cref{eqn:LEFTStr}, were essentially also derived in Ref.~\cite{Henning:2016lyp}.
In particular, Eq.~(2.37) in Ref.~\cite{Henning:2016lyp} summarized the total one-loop matching result as
\begin{equation}
\int \dd^d x\,\L_\text{EFT}^\text{(1-loop)} =
\frac{i}{2} \log \Sdet \left( - \frac{\delta^2\S_\text{UV}[\varphi]}{\delta\varphi^2} \biggr|_{\Phi=\Phi_\text{c}[\phi]} \right)
- \frac{i}{2} \log \text{Sdet} \Biggl( -\frac{\delta^2 \S_\text{EFT}^{(0)}[\phi]}{\delta\phi^2} \Biggr) \,.
\label{eqn:237}
\end{equation}
It was also stated that the first term gives the full UV 1LPI effective action $\Gamma_\text{L,UV}^\text{(1-loop)}[\phi]$; see Eq.~(2.32b) therein.
However, it was not articulated in Ref.~\cite{Henning:2016lyp} that with the method of regions, the second term can be identified with the soft region contribution, and hence the subtraction leaves us with the hard region.
This conceptual insight was highlighted in Refs.~\cite{Fuentes-Martin:2016uol,Zhang:2016pja}, which our present work inherits.

Apart from the above conceptual improvement, a more important development in the current work is to provide a simplified calculating procedure onward from \cref{eqn:LEFTStr}.
For historical reasons, Ref. \cite{Henning:2016lyp} was focused on explaining the meaning and origin of mixed heavy-light contributions.
This motivated an effort to separate the two in \cref{eqn:237}.
In particular, significant manipulations were performed to further split the first term in \cref{eqn:237} (see Eq.~(3.2) and App.~B in Ref.~\cite{Henning:2016lyp}):
\begin{align}
\frac{i}{2} \log \Sdet \left( - \frac{\delta^2\mathcal{S}_\text{UV}[\varphi]}{\delta\varphi^2} \biggr|_{\Phi=\Phi_\text{c}[\phi]} \right) =&\;
\frac{i}{2} \log \Sdet \left( - \frac{\delta^2\mathcal{S}_\text{UV}[\varphi]}{\delta\Phi^2} \biggr|_{\Phi=\Phi_\text{c}[\phi]} \right) \notag\\[8pt]
& + \frac{i}{2} \log \Sdet \left( - \frac{\delta^2\mathcal{S}_\text{UV}[\varphi]|_{\Phi=\Phi_\text{c}[\phi]}}{\delta\phi^2} \right) \,.
\label{eqn:32}
\end{align}
The first part of the RHS was then identified with the heavy-only contributions, and the second (after subtracting the soft region contribution) with mixed heavy-light contributions.
This decomposition was helpful for explicitly showing that mixed heavy-light contributions can be accounted for with functional matching.
However, using the RHS of \cref{eqn:32} for practical evaluation, as proposed in Ref.~\cite{Henning:2016lyp}, introduces unnecessary complications.
This is because in the second term of Eq.~\eqref{eqn:32}, one needs to substitute in the EOM solution $\Phi_\text{c}[\phi]$ before taking the functional variation with respect to the light fields $\phi$.
Furthermore, it is crucial to keep $\Phi_\text{c}[\phi]$ ``non-local'' (like in Eqs.~(3.23) and (3.24) in Ref.~\cite{Henning:2016lyp}) throughout this functional variation procedure.
This is quite tedious for UV theories with $\Phi$ interactions beyond quadratic order.
In the prescription presented in the current paper, we proceed using the LHS of \cref{eqn:32}, where everything is ``local'' at the stage of taking the functional variations.
Furthermore, the EOM solution $\Phi_\text{c}[\phi]$ can be kept implicit as in $U_{ij}[\phi]$ and $Z_{ij}^\mu[\phi]$ in our \cref{eq:Xmat_expand} until the very end of the calculation, and one never needs to use its non-local form.

\subsubsection*{Relation to Covariant Diagrams}

In Ref.~\cite{Zhang:2016pja}, a slightly different route was taken when computing the functional supertrace $\text{Str}\log(\bm{K} -\bm{X})$: the ``functional part'' of the supertrace was evaluated at the very beginning, and the expansion of the logarithm came afterward.
In contrast, in this paper, we first make the separation
\begin{equation}
\text{STr}\log(\bm{K} -\bm{X}) = \text{STr}\log \bm{K} + \text{STr}\log( \bm{1}-\bm{K}^{-1}\bm{X})\,,
\end{equation}
and expand the logarithm in the second term as in Eq.~\eqref{eq:separate_STr}, while postponing the evaluation to a later stage.
Ultimately, the two approaches produce the same operator expansion, as they must, but the key improvement here is a clean separation between the enumeration and evaluation steps that makes the calculation more compact.
Concretely, evaluating the ``functional part'' of the supertrace obviously generates more terms in the subsequent expansion.
Since the goal of covariant diagrams in Ref.~\cite{Zhang:2016pja} is to keep track of all these resulting terms, the number of diagrams can easily grow large.
In our present approach, the graphical enumeration is carried out before evaluating any part of the supertrace, and the number of graphs is therefore reduced.
For example, our results in Sec.~\ref{sec:Singlet} for the one-loop matching calculation of the singlet scalar extended SM can be fully reproduced by computing more than 40 covariant diagrams, as opposed to just 16 graphs here.

In a sense, our present approach corresponds to an efficient packaging of covariant diagrams.
Technically, each functional supertrace here evaluates to the same results as an infinite series of covariant diagrams.
In particular, a log-type supertrace is reproduced by the infinite sum of covariant diagrams with an even number (4, 6, $\dots$) of $P$ insertions and no $U$ or $Z$ insertions, while a power-type supertrace represented graphically as in Eq.~\eqref{eqn:CovariantGraph} --- more precisely, each term in a power-type supertrace following the expansion in Eq.~\eqref{eq:Xmat_expand} --- is reproduced by summing over covariant diagrams with the same structure with regard to $U$ and $Z$ insertions, but additionally allowing for an arbitrary number of $P$ and light mass insertions (filled nodes and crosses, respectively, in the notation of Ref.~\cite{Zhang:2016pja}), with the Lorentz indices contracted in all possible ways (represented by dotted lines).
As a concrete example, the results of evaluating the single supertrace in Eq.~\eqref{eq:Tr_SH} up to dimension six (as shown in Eq.~\eqref{eq:Tr_SH_evaluate}) is reproduced by a set of nine covariant diagrams:
\begin{fmffile}{SH-cd}
	\begin{align}
	&
	\parbox[c][60pt][c]{60pt}{\centering
		\begin{fmfgraph}(40,40)
		\fmfsurround{u2,u1}
		\fmf{dbl_dashes,left=1}{u1,u2}
		\fmf{dashes,left=1}{u2,u1}
		\fmfv{decor.shape=circle,decor.filled=empty,decor.size=2.5thick}{u1,u2}
		\end{fmfgraph}
	} \;\;\text{in this work}
	\notag\\
	=\;&
	\Biggl\{
	\parbox[c][60pt][c]{60pt}{\centering
		\begin{fmfgraph}(40,40)
		\fmfsurround{u2,u1}
		\fmf{dbl_dashes,left=1}{u1,u2}
		\fmf{dashes,left=1}{u2,u1}
		\fmfv{decor.shape=circle,decor.filled=empty,decor.size=2.5thick}{u1,u2}
		\end{fmfgraph}
	}
	\hspace{-4pt}+
	\parbox[c][60pt][c]{60pt}{\centering
		\begin{fmfgraph}(40,40)
		\fmfsurround{u2,o,u1,m}
		\fmf{dbl_dashes,left=0.414}{u1,o,u2}
		\fmf{dashes,left=0.414}{u2,m,u1}
		\fmfv{decor.shape=circle,decor.filled=empty,decor.size=2.5thick}{u1,u2}
		\fmfv{decor.shape=cross,decor.size=2.5thick}{m}
		\end{fmfgraph}
	}
	\hspace{-4pt}+
	\parbox[c][60pt][c]{60pt}{\centering
		\begin{fmfgraph}(40,40)
		\fmfsurround{u2,o2,o1,u1,m2,m1}
		\fmf{dbl_dashes,left=0.268}{u1,o1,o2,u2}
		\fmf{dashes,left=0.268}{u2,m1,m2,u1}
		\fmfv{decor.shape=circle,decor.filled=empty,decor.size=2.5thick}{u1,u2}
		\fmfv{decor.shape=cross,decor.size=2.5thick,decor.angle=45}{m1,m2}
		\end{fmfgraph}
	}
	\hspace{-4pt}+
	\parbox[c][60pt][c]{60pt}{\centering
		\begin{fmfgraph}(40,40)
		\fmfsurround{u2,p1,u1,p2}
		\fmf{dbl_dashes,left=0.414}{u1,p1,u2}
		\fmf{dashes,left=0.414}{u2,p2,u1}
		\fmfv{decor.shape=circle,decor.filled=empty,decor.size=2.5thick}{u1,u2}
		\fmfv{decor.shape=circle,decor.filled=full,decor.size=2.5thick}{p1,p2}
		\fmf{dots,width=1thick}{p1,p2}
		\end{fmfgraph}
	}
	\hspace{-4pt}+
	\parbox[c][60pt][c]{60pt}{\centering
		\begin{fmfgraph}(40,40)
		\fmfsurround{u2,o2,p1,o1,u1,o3,p2,m}
		\fmf{dbl_dashes,left=0.2}{u1,o1,p1,o2,u2}
		\fmf{dashes,left=0.2}{u2,m,p2,o3,u1}
		\fmfv{decor.shape=circle,decor.filled=empty,decor.size=2.5thick}{u1,u2}
		\fmfv{decor.shape=circle,decor.filled=full,decor.size=2.5thick}{p1,p2}
		\fmf{dots,width=1thick}{p1,p2}
		\fmfv{decor.shape=cross,decor.size=2.5thick,decor.angle=45}{m}
		\end{fmfgraph}
	} \nonumber\\
	&
	\quad+
	\parbox[c][60pt][c]{60pt}{\centering
		\begin{fmfgraph}(40,40)
		\fmfsurround{u2,p2,p1,u1,p4,p3}
		\fmf{dbl_dashes,left=0.268}{u1,p1,p2,u2}
		\fmf{dashes,left=0.268}{u2,p3,p4,u1}
		\fmfv{decor.shape=circle,decor.filled=empty,decor.size=2.5thick}{u1,u2}
		\fmfv{decor.shape=circle,decor.filled=full,decor.size=2.5thick}{p1,p2,p3,p4}
		\fmf{dots,width=1thick}{p1,p3}
		\fmf{dots,width=1thick}{p2,p4}
		\end{fmfgraph}
	}
	\hspace{-4pt}+
	\parbox[c][60pt][c]{60pt}{\centering
		\begin{fmfgraph}(40,40)
		\fmfsurround{u2,p2,p1,u1,p4,p3}
		\fmf{dbl_dashes,left=0.268}{u1,p1,p2,u2}
		\fmf{dashes,left=0.268}{u2,p3,p4,u1}
		\fmfv{decor.shape=circle,decor.filled=empty,decor.size=2.5thick}{u1,u2}
		\fmfv{decor.shape=circle,decor.filled=full,decor.size=2.5thick}{p1,p2,p3,p4}
		\fmf{dots,width=1thick,left=0.25}{p1,p4}
		\fmf{dots,width=1thick,right=0.25}{p2,p3}
		\end{fmfgraph}
	}
	\hspace{-4pt}+
	\parbox[c][60pt][c]{60pt}{\centering
		%\fmfset{dot_len}{1.8mm}
		\begin{fmfgraph}(40,40)
		\fmfsurround{u2,o2,p1,o1,u1,p4,p3,p2}
		\fmf{dbl_dashes,left=0.268}{u1,o1,p1,o2,u2}
		\fmf{dashes,left=0.268}{u2,p2,p3,p4,u1}
		\fmfv{decor.shape=circle,decor.filled=empty,decor.size=2.5thick}{u1,u2}
		\fmfv{decor.shape=circle,decor.filled=full,decor.size=2.5thick}{p1,p2,p3,p4}
		\fmf{dots,width=1thick}{p1,p3}
		\fmf{dots,width=1thick,right=0.5}{p2,p4}
		\end{fmfgraph}
	}
	\hspace{-4pt}+
	\parbox[c][60pt][c]{60pt}{\centering
		\fmfset{dot_len}{1.5mm}
		\begin{fmfgraph}(40,40)
		\fmfsurround{u2,o4,o3,o2,o1,u1,p4,p3,p2,p1}
		\fmf{dbl_dashes,left=0.2}{u1,o1,o2,o3,o4,u2}
		\fmf{dashes,left=0.2}{u2,p1,p2,p3,p4,u1}
		\fmfv{decor.shape=circle,decor.filled=empty,decor.size=2.5thick}{u1,u2}
		\fmfv{decor.shape=circle,decor.filled=full,decor.size=2.5thick}{p1,p2,p3,p4}
		\fmf{dots,width=1thick,right=0.4}{p1,p3}
		\fmf{dots,width=1thick,right=0.4}{p2,p4}
		\end{fmfgraph}
	}
	\Biggr\}
	\;\;\text{in Ref.~\cite{Zhang:2016pja}}
	.\;
	\end{align}
\end{fmffile}

\subsubsection*{Relation to the UOLEA}

The central goal of the UOLEA program is to derive a master formula for one-loop matching, in a form that expresses $\L_\text{EFT}^\text{(1-loop)}$ in terms of $P_\mu$, $U_{ij}$, $Z^\mu_{ij}$, etc.\ for generic relativistic UV theories.
The derivation is carried out once and for all; once such a master formula is available, all one needs to do to obtain $\L_\text{EFT}^\text{(1-loop)}$ for a specific UV theory is derive the concrete expressions of $U_{ij}$, $Z^\mu_{ij}$, etc.\ (as in Step 2 of our prescription) and directly substitute them into the formula.
We refer the reader to Ref.~\cite{Ellis:2020ivx} for a detailed review of the development and current status of the UOLEA program, and only give a brief summary here.

The UOLEA program began with a focus on the minimal case of heavy-only bosonic loops with no open covariant derivatives, where the results are quite simple, with only 19 terms up to dimension six~\cite{Henning:2014wua,Drozd:2015rsp}.
These 19 terms suffice for some matching calculations of phenomenological interest~\cite{Henning:2014wua,Drozd:2015rsp,Chiang:2015ura,Huo:2015exa,Huo:2015nka,Han:2017cfr}.
However, the limited scope of UV theories that these 19 terms can cover motivated efforts to explicitly compute mixed heavy-light bosonic loop and heavy fermionic loop contributions to the UOLEA (still in the absence of open covariant derivatives), and they yield many more terms~\cite{Ellis:2017jns,Ellis:2020ivx}.
In particular, when fermion couplings involving different gamma matrix structures are written out explicitly, the combinatorics give rise to more than two thousand terms~\cite{Ellis:2020ivx}.
At this point, we have gone a long way from the initial simple formula with 19 terms.
While the range of applicability of the UOLEA has been significantly expanded, the plethora of terms makes its application cumbersome beyond simple cases.

The prescription we have developed in this work shares the UOLEA spirit to some degree, in that we have isolated part of the calculation that can be done once and for all, so as to simplify the task an EFT practioner has to perform in a matching calculation.
In fact, the universal results of evaluating the log-type supertraces in Table~\ref{tab:LogTypeEval} readily form part of the UOLEA.
Also, each power-type supertrace can be evaluated once and for all, assuming generic functionals $U$, $Z$, $\bar Z$.
For example, the results in App.~\ref{appsec:TraceList} are useful beyond the singlet scalar example we worked out in Sec.~\ref{sec:Singlet}; if one is to perform a matching calculation for a different UV theory, it is likely that they would encounter some power-type supertraces that have the same form as those in App.~\ref{appsec:TraceList}.
Technically, the power-type supertraces in App.~\ref{appsec:TraceList} that do not involve fermionic propagators or open covariant derivatives reproduce many of the UOLEA terms previously computed in Refs.~\cite{Drozd:2015rsp,Ellis:2017jns}, whereas the those that involve both bosonic and fermionic propagators, as well as those that involve open covariant derivatives, essentially produce terms in the part of the UOLEA that has not been computed yet.

Given the complexity of the UOLEA in the most general case (including mixed bosonic-fermionic loops and open covariant derivatives), we believe our prescription offers the best alternative beyond the minimal cases where one can directly compute the EFT Lagrangian using a small number of terms in the UOLEA.
Also, our prescription (especially with our supertrace evaluation package~\cite{TraceCode}) offers the flexibility to go beyond dimension-six operators if desired, for which a general UOLEA would be too cumbersome to present explicitly.

%%%%%%%%%%%%%%%%%%%%%%%%%%%%%%%%%%%%%%%%%%%%%%%%%%%%%%%%%%%%%%%%%%%%%%%%%%%%%%%%
\subsection{Interaction Matrix for the Singlet Scalar Model}
\label{appsec:Xmatrix}
%%%%%%%%%%%%%%%%%%%%%%%%%%%%%%%%%%%%%%%%%%%%%%%%%%%%%%%%%%%%%%%%%%%%%%%%%%%%%%%%

Here we provide the explicit expressions for the interaction matrix $\bm{X}$ of the singlet scalar extended SM considered in \cref{sec:Singlet}.
Recall that the $\bm{X}$ matrix is derived by taking the second variation of $\L_\text{UV}$ as in Eq.~\eqref{eqn:LXvariation}, separating out the inverse propagator matrix $\bm{K}$ that takes the form of Eq.~\eqref{eq:Kphi}, and setting the heavy fields to their EOM solutions ($S=S_\c$ given by Eq.~\eqref{eq:Sc} in the present case).
As a simple example, $X_{SS}$ is obtained from the terms in $\L_\text{UV}$ with at least two powers of $S$:
\begin{eqnarray}
\delta^2 \L_\text{UV} &\supset& \delta^2\, \biggl[\frac{1}{2}\, S\, (P^2-M^2)\, S -\frac{1}{2}\,\kappa\, |H|^2S^2 -\frac{1}{3!}\, \mu_S^{}\, S^3 -\frac{1}{4!}\, \lambda_S\, S^4\biggr] \notag\\[3pt]
&\supset& \delta S\, ( P^2-M^2) \,\delta S - \delta S\,\biggl(\kappa\, |H|^2 +\mu_S^{}\, S +\frac{1}{2}\,\lambda_S\, S^2 \biggr)\, \delta S \,.
\end{eqnarray}
The expression in parentheses in the second term, with $S$ set to $S_\c$, is therefore identified with $X_{SS} = U_{SS}$.

As a second, less trivial example, let us work out $X_{SH}$ and $X_{HS}$. They come from the terms in $\L_\text{UV}$ with at least one power of $S$ and one power of $H$:
\begin{eqnarray}
\delta^2 \L_\text{UV} &\supset& \delta^2\, \biggl[ -A\,|H|^2 S-\frac{1}{2}\,\kappa\, |H|^2S^2\biggr] \nonumber\\[8pt]
&\supset& -2\,(\delta H^\dagger H +H^\dagger \delta H)\,(A +\kappa\,S)\, \delta S \nonumber\\[8pt]
&=& -(\delta H^\dagger H + H^T \delta H^* + H^\dagger \delta H +\delta H^T H^*)\,(A +\kappa\,S)\, \delta S \nonumber\\[8pt]
&=& -\delta S
\begin{pmatrix}
(A +\kappa\,S)\, H^\dagger \;\;&\;\; (A +\kappa\,S)\, H^T
\end{pmatrix}
\begin{pmatrix}
\delta H \\ \delta H^*
\end{pmatrix} \nonumber\\[3pt]
&& -\begin{pmatrix}
\delta H^\dagger \;\;&\;\; \delta H^T
\end{pmatrix}
\begin{pmatrix}
(A +\kappa\,S)\, H \\ (A +\kappa\,S)\, H^*
\end{pmatrix} \delta S\,.
\end{eqnarray}
The $1\times2$ matrix in the first term and the $2\times 1$ matrix in the second term are identified with $X_{SH} = U_{SH}$ and $X_{HS} = U_{HS}$, respectively, upon setting $S=S_\c$.
Importantly, we have rewritten the variation in a symmetric form between $H$ and $H^*$ (viewed as column vectors in gauge representation space), such that $U_{SH}$ and $U_{HS}$ are simply related by Hermitian conjugation.
Similarly, when dealing with terms in $\L_\text{UV}$ that involve fermions, we write the variation in a symmetric form between $f$ and $f^\c$ using $ \bar f_1\,\Gamma f_2= \pm\,\bar f_2^\c\, \Gamma f_1^\c$, where the $+$ ($-$) sign applies when $\Gamma$ is a product of an even (odd) number of $\gamma$ matrices.

As a final example, we consider $X_{HW}$ and $X_{WH}$. They are derived from the Higgs boson's gauge interactions:
\begin{eqnarray}
\delta^2 \L_\text{UV} &\supset& \delta^2\bigl(|D_\mu H|^2\bigr) \nonumber\\[8pt]
&=& \bigl[\delta^2 (D_\mu H)^\dagger\bigr] (D^\mu H) +(D_\mu H^\dagger) \bigl[\delta^2(D^\mu H)\bigr] +2\,\bigl[\delta\,(D_\mu H)^\dagger\bigr] \bigl[\delta\,(D^\mu H)\bigr] \nonumber\\[8pt]
&\supset& ig_2\,\delta W_\mu^{\wi{I}}\, \delta H^\dagger \,\sigma^{\wi{I}}\,(D^\mu H) -ig_2\, (D_\mu H)^\dagger \,\sigma^{\wi{I}}\,\delta H\, \delta B^\mu \nonumber\\
&& -ig_2\,(D_\mu \delta H)^\dagger \,\sigma^{\wi{I}}\,H\,\delta W_\mu^{\wi{I}} +ig_2\, \delta W_\mu^{\wi{I}}\,H^\dagger\,\sigma^{\wi{I}}\, (D^\mu \delta H)  \nonumber\\[8pt]
&=& -\frac{g_2}{2}\,\delta W_\mu^{\wi{I}} \,\bigl[ -i\, \delta H^\dagger \,\sigma^{\wi{I}}\,(D^\mu H) -i\,(D^\mu H)^T\, \sigma^{\wi{I}*}\,\delta H^* \nonumber\\
&&\qquad\qquad\;\;\, +i\,(D_\mu H)^\dagger \,\sigma^{\wi{I}}\,\delta H +i\,\delta H^T\, \sigma^{\wi{I}*} \, (D_\mu H)^*\nonumber\\[2pt]
&&\qquad\qquad\;\;\, +i\,(D_\mu \delta H)^\dagger \,\sigma^{\wi{I}}\,H +i\,H^T\, \sigma^{\wi{I}*} \, (D_\mu \delta H)^* \nonumber\\[2pt]
&&\qquad\qquad\;\;\, -i\, H^\dagger\,\sigma^{\wi{I}}\, (D^\mu \delta H) -i\, (D^\mu \delta H)^T\, \sigma^{\wi{I}*} \,H^* \bigr] \nonumber\\[8pt]
&\overset{\text{IBP}}{=}&
-\begin{pmatrix}
\delta H^\dagger & & \delta H^T
\end{pmatrix}
\left[
\begin{pmatrix}
-\frac{ig_2}{2}\,\sigma^{\wi{J}}\,(D^{\li{\nu}} H) \\
\frac{ig_2}{2}\,\sigma^{\wi{J}^*}\, (D^{\li{\nu}} H)^*
\end{pmatrix}
+iD_\rho
\begin{pmatrix}
-\eta^{\rho\li{\nu}}\,\frac{g_2}{2}\,\sigma^{\wi{J}}\,H \\
\eta^{\rho\li{\nu}}\,\frac{g_2}{2}\,\sigma^{\wi{J}^*}\,  H^*
\end{pmatrix}
\right]
\,\delta W_{\li{\nu}}^{\wi{J}} \nonumber\\[5pt]
&&-\,\delta W_{\li{\mu}}^{\wi{I}}
\biggl[
\begin{pmatrix}
\frac{ig_2}{2}\, (D^{\li{\mu}} H)^\dagger \,\sigma^{\wi{I}} \;\;&\;\;
-\frac{ig_2}{2}\, (D^{\li{\mu}} H)^T \sigma^{\wi{I}*}
\end{pmatrix}
\nonumber\\[-5pt]
&&\qquad\quad\;\;
+\begin{pmatrix}
-\eta^{\rho\li{\mu}}\,\frac{g_2}{2}\, H^\dagger\,\sigma^{\wi{I}}  \;\;&\;\;
\eta^{\rho\li{\mu}}\,\frac{g_2}{2}\, H^T\,\sigma^{\wi{I}*}
\end{pmatrix}
iD_\rho
\biggr]
\begin{pmatrix}
\delta H \\ \delta H^*
\end{pmatrix}\,,
\end{eqnarray}
where we have again symmetrized the variation between $H$ and $H^*$.
In the last equation, we can identify the expressions in brackets as $X_{HW}^{\li{\nu}\wi{J}}$ and $X_{WH}^{\li{\mu}\wi{I}}$, where $\li{\mu}$ ($\li{\nu}$) and $\wi{I}$ ($\wi{J}$) are the Lorentz and gauge indices of the $W$ fluctuation field on the left (right).
In this case, the series in \cref{eq:Xmat_expand} truncates after the second order:
\begin{equation}
X_{HW}^{\li{\nu}\wi{J}} = U_{HW}^{\li{\nu}\wi{J}} + P_\rho Z_{HW}^{\rho\,\li{\nu}\wi{J}} \,,\qquad
X_{WH}^{\li{\mu}\wi{I}} = U_{WH}^{\li{\mu}\wi{I}} + \bar Z_{WH}^{\rho\,\li{\mu}\wi{I}} P_\rho \,.
\label{eq:XHV}
\end{equation}
The $X_{HB}$ and $X_{BH}$ blocks are analogous.
The blocks between the SM Higgs $H$ and electroweak gauge bosons $W$, $B$ are the only blocks in the $\bm{X}$ matrix where open covariant derivatives $P_\rho = iD_\rho$ appear.
For all the other blocks, $X_{ij}=U_{ij}$.

In what follows, we present the nonzero $U_{ij}$, $Z_{ij}$, $\bar Z_{ij}$ blocks.
We will keep the adjoint $\ci{SU(3)_C}$ and $\wi{SU(2)_L}$ indices explicit, using $\ci{A}$ and $\wi{I}$ for the conjugate fields $\bar\varphi_i$ on the left, and $\ci{B}$ and $\wi{J}$ for the fields $\varphi_j$ on the right.
The (anti-)fundamental representation indices are mostly suppressed, with the understanding that the fields can be thought of as column and row vectors, and we write the results in matrix form in these gauge representation spaces whenever possible, \eg\ $\cid$, $\wid$, $\weps$ matrices appear in some of the equations.
In a few cases where all the index contractions cannot be unambiguously written in the matrix multiplication form, we make the $\wi{SU(2)_L}$ (anti-)fundamental indices explicit, using $\wi{\alpha}$, $\wi{\bar\alpha}$ for the conjugate fields $\bar\varphi_i$ on the left, and $\wi{\beta}$, $\wi{\bar\beta}$ for the fields $\varphi_j$ on the right.
When $\bar\varphi_i$ ($\varphi_j$) is a vector boson, it also carries a \li{Lorentz} index, for which we use $\li{\mu}$ ($\li{\nu}$); to avoid notation clashes, we use $\rho$ instead of $\mu$ for the additional Lorentz indices carried by $Z$, $\bar Z$ in this appendix, as in \cref{eq:XHV} above.

It is worth emphasizing that most of the results presented in this appendix are in fact derived from the renormalizable SM Lagrangian. Concretely, these include the term proportional to $\lambda_H$ in \cref{eq:U_HH}, and all of Eqs.~\eqref{eq:U_qu}-\eqref{eq:U_eB}. The utility of these results therefore extends beyond the singlet scalar extended SM, as they serve as a common reference for future SMEFT matching calculations.

\paragraph{Scalar sector entries}
~\\
\begin{equation}
U_{SS} = \kappa\, |H|^2 +\mu_S^{}\, S_\c +\frac{1}{2}\,\lambda_S^{} \,S_\c^2 \,.
\end{equation}
\begin{equation}
U_{SH} = (A+\kappa\, S_\c)
\begin{pmatrix}
H^\dagger & & H^T
\end{pmatrix}\,,\quad
U_{HS} = (A+\kappa\, S_\c)
\begin{pmatrix}
H \\ H^*
\end{pmatrix}\,.
\label{eq:U_SH}
\end{equation}
\vspace{5pt}
\begin{equation}
U_{HH} = \left( A\,S_\c +\frac12\,\kappa\, S_\c^2 \right) \mqty( \wid & 0 \\ 0 & \wid )
+ \lambda_H\, \mqty ( |H|^2\wid + H H^\dagger & H H^T \\ H^* H^\dagger & |H|^2\wid +H^* H^T ) \,.
\label{eq:U_HH}
\end{equation}
~\\[-20pt]

%\vspace{5pt}
\paragraph{Fermion-fermion entries}
~\\[-5pt]
\begin{alignat}{2}
\label{eq:U_qu}
U_{qu} &= \mqty( \cid\,\mat{y}_u \, \frac{1+\gamma^5}{2}\,\widetilde H & 0 \\
0 & \cid\,\mat{y}_u^* \, \frac{1-\gamma^5}{2}\,\widetilde H^* ) \,,\quad
&
U_{uq} &= \mqty( \cid\,\mat{y}_u^\dagger \, \frac{1-\gamma^5}{2}\,\widetilde H^\dagger & 0 \\
0 & \cid\,\mat{y}_u^T  \,\frac{1+\gamma^5}{2}\,\widetilde H^T ) \,. \\[25pt]
U_{qd} &= \mqty( \cid\,\mat{y}_d\,\frac{1+\gamma^5}{2} \, H  & 0 \\
0 & \cid\,\mat{y}_d^*\,\frac{1-\gamma^5}{2}\, H^{*}) \,,\quad
&
U_{dq} &= \mqty( \cid\,\mat{y}_d^\dagger\,\frac{1-\gamma^5}{2}\, H^\dagger  & 0 \\
0 & \cid\,\mat{y}_d^T\, \frac{1+\gamma^5}{2}\, H^T ) \,. \\[25pt]
U_{le} &= \mqty( \mat{y}_e\, \frac{1+\gamma^5}{2}\, H & 0 \\ 0 & \mat{y}_e^*\, \frac{1-\gamma^5}{2}\, H^{*} ) \,,\quad
&
U_{el} &= \mqty( \mat{y}_e^\dagger\,\frac{1-\gamma^5}{2}\, H^\dagger  & 0 \\ 0 & \mat{y}_e^T\,\frac{1+\gamma^5}{2}\, H^T ) \,.
\end{alignat}

~\\[-20pt]

\paragraph{Vector-vector entries}
%
%~\\[-5pt]
\begin{align}
U_{GG}^{\li{\mu}\ci{A},\,\li{\nu}\ci{B}}
&= 2\,g_3\, f^{\ci{ABC}} G^{\ci{C}\li{\mu\nu}} \,. \\[5pt]
U_{WW}^{\li{\mu}\wi{I},\,\li{\nu}\wi{J}}
&= 2\,g_2\,\epsilon^{\wi{IJK}} W^{\wi{K}\li{\mu\nu}} -\frac{g_2^2}{2}\, \eta^{\li{\mu\nu}}\, \delta^{\wi{IJ}}\, |H|^2 \,. \\[5pt]
U_{BB}^{\li{\mu},\,\li{\nu}}
&= -\frac{g_1^2}{2}\, \eta^{\li{\mu\nu}}\, |H|^2 \,.
\end{align}
\begin{equation}
U_{WB}^{\li{\mu}\wi{I},\,\li{\nu}} = -\frac{g_1g_2}{2} \,\eta^{\li{\mu\nu}} \,H^\dagger \sigma^{\wi{I}}H \,,\qquad
U_{BW}^{\li{\mu},\,\li{\nu}\wi{J}} = -\frac{g_1g_2}{2} \,\eta^{\li{\mu\nu}} H^\dagger \sigma^{\wi{J}}H \,.
\end{equation}

~\\[-20pt]

\paragraph{Higgs-fermion entries}
~\\[-5pt]
%\begin{footnotesize}
	\begin{alignat}{2}
	U_{Hq} &= 
	\begin{footnotesize}
	\begin{pmatrix}
	\wid \,\bar d \,\mat{y}_d^\dagger \,\frac{1-\gamma^5}{2} &&
	-\weps \,\bar u^\c \,\mat{y}_u^T \,\frac{1+\gamma^5}{2} \\
	-\weps \,\bar u \,\mat{y}_u^\dagger \,\frac{1-\gamma^5}{2} &&
	\wid \,\bar d^\c \,\mat{y}_d^T\,\frac{1+\gamma^5}{2}
	\end{pmatrix} 
	\end{footnotesize}
	\,,\quad
	&
	U_{qH} &=
	\begin{footnotesize}
	\begin{pmatrix}
	\mat{y}_d \,\frac{1+\gamma^5}{2}\,d\, \wid  &&
	\mat{y_u} \,\frac{1+\gamma^5}{2}\,u \,\weps\\
	\mat{y}_u^* \,\frac{1-\gamma^5}{2}\,u^{\c}\,\weps &&
	\mat{y}_d^* \,\frac{1-\gamma^5}{2} \,d^{\c}\,\wid
	\end{pmatrix} 
	\end{footnotesize}
	\,. 
	\label{eq:U_Hq} \\[15pt]
	U_{Hu} &= 
	\begin{footnotesize}
	\mqty( (\bar q \,\weps)_{\wi{\bar\alpha}} \,\mat{y}_u\,\frac{1+\gamma^5}{2} & 0 \\
	0 & (\bar q^\c \,\weps)_{\wi{\alpha}} \,\mat{y}_u^*\,\frac{1-\gamma^5}{2}) 
	\end{footnotesize}
	\,,\quad
	&
	U_{uH} &= 
	\begin{footnotesize}
	\mqty(-\mat{y}_u^\dagger\,\frac{1-\gamma^5}{2}\, (\weps\, q)_{\wi{\beta}} & 0 \\
	0 & - \mat{y}_u^T\,\frac{1+\gamma^5}{2}\, (\weps\, q^\c)_{\wi{\bar\beta}}) 
	\end{footnotesize}
	\,. \\[15pt]
	U_{Hd} &= 
	\begin{footnotesize}
	\mqty( 0 & \bar q^\c_{\wi{\bar\alpha}}\, \mat{y}_d^*\,\frac{1-\gamma^5}{2} \\
	\bar q_{\wi{\alpha}}\, \mat{y}_d \,\frac{1+\gamma^5}{2} & 0 ) 
	\end{footnotesize}
	\,,\quad
	&
	U_{dH} &= 
	\begin{footnotesize}
	\mqty( 0 & \mat{y}_d^\dagger\, \frac{1-\gamma^5}{2}\, q_{\wi{\bar\beta}} \\
	\mat{y}_d^T\, \frac{1+\gamma^5}{2}\, q^{\c}_{\wi{\beta}} & 0 ) 
	\end{footnotesize}
	\,. \\[15pt]
	U_{Hl} &= 
	\begin{footnotesize}
	\mqty( \wid \,\bar e\,\mat{y}_e^\dagger \,\frac{1-\gamma^5}{2} & 0 \\
	0 & \wid \,\bar e^\c \,\mat{y}_e^T\,\frac{1+\gamma^5}{2} ) 
	\end{footnotesize}
	\,,\quad
	&
	U_{lH} &= 
	\begin{footnotesize}
	\mqty( \mat{y}_e \,\frac{1+\gamma^5}{2}\, e \,\wid  & 0 \\
	0 & \mat{y}_e^* \,\frac{1-\gamma^5}{2} \,e^\c \,\wid ) 
	\end{footnotesize}
	\,. \\[15pt]
	U_{He} &= 
	\begin{footnotesize}
	\mqty( 0 & \bar l_{\wi{\bar\alpha}}^\c\, \mat{y}_e^*\,\frac{1-\gamma^5}{2} \\
	\bar l_{\wi{\alpha}}\,\mat{y}_e \,\frac{1+\gamma^5}{2} & 0 ) 
	\end{footnotesize}
	\,,\quad
	&
	U_{eH} &= 
	\begin{footnotesize}
	\mqty( 0 &  \mat{y}_e^\dagger\, \frac{1-\gamma^5}{2}\,l_{\wi{\bar\beta}} \\
	\mat{y}_e^T\, \frac{1+\gamma^5}{2}\, l^{\c}_\wi{\beta} & 0 )
	\end{footnotesize}
	 \,.
	\end{alignat}
%\end{footnotesize}\noindent

~\\[-20pt]
\paragraph{Higgs-vector entries}
~\\[-5pt]
\begin{alignat}{2}
U_{HW}^{\li{\nu}\wi{J}} &= \frac{ig_2}{2}\,
\mqty( - \sigma^{\wi{J}} (D^{\li{\nu}} H) \\
\sigma^{\wi{J}*}\, (D^{\li{\nu}} H)^* ) \,,\quad
&
U_{WH}^{\li{\mu}\wi{I}} &= \frac{ig_2}{2}\,
\mqty( (D^{\li{\mu}} H)^\dagger\,\sigma^{\wi{I}} \;\;&\;\;
- (D^{\li{\mu}} H)^T\, \sigma^{\wi{I}*} ) \,. \\[10pt]
Z_{HW}^{\rho\,\li{\nu}\wi{J}} &= \eta^{\rho\li{\nu}} \,\frac{g_2}{2}\,
\mqty( - \sigma^{\wi{J}} H \\
\sigma^{\wi{J}*}\, H^* ) \,,\quad
&
\bar Z_{WH}^{\rho\,\li{\mu}\wi{I}} &= \eta^{\rho\li{\mu}} \,\frac{g_2}{2}\,
\mqty( - H^\dagger\sigma^{\wi{I}} \;\;&\;\;
H^T\, \sigma^{\wi{I}*} ) \,. \label{eq:Z_HW} \\[10pt]
U_{HB}^{\li{\nu}} &= \frac{ig_1}{2}\,
\mqty( - D^{\li{\nu}} H \\
(D^{\li{\nu}} H)^{*} ) \,,\quad
&
U_{BH}^{\li{\mu}} &= \frac{ig_1}{2}\,
\mqty( (D^{\li{\mu}} H)^\dagger \;\;&\;\;
- (D^{\li{\mu}} H)^T ) \,. \\[10pt]
Z_{HB}^{\rho\,\li{\nu}} &= \eta^{\rho\li{\nu}} \,\frac{g_1}{2}\,
\mqty( - H \\ H^{*} ) \,,\quad
&
\bar Z_{BH}^{\rho\,\li{\mu}} &= \eta^{\rho\li{\mu}} \,\frac{g_1}{2}\,
\mqty( - H^\dagger \;\;&\;\; H^T ) \,. \label{eq:Z_HB}
\end{alignat}

%~\\
\paragraph{Fermion-vector entries}
~\\
\begin{alignat}{2}
U_{qG}^{\li{\nu}\ci{B}} &= \frac{g_3}{2}
\begin{pmatrix}
- \gamma^{\li{\nu}}\lambda^{\ci{B}} q \\
\gamma^{\li{\nu}}\lambda^{\ci{B}*}\, q^\c
\end{pmatrix}
,&\qquad
U_{Gq}^{\li{\mu}\ci{A}} &=
\frac{g_3}{2}
\begin{pmatrix}
-\bar q \,\gamma^{\li{\mu}}\lambda^{\ci{A}} \;\;&\;\;
\bar q^\c \,\gamma^{\li{\mu}}\lambda^{\ci{A}*}
\end{pmatrix}
\,, \\[15pt]
U_{uG}^{\li{\nu}\ci{B}} &=
\frac{g_3}{2}
\begin{pmatrix}
- \gamma^{\li{\nu}} \lambda^{\ci{B}} u \\
\gamma^{\li{\nu}} \lambda^{\ci{B}*}\, u^\c
\end{pmatrix}
,&\qquad
U_{Gu}^{\li{\mu}\ci{A}} &=
\frac{g_3}{2}
\begin{pmatrix}
- \bar u \,\gamma^{\li{\mu}}\lambda^{\ci{A}} \;\;&\;\;
\bar u^\c \,\gamma^{\li{\mu}}\lambda^{\ci{A}*}
\end{pmatrix}
\,, \\[15pt]
U_{dG}^{\li{\nu}\ci{B}} &=
\frac{g_3}{2}
\begin{pmatrix}
- \gamma^{\li{\nu}} \lambda^{\ci{B}} d \\
\gamma^{\li{\nu}} \lambda^{\ci{B}*}\, d^\c
\end{pmatrix}
,&\qquad
U_{Gd}^{\li{\mu}\ci{A}} &=
\frac{g_3}{2}
\begin{pmatrix}
- \bar d \,\gamma^{\li{\mu}}\lambda^{\ci{A}} \;\;&\;\;
\bar d^\c \,\gamma^{\li{\mu}}\lambda^{\ci{A}*}
\end{pmatrix}
\,, \\[15pt]
U_{qW}^{\li{\nu}\wi{J}} &= \frac{g_2}{2}
\begin{pmatrix}
- \gamma^{\li{\nu}}\sigma^{\wi{J}} q \\
\gamma^{\li{\nu}}\sigma^{\wi{J}*}\, q^\c
\end{pmatrix}
,&\qquad
U_{Wq}^{\li{\mu}\wi{I}} &=
\frac{g_2}{2}
\begin{pmatrix}
-\bar q \,\gamma^{\li{\mu}}\sigma^{\wi{I}} \;\;&\;\;
\bar q^\c \,\gamma^{\li{\mu}}\sigma^{\wi{I}*}
\end{pmatrix}
\,, \\[15pt]
U_{lW}^{\li{\nu}\wi{J}} &= \frac{g_2}{2}
\begin{pmatrix}
- \gamma^{\li{\nu}}\sigma^{\wi{J}} l \\
\gamma^{\li{\nu}}\sigma^{\wi{J}*}\, l^\c
\end{pmatrix}
,&\qquad
U_{Wl}^{\li{\mu}\wi{I}} &=
\frac{g_2}{2}
\begin{pmatrix}
-\bar l \,\gamma^{\li{\mu}}\sigma^{\wi{I}} \;\;&\;\;
\bar l^\c \,\gamma^{\li{\mu}}\sigma^{\wi{I}*}
\end{pmatrix}
\,, \\[15pt]
U_{qB}^{\li{\nu}} &= \frac{g_1}{6}
\begin{pmatrix}
- \gamma^{\li{\nu}} q \\
\gamma^{\li{\nu}} q^\c
\end{pmatrix}
,&\qquad
U_{Bq}^{\li{\mu}} &=
\frac{g_1}{6}
\begin{pmatrix}
-\bar q \,\gamma^{\li{\mu}} \;\;&\;\;
\bar q^\c \,\gamma^{\li{\mu}}
\end{pmatrix}
\,, \\[15pt]
U_{uB}^{\li{\nu}} &= \frac{2g_1}{3}
\begin{pmatrix}
- \gamma^{\li{\nu}} u \\
\gamma^{\li{\nu}} u^\c
\end{pmatrix}
,&\qquad
U_{Bu}^{\li{\mu}} &=
\frac{2g_1}{3}
\begin{pmatrix}
-\bar u \,\gamma^{\li{\mu}} \;\;&\;\;
\bar u^\c \,\gamma^{\li{\mu}}
\end{pmatrix}
\,, \\[15pt]
U_{dB}^{\li{\nu}} &= -\frac{g_1}{3}
\begin{pmatrix}
- \gamma^{\li{\nu}} d \\
\gamma^{\li{\nu}} d^\c
\end{pmatrix}
,&\qquad
U_{Bd}^{\li{\mu}} &=
-\frac{g_1}{3}
\begin{pmatrix}
-\bar d \,\gamma^{\li{\mu}} \;\;&\;\;
\bar d^\c \,\gamma^{\li{\mu}}
\end{pmatrix}
\,, \\[15pt]
U_{lB}^{\li{\nu}} &= -\frac{g_1}{2}
\begin{pmatrix}
- \gamma^{\li{\nu}} l \\
\gamma^{\li{\nu}} l^\c
\end{pmatrix}
,&\qquad
U_{Bl}^{\li{\mu}} &=
-\frac{g_1}{2}
\begin{pmatrix}
-\bar l \,\gamma^{\li{\mu}} \;\;&\;\;
\bar l^\c \,\gamma^{\li{\mu}}
\end{pmatrix}
\,, \\[15pt]
U_{eB}^{\li{\nu}} &= -g_1
\begin{pmatrix}
- \gamma^{\li{\nu}} e \\
\gamma^{\li{\nu}} e^\c
\end{pmatrix}
,&\qquad
U_{Be}^{\li{\mu}} &=
-g_1
\begin{pmatrix}
-\bar e \,\gamma^{\li{\mu}} \;\;&\;\;
\bar e^\c \,\gamma^{\li{\mu}}
\end{pmatrix}
\,. \label{eq:U_eB}
\end{alignat}

\vspace{4pt}

\newpage
%%%%%%%%%%%%%%%%%%%%%%%%%%%%%%%%%%%%%%%%%%%%%%%%%%%%%%%%%%%%%%%%%%%%%%%%%%%%%%%%
\subsection{Evaluated Supertraces for the Singlet Scalar Model}
\label{appsec:TraceList}
%%%%%%%%%%%%%%%%%%%%%%%%%%%%%%%%%%%%%%%%%%%%%%%%%%%%%%%%%%%%%%%%%%%%%%%%%%%%%%%%

\vspace{10pt}

In this appendix, we present formulae for the set of power-type supertraces used in \cref{sec:Singlet}.
These are derived with the CDE technique for generic $U$, $Z$, $\bar Z$, which we simply write as $U_1$, $U_2$, etc.
In Eq.~\eqref{eq:mt_4-8}, we set $Z^{\rho\li{\nu}}=\eta^{\rho\li{\nu}} Z$ and $\bar Z^{\rho\li{\mu}}=\eta^{\rho\li{\mu}} \bar Z$, which is the case we actually need in Sec.~\ref{sec:Singlet}, see Eqs.~\eqref{eq:Tr_SHVH-3}, \eqref{eq:Z_HW} and \eqref{eq:Z_HB}.
When computing the loop integrals, we use the $\overline{\text{MS}}$ scheme, with matching scale $\mu$, \ie, the renormalization scale where we connect the UV and EFT parameters.
We drop the $\frac{1}{\epsilon}$ poles and the associated finite terms accompanying the logarithms that will eventually be cancelled by the $\MSbar$ counterterms; they can be easily restored when needed. 
Each supertrace evaluates to an infinite series of effective operators, which we truncate at operator dimension six.
Results for these supertraces at higher operator dimensions, as well as for any other supertraces one could encounter in general relativistic matching calculations, can be obtained with our CDE evaluation package, to be presented in Ref.~\cite{TraceCode}.

We group the supertraces by the number of propagators in what follows.
In the propagators, $M$ is a heavy mass and $m$ is a light mass.
As in Sec.~\ref{sec:Singlet}, we put superscripts ``$[\cdot]$'' on the functionals $U$, $Z$, $\bar Z$ to indicate their minimum operator dimensions.
All covariant derivatives on the RHS are enclosed in parentheses, meaning they are closed, see the discussion below \cref{eq:Xmat_expand}.
We also use the notation $\G_{\mu\nu}\equiv -i\, [P_\mu, P_\nu]$ as in Sec.~\ref{sec:Evaluating}; they come from the propagators and always appear between the $U$, $Z$, $\bar Z$ functionals, and the gauge representation is determined by the field associated with the propagator.
In the case of the SMEFT, $\G_{\mu\nu}=g_3 G_{\mu\nu}^{\ci{A}}T^{\ci{A}} +g_2 W_{\mu\nu}^{\wi{I}} t^{\wi{I}} +g_1 B_{\mu\nu} Y$, with $T^{\ci{A}}$ and $t^{\wi{I}}$ the $\ci{SU(3)_C}$ and $\wi{SU(2)_L}$ generators in the representation that the corresponding field transforms under, and $Y$ is its $U(1)_Y$ hypercharge.
For the fundamental representation, $T^{\ci{A}} = \frac{\lambda^{\ci{A}}}{2}$ where $\lambda^{\ci{A}}$ are the Gell-Mann matrices, and $t^{\wi{I}}=\frac{\sigma^{\wi{I}}}{2}$ where $\sigma^{\wi{I}}$ are the Pauli matrices; for the adjoint representation, $(T^{\ci{A}})^{\ci{BC}} = -if^{\ci{ABC}}$ and $(t^{\wi{I}})^{\wi{JK}} = -i\epsilon^{\wi{IJK}}$ are given by the structure constants.
For a gauge singlet, \eg\ the singlet scalar field $S$, we have $\G_{\mu\nu}=0$, so \eg\ $\tr(\G_{\mu\nu} \G^{\mu\nu} U_{SH}U_{HS})=0$, since $\G_{\mu\nu}$ inherits the representation of $S$.

~\\[-5pt]

\paragraph{1-propagator supertrace}
~\\[-5pt]
\begin{flalign}
- i\, \text{STr} &\Bigl[\tfrac{1}{P^2-M^2} \,U_1^{[2]}\Bigr] \Bigr|_\text{hard}
&\notag\\[5pt]
&=\int \dd^dx\, \tfrac{1}{16\s\pi^2}\,\tr \Bigl[
M^2 \bigl( 1-\log\tfrac{M^2}{\mu^2}\bigr) \, U_1
+\tfrac{1}{12M^2}\, \G_{\mu\nu}\G^{\mu\nu}\, U_1 \Bigr]
\,.&\label{eq:mt_1}
\end{flalign}
%
%\vspace{20pt}
\newpage
\paragraph{2-propagator supertraces}
~\\[-5pt]
\begin{flalign}
-i\, \text{STr} &\Bigl[\tfrac{1}{P^2-M^2} \, U_1^{[2]} \tfrac{1}{P^2-M^2}\, U_2^{[2]}\Bigr] \Bigr|_\text{hard}
&\notag\\[5pt]
&=\int \dd^dx\, \tfrac{1}{16\s\pi^2}\,\tr \Bigl[ \bigl(-\log\tfrac{M^2}{\mu^2}\bigr)\, U_1 U_2 -\tfrac{1}{6M^2}\, (D^2 U_1)U_2 \Bigr]
\,,&\label{eq:mt_2-1}\\[15pt]
%%%
%%%
-i\, \text{STr} &\Bigl[\tfrac{1}{P^2-M^2} \, U_1^{[1]} \tfrac{1}{P^2-m^2}\, U_2^{[1]}\Bigr] \Bigr|_\text{hard}
&\notag\\[5pt]
&=\int \dd^dx\, \tfrac{1}{16\s\pi^2}\, \tr \bigg\{ \Bigl(1+\tfrac{m^2}{M^2} +\tfrac{m^4}{M^4}\Bigr)\bigl(1-\log\tfrac{M^2}{\mu^2}\bigr)\, U_1 U_2
&\notag\\
&\hspace{90pt}
+\Bigl[\tfrac{1}{2M^2} +\tfrac{m^2}{M^4}\bigl(\tfrac{5}{2}-\log\tfrac{M^2}{\mu^2}\bigr) \Bigr]\, (D_\mu U_1) (D^\mu U_2)
&\notag\\[5pt]
&\hspace{90pt}
+\tfrac{1}{6M^4}\, (D^2 U_1) (D^2 U_2)
+\tfrac{i}{3M^4}\, \G^{\mu\nu} (D_\mu U_1) (D_\nu U_2)
&\notag\\[5pt]
&\hspace{90pt}
-\tfrac{7i}{18M^4}\, (D_\nu \G^{\mu\nu}) U_1 (D_\mu U_2)
&\notag\\[5pt]
&\hspace{90pt}
-\tfrac{i}{3M^4} \bigl(\tfrac{17}{6}-\log\tfrac{M^2}{\mu^2}\bigr)\, U_1 (D_\nu \G^{\mu\nu}) (D_\mu U_2)
&\notag\\
&\hspace{90pt}
-\tfrac{1}{4M^4}\, \G_{\mu\nu} \G^{\mu\nu} U_1 U_2
+\tfrac{1}{3M^4}\, U_1\,\G_{\mu\nu} \G^{\mu\nu} U_2
\bigg\}
\,.&\label{eq:mt_2-2}
\end{flalign}

\vspace{2pt}

\paragraph{3-propagator supertraces}
~\\[-5pt]
\begin{flalign}
-i\, \text{STr} &\Bigl[\tfrac{1}{P^2-M^2} \, U_1^{[2]} \tfrac{1}{P^2-M^2}\, U_2^{[2]} \tfrac{1}{P^2-M^2}\, U_3^{[2]}\Bigr] \Bigr|_\text{hard}
&\notag\\[5pt]
&= \int \dd^dx\, \tfrac{1}{16\s\pi^2}\, \tr \Bigl[ -\tfrac{1}{2M^2}\, U_1 U_2 U_3 \Bigr]
\,,&\label{eq:mt_3-1}\\[15pt]
%%%
%%%
-i\, \text{STr} &\Bigl[\tfrac{1}{P^2-M^2} \, U_1^{[2]} \tfrac{1}{P^2-M^2}\, U_2^{[1]} \tfrac{1}{P^2-m^2}\, U_3^{[1]}\Bigr] \Bigr|_\text{hard}
&\notag\\[5pt]
&=\int \dd^dx\, \tfrac{1}{16\s\pi^2}\, \tr \bigg\{ -\tfrac{1}{M^2}\Bigl[ 1+\tfrac{m^2}{M^2}\bigl(2-\log\tfrac{M^2}{\mu^2}\bigr)\Bigr]\, U_1 U_2 U_3
&\notag\\
&\hspace{90pt}
-\tfrac{1}{2M^4}\, U_1 (D^\mu U_2) (D_\mu U_3) + \tfrac{1}{3M^4}\, (D^2 U_1) U_2 U_3 \bigg\}
\,,&\label{eq:mt_3-2}\\[15pt]
%%%
%%%
-i\, \text{STr} &\Bigl[\tfrac{1}{P^2-M^2} \, U_1^{[1]} \tfrac{1}{P^2-m^2}\, U_2^{[2]} \tfrac{1}{P^2-m^2}\, U_3^{[1]}\Bigr] \Bigr|_\text{hard}
&\notag\\[5pt]
&=\int \dd^dx\, \tfrac{1}{16\s\pi^2}\, \tr \bigg\{ \tfrac{1}{M^2} \bigl(1+\tfrac{2m^2}{M^2}\bigr) \bigl(1-\log\tfrac{M^2}{\mu^2}\bigr)\, U_1\,U_2\,U_3 -\tfrac{1}{2M^4}\, U_1 (D^2 U_2) U_3
&\notag\\
&\hspace{90pt}
+\tfrac{1}{M^4}\bigl(\tfrac{5}{2}-\log\tfrac{M^2}{\mu^2}\bigr)\, (D_\mu U_1) U_2 (D^\mu U_3) \bigg\}
\,.&\label{eq:mt_3-3}
\end{flalign}

%\vspace{10pt}

\paragraph{4-propagator supertraces}
~\\[-10pt]
\begin{flalign}
-i\, \text{STr} &\Bigl[\tfrac{1}{P^2-M^2} \, U_1^{[2]} \tfrac{1}{P^2-M^2}\, U_2^{[2]} \tfrac{1}{P^2-M^2}\, U_3^{[1]} \tfrac{1}{P^2-m^2}\, U_4^{[1]}\Bigr] \Bigr|_\text{hard}
&\notag\\[5pt]
&=\int\dd^dx\, \tfrac{1}{16\s\pi^2}\, \tr \Big[ \tfrac{1}{2M^4}\, U_1\, U_2\, U_3\, U_4 \Big]
\,,&\label{eq:mt_4-1}\\[12pt]
%%%
%%%
-i\, \text{STr} &\Bigl[\tfrac{1}{P^2-M^2} \, U_1^{[2]} \tfrac{1}{P^2-M^2}\, U_2^{[1]} \tfrac{1}{P^2-m^2}\, U_3^{[2]} \tfrac{1}{P^2-m^2}\, U_4^{[1]}\Bigr] \Bigr|_\text{hard}
&\notag\\[5pt]
&=\int\dd^dx\, \tfrac{1}{16\s\pi^2}\, \tr \Big[ -\tfrac{1}{M^4} \bigl(2-\log\tfrac{M^2}{\mu^2}\bigr)\, U_1\, U_2\, U_3\, U_4 \Big]
\,,&\label{eq:mt_4-2}\\[12pt]
%%%
%%%
-i\, \text{STr} &\Bigl[\tfrac{1}{P^2-M^2} \, U_1^{[1]} \tfrac{1}{P^2-m^2}\, U_2^{[1]} \tfrac{1}{P^2-M^2}\, U_3^{[1]} \tfrac{1}{P^2-m^2}\, U_4^{[1]}\Bigr] \Bigr|_\text{hard}
&\notag\\[5pt]
&=\int\dd^dx\, \tfrac{1}{16\s\pi^2}\, \tr \bigg\{ -\Bigl[\tfrac{1}{M^4}\bigl(2-\log\tfrac{M^2}{\mu^2}\bigr) +\tfrac{4m^2}{M^6}\bigl(\tfrac{3}{2}-\,\log\tfrac{M^2}{\mu^2}\bigr)\Bigr]\, U_1\, U_2\, U_3\, U_4
&\notag\\[2pt]
&\hspace{90pt}
+\tfrac{1}{M^6}\bigl(\tfrac{8}{3}-\log\tfrac{M^2}{\mu^2}\bigr)\, (D_\mu U_1) (D^\mu U_2) \, U_3\, U_4
&\notag\\[5pt]
&\hspace{90pt}
+\tfrac{1}{M^6}\bigl( 3-\log\tfrac{M^2}{\mu^2}\bigr)\, (D_\mu U_1) \,U_2 \,  (D^\mu U_3)\, U_4
&\notag\\[5pt]
&\hspace{90pt}
-\tfrac{1}{M^6} \bigl( 1-\log\tfrac{M^2}{\mu^2}\bigr)\, U_1 \,(D_\mu U_2) (D^\mu U_3)\, U_4
&\notag\\[5pt]
&\hspace{90pt}
+\tfrac{1}{M^6}\bigl(\tfrac{17}{6}-\log\tfrac{M^2}{\mu^2}\bigr)\, (D^2 U_1)\, U_2\, U_3\, U_4
&\notag\\
&\hspace{90pt}
+\tfrac{5}{6M^6}\, U_1\, (D^2 U_2)\, U_3\, U_4
+\tfrac{1}{M^6}\, U_1\, U_2\, (D^2 U_3)\, U_4
\bigg\}
\,,&\label{eq:mt_4-3}\\[12pt]
%%%
%%%
-i\, \text{STr} &\Bigl[\tfrac{1}{P^2-M^2} \, U_1^{[1]} \tfrac{1}{P^2-m^2}\, U_2^{[2]} \tfrac{1}{P^2-m^2}\, U_3^{[2]} \tfrac{1}{P^2-m^2}\, U_4^{[1]}\Bigr] \Bigr|_\text{hard}
&\notag\\[5pt]
&=\int\dd^dx\, \tfrac{1}{16\s\pi^2}\, \tr \Big[ \tfrac{1}{M^4}\bigl(1-\log\tfrac{M^2}{\mu^2}\bigr)\, U_1\, U_2\, U_3\, U_4 \Big]
\,,&\label{eq:mt_4-4}\\[12pt]
%%%
%%%
-i\, \text{STr} &\Bigl[\tfrac{1}{P^2-M^2} \, U_1^{[1]} \tfrac{1}{P^2-m^2}\, U_2^{[3/2]} \tfrac{1}{\Psl}\, U_3^{[3/2]} \tfrac{1}{P^2-m^2}\, U_4^{[1]}\Bigr] \Bigr|_\text{hard}
&\notag\\[5pt]
&=\int\dd^dx\, \tfrac{1}{16\s\pi^2}\, \tr \bigg\{
\tfrac{i}{2M^4}\bigl(\tfrac{3}{2}-\log\tfrac{M^2}{\mu^2}\bigr)\, (D_\mu U_1)\, U_2\, \gamma^\mu\, U_3\, U_4
&\notag\\
&\hspace{90pt}
-\tfrac{i}{2M^4} \, U_1\, U_2\, \gamma^\mu\, U_3\, (D_\mu U_4)
&\notag\\
&\hspace{90pt}
+\tfrac{i}{2M^4}\bigl(\tfrac{1}{2}-\log\tfrac{M^2}{\mu^2}\bigr)\, U_1\, U_2\, \gamma^\mu\, (D_\mu U_3)\, U_4
\bigg\}
\,,&\label{eq:mt_4-5}\\[12pt]
%%%
%%%
-i\, \text{STr} &\Bigl[\tfrac{1}{P^2-M^2} \, U_1^{[1]} \tfrac{1}{P^2-m^2}\, U_2^{[2]} \tfrac{1}{P^2}\, U_3^{[2]} \tfrac{1}{P^2-m^2}\, U_4^{[1]}\Bigr] \Bigr|_\text{hard}
&\notag\\[5pt]
&=\int\dd^dx\, \tfrac{1}{16\s\pi^2}\, \tr \Big[ \tfrac{1}{M^4}\bigl(1-\log\tfrac{M^2}{\mu^2}\bigr)\, U_1\, U_2\, U_3\, U_4 \Big]
\,,&\label{eq:mt_4-6}\\[12pt]
%%%
%%%
-i\, \text{STr} &\Bigl[\tfrac{1}{P^2-M^2} \, U_1^{[1]} \tfrac{1}{P^2-m^2}\, P_\mu \,Z_2^{\mu[1]} \tfrac{1}{P^2}\, U_3^{[2]}  \tfrac{1}{P^2-m^2}\, U_4^{[1]}\Bigr] \Bigr|_\text{hard}
&\notag\\[5pt]
&=\int\dd^dx\, \tfrac{1}{16\s\pi^2}\, \tr \bigg\{ \tfrac{i}{2M^4} \bigl(\tfrac{3}{2}-\log\tfrac{M^2}{\mu^2}\bigr) \Bigl[ U_1\, (D_\mu Z_2^\mu)\, U_3\, U_4 - U_1\, Z_2^\mu\, U_3\, (D_\mu U_4) \Bigr]
&\notag\\
&\hspace{90pt}
+\tfrac{i}{2M^4}\, (D_\mu U_1) \,Z_2^\mu\, U_3\,  U_4 \bigg\}
\,,&\label{eq:mt_4-7}\\[12pt]
%%%
%%%
-i\, \text{STr} &\Bigl[\tfrac{1}{P^2-M^2} \, U_1^{[1]} \tfrac{1}{P^2-m^2}\, P_\mu\, Z_2^{[1]} \tfrac{1}{P^2}\, \bar Z_3^{[1]} P^\mu\, \tfrac{1}{P^2-m^2}\, U_4^{[1]}\Bigr] \Bigr|_\text{hard}
&\notag\\[5pt]
&=\int\dd^dx\, \tfrac{1}{16\s\pi^2}\, \tr \bigg\{ \tfrac{1}{M^2} \left(1 + \tfrac{2m^2}{M^2} \right) \left(1 - \log\tfrac{M^2}{\mu^2}\right) U_1\, Z_2\, \bar Z_3\, U_4
&\notag\\[2pt]
&\hspace{90pt}
-\tfrac{1}{2M^4} \bigl(\tfrac{1}{2}-\log\tfrac{M^2}{\mu^2}\bigr) \left(D_\mu U_1\right) \left(D^\mu Z_2\right) \bar Z_3\, U_4
&\notag\\[5pt]
&\hspace{90pt}
-\tfrac{1}{2M^4} \bigl(\tfrac{1}{2}-\log\tfrac{M^2}{\mu^2}\bigr)\, U_1\, Z_2 \left(D_\mu \bar Z_3\right) \left(D^\mu U_4\right)
&\notag\\[5pt]
&\hspace{90pt}
+\tfrac{1}{M^4} \Big[ \left(D_\mu U_1\right) Z_2 \left(D^\mu \bar Z_3\right) U_4 + U_1 \left(D_\mu Z_2\right) \bar Z_3 \left(D^\mu U_4\right) \Big]
&\notag\\[5pt]
&\hspace{90pt}
+\tfrac{1}{2M^4}\, U_1 \left(D_\mu Z_2\right) \left(D^\mu \bar Z_3\right) U_4
&\notag\\
&\hspace{90pt}
+\tfrac{1}{M^4} \bigl(\tfrac{5}{2}-\log\tfrac{M^2}{\mu^2}\bigr) \left(D_\mu U_1\right) Z_2\, \bar Z_3 \left(D^\mu U_4\right)
\bigg\}
\,.&\label{eq:mt_4-8}
\end{flalign}

%\vspace{10pt}

\paragraph{5-propagator supertraces}
~\\[-10pt]
\begin{flalign}
-i\, \text{STr} &\Bigl[\tfrac{1}{P^2-M^2} \, U_1^{[2]} \tfrac{1}{P^2-M^2}\, U_2^{[1]} \tfrac{1}{P^2-m^2}\, U_3^{[1]} \tfrac{1}{P^2-M^2}\, U_4^{[1]} \tfrac{1}{P^2-m^2}\, U_5^{[1]}\Bigr] \Bigr|_\text{hard}
&\notag\\[5pt]
&=\int \dd^dx\, \tfrac{1}{16\s\pi^2}\, \tr \Big[ \tfrac{1}{M^6}\bigl( \tfrac{5}{2} -\log\tfrac{M^2}{\mu^2}\bigr)\, U_1\, U_2\, U_3\, U_4\, U_5 \Big]
\,,&\label{eq:mt_5-1}\\[12pt]
%%%
%%%
-i\, \text{STr} &\Bigl[\tfrac{1}{P^2-M^2} \, U_1^{[1]} \tfrac{1}{P^2-m^2}\, U_2^{[1]} \tfrac{1}{P^2-M^2}\, U_3^{[1]} \tfrac{1}{P^2-m^2}\, U_4^{[2]} \tfrac{1}{P^2-m^2}\, U_5^{[1]}\Bigr] \Bigr|_\text{hard}
&\notag\\[5pt]
&=\int \dd^dx\, \tfrac{1}{16\s\pi^2}\, \tr \Big[ -\tfrac{1}{M^6}\bigl( 3 -2\,\log\tfrac{M^2}{\mu^2}\bigr)\, U_1\, U_2\, U_3\, U_4\, U_5 \Big]
\,,&\label{eq:mt_5-2}\\[12pt]
%%%
%%%
-i\, \text{STr} &\Bigl[\tfrac{1}{P^2-M^2} \, U_1^{[1]} \tfrac{1}{P^2-m^2}\, U_2^{[3/2]} \tfrac{1}{\Psl}\, U_3^{[1]} \tfrac{1}{\Psl}\, U_4^{[3/2]} \tfrac{1}{P^2-m^2}\, U_5^{[1]}\Bigr] \Bigr|_\text{hard}
&\notag\\[5pt]
&=\int \dd^dx\, \tfrac{1}{16\s\pi^2}\, \tr \Big[ \tfrac{1}{4M^4}\bigl(\tfrac{3}{2}-\log\tfrac{M^2}{\mu^2}\bigr)\, U_1\, U_2\, \gamma^\mu\, U_3\, \gamma_\mu\, U_4\, U_5 \Big]
\,.&\label{eq:mt_5-3}
\end{flalign}

%\vspace{10pt}

\paragraph{6-propagator supertrace}
~\\[-10pt]
\begin{flalign}
-i\, \text{STr} &\Bigl[\tfrac{1}{P^2-M^2} \, U_1^{[1]} \tfrac{1}{P^2-m^2}\, U_2^{[1]} \tfrac{1}{P^2-M^2} \, U_3^{[1]} \tfrac{1}{P^2-m^2}\, U_4^{[1]} \tfrac{1}{P^2-M^2} \, U_5^{[1]} \tfrac{1}{P^2-m^2}\, U_6^{[1]}\Bigr] \Bigr|_\text{hard}
&\notag\\[5pt]
&=\int \dd^dx\, \tfrac{1}{16\s\pi^2}\, \tr\Big[ \tfrac{1}{M^8}\bigl(\tfrac{11}{2} -3\,\log\tfrac{M^2}{\mu^2}\bigr)\, U_1\, U_2\, U_3\, U_4\, U_5\, U_6 \Big]
\,.&\label{eq:mt_6}
\end{flalign}

%\vspace{10pt}

\newpage
%%%%%%%%%%%%%%%%%%%%%%%%%%%%%%%
\subsection{Summary of Results for the Singlet Scalar Model}
\label{sec:SingletSummary}
%%%%%%%%%%%%%%%%%%%%%%%%%%%%%%%

The SMEFT operators that we obtained in Eqs.~\eqref{eq:Tr_S_evaluate}-\eqref{eq:Tr_SHSHSH_evaluate} are not in any specific basis.
They are the raw output of functional matching.
Of course, one is free to post-process these results into a non-redundant complete basis.
We will not do so here, since this is not part of ``matching'' and the procedure is already well established, see \eg\ Refs.~\cite{Grzadkowski:2010es,Gripaios:2018zrz,Gherardi:2020det}.
Rather, the idea is to present the EFT Lagrangian in a form that reflects its origin from a pure-UV calculation as much as possible, without further field redefinitions within the EFT.
Nevertheless, there are some trivial simplifications that we shall implement to shorten some of the expressions in Eqs.~\eqref{eq:Tr_S_evaluate}-\eqref{eq:Tr_SHSHSH_evaluate}.
These include applying the product rule and IBP, {\it e.g.}
\begin{subequations}
	\begin{align}
	%\bigl[ H^\dagger (D_\mu H) + (D_\mu H)^\dagger H\bigr]^2 &=
	H^\dagger(D^2 H) + (D^2 H)^\dagger H &\,\,=\; (\partial^2 |H|^2) -2 |D_\mu H|^2 \,,\\[3pt]
	\bigl(\partial_\mu |H|^2\bigr)^2 &\overset{\text{IBP}}{\longrightarrow}\; -|H|^2 (\partial^2|H|^2) \,,
	\end{align}
\end{subequations}
and using group theoretic identities, {\it e.g.}
\begin{subequations}
	\begin{align}
	\epsilon_{\wi{\alpha\alpha'}} \epsilon_{\wi{\beta\beta'}} &=
	\frac{1}{2}\, \Bigl( \delta_{\wi{\alpha}\wi{\beta}} \delta_{\wi{\alpha'}\wi{\beta'}} -\sigma^{\wi{I}}_\wi{\alpha\beta} \sigma^{\wi{I}}_\wi{\alpha'\beta'} \Bigr) \,,\\[5pt]
	\delta_{\wi{\alpha}\wi{\beta'}} \delta_{\wi{\alpha'}\wi{\beta}} &= \frac{1}{2}\, \Bigl( \delta_{\wi{\alpha}\wi{\beta}} \delta_{\wi{\alpha'}\wi{\beta'}} +\sigma^{\wi{I}}_\wi{\alpha\beta} \sigma^{\wi{I}}_\wi{\alpha'\beta'} \Bigr) \,.
	\end{align}
\end{subequations}
Finally, Eq.~\eqref{eq:Tr_SHffH_evaluate} contains an $\O(\epsilon)$ terms that comes from $\gamma^\mu\gamma_\mu = (4-\epsilon)\, \bm{1}$, and we need to restore the $\frac{1}{\epsilon}$ poles to obtain the additional finite pieces, {\it e.g.}
\begin{equation}
(4-\epsilon) \biggl(\frac{3}{2}-\log\frac{\MS^2}{\mu^2}\biggr) \to
(4-\epsilon) \biggl(\frac{2}{\epsilon} + \frac{3}{2}-\log\frac{\MS^2}{\mu^2}\biggr) \to
4\, \biggl(1-\log\frac{\MS^2}{\mu^2}\biggr)\,.
\end{equation}

We summarize the results after these simplifications in Tables~\ref{tab:renormalizable}, \ref{tab:higgs} and \ref{tab:fermions}.
The EFT Lagrangian, up to one-loop level, is the sum of $\L_\text{EFT}^\text{(tree)}$, given in Eq.~\eqref{eq:L_EFT_tree_singlet}, and $\L_\text{EFT}^\text{(1-loop)}$, obtained by adding up all the entries in these tables.
For an operator labeled with ``(+h.c.)''\ it is understood that one should multiply the operator and coefficient as listed in the table and then add the Hermitian conjugate term.
Also, note that the fermion fields carry generation indices that we have left implicit, so coefficients of operators involving fermion bilinears are $3\times 3$ matrices in generation space.

A few remarks are in order regarding the renormalizable operators generated by matching. First, from Table~\ref{tab:renormalizable} we see that
\begin{equation}
\L_\text{EFT}^\text{(1-loop)} \supset \delta Z_H |D_\mu H|^2\,,
\end{equation}
with
\begin{equation}
\delta Z_H = \frac{1}{16\s\pi^2} \biggl[\frac{A^2}{2\MS^2}+\frac{A^2\mH^2}{\MS^4}\biggl(\frac{5}{2}-\log \frac{\MS^2}{\mu^2}\biggr)\biggr]\,.
\end{equation}
When calculating observables in the EFT, it is convenient to canonically normalize the kinetic terms. This is achevied by rescaling the Higgs field,
\begin{equation}
H\to \biggl(1-\frac{1}{2}\,\delta Z_H\biggr) H\,.
\label{eq:H_rescale}
\end{equation}
As a result, when written in terms of canonically normalized fields, the dimension-six SMEFT Lagrangian contains the following additional terms at one loop order:
\begin{equation}
\Delta \L_\text{EFT}^\text{(1-loop)} = \delta Z_H\biggl[
%-\frac{A^2}{\MS^2} |H|^4 +
\frac{A^2}{\MS^4} |H|^2 \bigl(\partial^2 |H|^2\bigr) +\frac{3A^2}{2\MS^4} \biggl(\kappa -\frac{\mu_S^{} A}{3\MS^2}\biggr) |H|^6 \biggr]\,,
\end{equation}
which come from applying \cref{eq:H_rescale} to the tree-level Lagrangian, \cref{eq:L_EFT_tree_singlet}.

The rescaling of $H$ also introduces additional one-loop-level contributions to the coefficients of the renormalizable operators involving the $H$ field.
These, together with the $|H|^4$ term in $\L_\text{EFT}^\text{(tree)}$ in \cref{eq:L_EFT_tree_singlet} and the $|H|^2$, $|H|^4$ terms in $\L_\text{EFT}^\text{(1-loop)}$ in \cref{tab:renormalizable}, contribute to the threshold corrections, \ie, the differences between the renormalizable EFT parameters and the corresponding UV theory parameters; see Sec.~2.1 of Ref.~\cite{Wells:2017vla} for a recent discussion in the context of functional matching.
In order to compare our results to Refs.~\cite{Jiang:2018pbd,Haisch:2020ahr}, threshold correction to the SM Higgs quartic $\lambda_H$ must be taken into account.
We have additionally kept the $\mH^2$ suppressed terms in the coefficients of the renormalizable operators in Table~\ref{tab:renormalizable}, up to the order consistent with the truncation at dimension six.

~\\

%%%
\begin{table}[h]
	\begin{center}
		\renewcommand{\arraystretch}{2.5}
		\setlength{\arrayrulewidth}{.3mm}
		\begin{small}
			\begin{tabular}{cc}
				\rowcolor[HTML]{EFEFEF}
				\hline
				{\bf \;\;Operator\;\;} & {\bf Coefficient $\times\; 16\s\pi^2$}
				\Tstrut\Bstrut\\
				\hline
				$|H|^2$ &
				$\Bigl[\frac{1}{2} \bigl(\kappa \MS^2 -\mu_S^{} A\bigr) +A^2\Bigl(1+\frac{\mH^2}{\MS^2}+\frac{\mH^4}{\MS^4}\Bigr)\Bigr] \Bigl(1-\log\tfrac{\MS^2}{\mu^2}\Bigr)$
				\Tstrut\Bstrut\\
				\hline
				&
				$\frac{\kappa^2}{4}\Bigl(-\log\tfrac{\MS^2}{\mu^2}\Bigr) +\tfrac{\mu_S^{} A}{\MS^2}\Bigl(\tfrac{\kappa}{2}-\tfrac{\mu_S^{} A}{4\MS^2} +\tfrac{A^2}{\MS^2}\Bigr)$
				\Tstrut\Bstrut\\
				$|H|^4$ & $+\tfrac{A^2}{\MS^2} \Bigl[ \Bigl(\tfrac{\lambda_S}{4}+3\lambda_H\Bigr)\Bigl(1-\log\tfrac{\MS^2}{\mu^2}\Bigr) -2\Bigl(\kappa +\tfrac{A^2}{\MS^2}\Bigr) \Bigl(\tfrac{3}{2}-\log\tfrac{\MS^2}{\mu^2}\Bigr) \Bigr]$
				\Tstrut\Bstrut\\
				& $+\tfrac{\mH^2}{\MS^2}\tfrac{A^2}{\MS^2} \Bigl[
				6\lambda_H \Bigl(1-\log\tfrac{\MS^2}{\mu^2}\Bigr) -3\Bigl(\kappa +\tfrac{2A^2}{\MS^2}\Bigr) \Bigl(\tfrac{4}{3}-\log\tfrac{\MS^2}{\mu^2}\Bigr)
				+\tfrac{\mu_S^{} A}{\MS^2} \Bigl(2-\log\tfrac{\MS^2}{\mu^2}\Bigr) \Bigr]$
				\Tstrut\Bstrut\\
				\hline
				$|D_\mu H|^2$ &
				$\frac{A^2}{2\MS^2}+\frac{A^2\mH^2}{\MS^4}\Bigl(\frac{5}{2}-\log\tfrac{\MS^2}{\mu^2}\Bigr)$
				\Tstrut\Bstrut\\
				\hline
			\end{tabular}	
		\end{small}
		\caption{\label{tab:renormalizable} {Corrections to renormalizable operators.}\Tstrut\Bstrut}
	\end{center}
\end{table}
%%%

%%%
\begin{table}[h!]
	\begin{center}
		\begin{small}
			\renewcommand{\arraystretch}{2.}
			\setlength{\arrayrulewidth}{.3mm}
			\begin{tabular}{cc}
				\rowcolor[HTML]{EFEFEF}
				\hline
				{\bf Operator} & {\bf Coefficient  $\times\; 16\s\pi^2$}
				\Tstrut\Bstrut\\
				\hline
				&
				$\frac{1}{\MS^2}\Bigl(-\tfrac{\kappa^3}{12} -\tfrac{\kappa^2\mu_S^{} A}{4\MS^2} +\tfrac{\kappa\mu_S^2 A^2}{2\MS^4} -\tfrac{\lambda_S A^4}{2\MS^4} -\tfrac{\mu_S^3 A^3}{6\MS^6} +\tfrac{\mu_S^2 A^4}{\MS^6}\Bigr)$
				\Tstrut\Bstrut\\
				& $+\tfrac{\kappa A^2}{\MS^4}\Bigl[3\kappa \Bigl(\frac{11}{6}-\log\tfrac{\MS^2}{\mu^2}\Bigr)
				-\tfrac{\lambda_S}{4} \Bigl(2-\log\tfrac{\MS^2}{\mu^2}\Bigr)\Bigr]$
				\Tstrut\Bstrut\\
				\multirow{2}{*}{$|H|^6$\Tstrut}
				& $+\tfrac{9\lambda_H A^2}{\MS^4}\Bigl[-\kappa \Bigl(\tfrac{4}{3}-\log\tfrac{\MS^2}{\mu^2}\Bigr)
				+\lambda_H\Bigl(1-\log\tfrac{\MS^2}{\mu^2}\Bigr)\Bigr]$
				\Tstrut\Bstrut\\
				& $+\tfrac{\mu_S^{} A^3}{\MS^6} \Bigl[-\kappa\Bigl(5-\log\tfrac{\MS^2}{\mu^2}\Bigr) +\tfrac{\lambda_S}{12} \Bigl(4-\log\tfrac{\MS^2}{\mu^2}\Bigr) +3\lambda_H \Bigl(2-\log\tfrac{\MS^2}{\mu^2}\Bigr)\Bigr]$
				\Tstrut\Bstrut\\
				& $+\tfrac{A^4}{\MS^6} \Bigl[\tfrac{21\kappa}{2}\Bigl(\frac{37}{21}-\log\tfrac{\MS^2}{\mu^2}\Bigr)
				-18\lambda_H \Bigl(\frac{4}{3}-\log\tfrac{\MS^2}{\mu^2}\Bigr) \Bigr]$
				\Tstrut\Bstrut\\
				& $-\tfrac{7\mu_S^{} A^5}{2\MS^8} \Bigl(\frac{15}{7}-\log\tfrac{\MS^2}{\mu^2}\Bigr) +\tfrac{9A^6}{\MS^8} \Bigl(\frac{43}{27}-\log\tfrac{\MS^2}{\mu^2}\Bigr)$
				\Tstrut\Bstrut\\
				\hline
				&
				$-\tfrac{\kappa^2}{24\MS^2}-\tfrac{5\kappa\mu_S^{} A}{12\MS^4}$
				\Tstrut\Bstrut\\
				$|H|^2 \bigl(\partial^2|H|^2\bigr)$ & $+\tfrac{A^2}{\MS^4} \Bigl[ 2\kappa\Bigl(\frac{17}{12}-\log\tfrac{\MS^2}{\mu^2}\Bigr)
				-\tfrac{\lambda_S}{2}\Bigl(1-\log\tfrac{\MS^2}{\mu^2}\Bigr)
				-\tfrac{\lambda_H}{2}\Bigl(\frac{9}{2}-\log\tfrac{\MS^2}{\mu^2}\Bigr)\Bigr]$
				\Tstrut\Bstrut\\
				& $+\tfrac{11\mu_S^2 A^2}{24\MS^6} -\tfrac{4\mu_S^{} A^3}{3\MS^6} +\tfrac{3A^4}{2\MS^6}\Bigl(\frac{20}{9}-\log\tfrac{\MS^2}{\mu^2}\Bigr) -\frac{3g_2^2 A^2}{8\MS^4}\Bigl(\frac{5}{6}-\log\tfrac{\MS^2}{\mu^2}\Bigr)$
				\Tstrut\Bstrut\\
				\hline
				$|H|^2|D_\mu H|^2$ &
				$\tfrac{A^2}{\MS^4} \Bigl[\Bigl(\lambda_H -\tfrac{A^2}{\MS^2}\Bigr)\Bigl(\frac{9}{2}-\log\tfrac{\MS^2}{\mu^2}\Bigr) -\tfrac{3\kappa}{2} +\tfrac{\mu_S^{} A}{2\MS^2}\Bigr] -\frac{3g_2^2 A^2}{2\MS^4}\Bigl(\frac{5}{6}-\log\tfrac{\MS^2}{\mu^2}\Bigr)$
				\Tstrut\Bstrut\\
				\hline
				$\frac{1}{2} \bigl(H^\dagger\overleftrightarrow{D}^\mu H\bigr)^2$ &
				$\frac{3g_1^2 A^2}{4\MS^4}\Bigl(\frac{5}{6}-\log\tfrac{\MS^2}{\mu^2}\Bigr)$ \Tstrut\Bstrut
				\Tstrut\Bstrut\\
				\hline
				$|D^2 H|^2$ &
				$\frac{A^2}{6\MS^4}$
				\Tstrut\Bstrut\\		
				\hline\\[10pt]
			\end{tabular}
			\begin{tabular}{cc}
				\rowcolor[HTML]{EFEFEF}
				\hline
				{\bf Operator} & {\bf Coefficient  $\times\; 16\s\pi^2$}
				\Tstrut\Bstrut\\
				\hline
				$ig_2(D^\mu H)^\dagger \sigma^{\wi{I}}(D^\nu H) \,W_{\mu\nu}^{\wi{I}}$ &
				$-\frac{A^2}{12\MS^4}$
				\Tstrut\Bstrut\\
				\hline
				$ig_1(D^\mu H)^\dagger (D^\nu H) \,B_{\mu\nu}$ &
				$-\frac{A^2}{12\MS^4}$
				\Tstrut\Bstrut\\
				\hline
				$\frac{ig_2}{2}\bigl(H^\dagger \sigma^{\wi{I}}\overleftrightarrow{D}^\mu H\bigr) (D^\nu W_{\mu\nu}^{\wi{I}})$ &
				$-\frac{A^2}{6\MS^4}\Bigl(\frac{7}{3}-\log\tfrac{\MS^2}{\mu^2}\Bigr)$
				\Tstrut\Bstrut\\
				\hline
				$\frac{ig_1}{2}\bigl(H^\dagger \overleftrightarrow{D}^\mu H\bigr) (\partial^\nu B_{\mu\nu})$ &
				$-\frac{A^2}{6\MS^4}\Bigl(\frac{7}{3}-\log\tfrac{\MS^2}{\mu^2}\Bigr)$
				\Tstrut\Bstrut\\
				\hline
				$|H|^2W_{\mu\nu}^{\wi{I}}W^{\wi{I}\mu\nu}$ &
				$\frac{g_2^2 A^2}{16\MS^4}$
				\Tstrut\Bstrut\\
				\hline
				$|H|^2B_{\mu\nu}B^{\mu\nu}$ &
				$\frac{g_1^2 A^2}{16\MS^4}$
				\Tstrut\Bstrut\\
				\hline
				$H^\dagger\sigma^{\wi{I}}H W_{\mu\nu}^{\wi{I}}B^{\mu\nu}$ &
				$\frac{g_1g_2 A^2}{8\MS^4}$
				\Tstrut\Bstrut\\
				\hline
			\end{tabular}	
		\end{small}
		\caption{\label{tab:higgs} {Dimension six bosonic operators.}\Tstrut\Bstrut}	
	\end{center}
\end{table}
%%%

%%%
\begin{table}[h!]
	\begin{center}
		\begin{small}
			\renewcommand{\arraystretch}{2.4}
			\setlength{\arrayrulewidth}{.3mm}
			\begin{tabular}{cc}
				\rowcolor[HTML]{EFEFEF}
				\hline
				{\bf Operator} & {\bf Coefficient  $\times\; 16\s\pi^2$}
				\Tstrut\Bstrut\\
				\hline
				$\bigl(H^\dagger \sigma^{\wi{I}}i \overleftrightarrow{D}_\mu H\bigr) \bigl(\bar q\sigma^{\wi{I}}\gamma^\mu q\bigr)$ &
				$\frac{A^2}{8\MS^4}\bigl(\mat{y}_u\mat{y}_u^\dagger+\mat{y}_d\mat{y}_d^\dagger\bigr)\Bigl(\frac{5}{2}-\log\tfrac{\MS^2}{\mu^2}\Bigr)$
				\Tstrut\Bstrut\\
				\hline
				$\bigl(H^\dagger \,i\overleftrightarrow{D}_\mu H\bigr) \bigl(\bar q\gamma^\mu q\bigr)$ &
				$-\frac{A^2}{8\MS^4}\bigl(\mat{y}_u\mat{y}_u^\dagger-\mat{y}_d\mat{y}_d^\dagger\bigr)\Bigl(\frac{5}{2}-\log\tfrac{\MS^2}{\mu^2}\Bigr)$
				\Tstrut\Bstrut\\
				\hline
				$\bigl(H^\dagger \,i\overleftrightarrow{D}_\mu H\bigr) \bigl(\bar u\gamma^\mu u\bigr)$ &
				$\frac{A^2}{4\MS^4}\,\mat{y}_u^\dagger\mat{y}_u\Bigl(\frac{5}{2}-\log\tfrac{\MS^2}{\mu^2}\Bigr)$
				\Tstrut\Bstrut\\
				\hline
				$\bigl(H^\dagger \,i\overleftrightarrow{D}_\mu H\bigr) \bigl(\bar d\gamma^\mu d\bigr)$ &
				$-\frac{A^2}{4\MS^4}\,\mat{y}_d^\dagger\mat{y}_d\Bigl(\frac{5}{2}-\log\tfrac{\MS^2}{\mu^2}\Bigr)$
				\Tstrut\Bstrut\\
				\hline
				$\bigl(H^\dagger \sigma^{\wi{I}} i\overleftrightarrow{D}_\mu H\bigr) \bigl(\bar l\sigma^{\wi{I}}\gamma^\mu l\bigr)$ &
				$\frac{A^2}{8\MS^4}\,\mat{y}_e\mat{y}_e^\dagger\Bigl(\frac{5}{2}-\log\tfrac{\MS^2}{\mu^2}\Bigr)$
				\Tstrut\Bstrut\\
				\hline
				$\bigl(H^\dagger \,i\overleftrightarrow{D}_\mu H\bigr) \bigl(\bar l\gamma^\mu l\bigr)$ &
				$\frac{A^2}{8\MS^4}\mat{y}_e\mat{y}_e^\dagger\Bigl(\frac{5}{2}-\log\tfrac{\MS^2}{\mu^2}\Bigr)$
				\Tstrut\Bstrut\\
				\hline
				$\bigl(H^\dagger\, i\overleftrightarrow{D}_\mu H\bigr) \bigl(\bar e\gamma^\mu e\bigr)$ &
				$-\frac{A^2}{4\MS^4}\,\mat{y}_e^\dagger\mat{y}_e\Bigl(\frac{5}{2}-\log\tfrac{\MS^2}{\mu^2}\Bigr)$
				\Tstrut\Bstrut\\
				\hline
				$i\,\bigl(\widetilde{H}^\dagger (D_\mu H)\bigr) \bigl(\bar u\gamma^\mu d\bigr)\; (+\text{h.c.})$ &
				$-\frac{A^2}{2\MS^4}\,\mat{y}_u^\dagger\mat{y}_d \Bigl(\frac{5}{2}-\log\tfrac{\MS^2}{\mu^2}\Bigr)$
				\Tstrut\Bstrut\\
				\hline
				$\bigl(H^\dagger \sigma^{\wi{I}} H\bigr) \bigl(\bar q\,\sigma^{\wi{I}}i\overleftrightarrow{\slashed{D}} q\bigr)$ &
				$-\frac{A^2}{8\MS^4}\bigl(\mat{y}_u\mat{y}_u^\dagger-\mat{y}_d\mat{y}_d^\dagger\bigr)\Bigl(\frac{1}{2}-\log\tfrac{\MS^2}{\mu^2}\Bigr)$
				\Tstrut\Bstrut\\
				\hline
				$|H|^2\bigl(\bar q\,i\overleftrightarrow{\slashed{D}} q\bigr)$ &
				$\frac{A^2}{8\MS^4}\bigl(\mat{y}_u\mat{y}_u^\dagger+\mat{y}_d\mat{y}_d^\dagger\bigr)\Bigl(\frac{1}{2}-\log\tfrac{\MS^2}{\mu^2}\Bigr)$
				\Tstrut\Bstrut\\
				\hline
				$|H|^2\bigl(\bar u\,i\overleftrightarrow{\slashed{D}} u\bigr)$ &
				$\frac{A^2}{4\MS^4}\,\mat{y}_u^\dagger\mat{y}_u\Bigl(\frac{1}{2}-\log\tfrac{\MS^2}{\mu^2}\Bigr)$
				\Tstrut\Bstrut\\
				\hline
				$|H|^2\bigl(\bar d\,i\overleftrightarrow{\slashed{D}} d\bigr)$ &
				$\frac{A^2}{4\MS^4}\,\mat{y}_d^\dagger\mat{y}_d\Bigl(\frac{1}{2}-\log\tfrac{\MS^2}{\mu^2}\Bigr)$
				\Tstrut\Bstrut\\
				\hline
				$\bigl(H^\dagger \sigma^{\wi{I}} H\bigr) \bigl(\bar l\,\sigma^{\wi{I}}i\overleftrightarrow{\slashed{D}} l\bigr)$ &
				$\frac{A^2}{8\MS^4}\,\mat{y}_e\mat{y}_e^\dagger\Bigl(\frac{1}{2}-\log\tfrac{\MS^2}{\mu^2}\Bigr)$
				\Tstrut\Bstrut\\
				\hline
				$|H|^2\bigl(\bar l\,i\overleftrightarrow{\slashed{D}} l\bigr)$ &
				$\frac{A^2}{8\MS^4}\,\mat{y}_e\mat{y}_e^\dagger\Bigl(\frac{1}{2}-\log\tfrac{\MS^2}{\mu^2}\Bigr)$
				\Tstrut\Bstrut\\
				\hline
				$|H|^2\bigl(\bar e\,i\overleftrightarrow{\slashed{D}} e\bigr)$ &
				$\frac{A^2}{4\MS^4}\,\mat{y}_e^\dagger\mat{y}_e\Bigl(\frac{1}{2}-\log\tfrac{\MS^2}{\mu^2}\Bigr)$
				\Tstrut\Bstrut\\
				\hline
				$|H|^2\,\bar q\, u\, \widetilde{H}\; (+\text{h.c.})$ &
				$\frac{A^2}{\MS^4}\,\mat{y}_u\mat{y}_u^\dagger \mat{y}_u \Bigl(1-\log\tfrac{\MS^2}{\mu^2}\Bigr)$
				\Tstrut\Bstrut\\
				\hline
				$|H|^2\,\bar q\, d\, H \;(+\text{h.c.})$ &
				$\frac{A^2}{\MS^4}\,\mat{y}_d\mat{y}_d^\dagger\mat{y}_d \Bigl(1-\log\tfrac{\MS^2}{\mu^2}\Bigr)$
				\Tstrut\Bstrut\\
				\hline
				$|H|^2\,\bar l\, e\, H \;(+\text{h.c.})$ &
				$\frac{A^2}{\MS^4}\,\mat{y}_e \mat{y}_e^\dagger \mat{y}_e \Bigl(1-\log\tfrac{\MS^2}{\mu^2}\Bigr)$
				\Tstrut\Bstrut\\
				\hline
			\end{tabular}
		\end{small}
		\caption{\label{tab:fermions} {Dimension six operators with fermions.}\Tstrut\Bstrut}	
	\end{center}
\end{table}
%%%

\clearpage

\end{spacing}

\begin{spacing}{1.09}
\addcontentsline{toc}{section}{\protect\numberline{}References}%
\bibliographystyle{utphys}
\bibliography{ref}
\end{spacing}

\end{document}